\documentclass[twocolumn,showpacs,preprintnumbers,amsmath,amssymb,prb,nofootinbib]{revtex4}

\usepackage{amsmath}
\usepackage{amsfonts}
\usepackage{mathrsfs} 
\usepackage{bbm} 
\usepackage{bm} 
\usepackage{dsfont} 

\usepackage{enumitem} 
\usepackage{dcolumn}
\usepackage{cases}  
\usepackage{mathtools} 

\usepackage{multirow} 

\usepackage{graphicx}
\usepackage[tight]{subfigure}
\usepackage{verbatim}
\usepackage{float}

\usepackage[usenames,dvipsnames]{color}

\usepackage{hyperref} 
\usepackage{xspace}   
\usepackage{etoolbox} 

\usepackage{cancel} 
\usepackage[normalem]{ulem} 
\newrobustcmd{\hl}[1]{{\textcolor{black}{#1}}}

\newrobustcmd{\one}{\mathbbm{1}}   
\newrobustcmd{\bra}[1]{\langle #1|}   
\newrobustcmd{\ket}[1]{|#1 \rangle}   
\newrobustcmd{\braket}[1]{\langle #1 \rangle}  

\newrobustcmd{\delR}{\vec{\nabla}}

\newrobustcmd{\Ket}[1]{ \pmb{|} {#1} \pmb{)} }
\newrobustcmd{\Bra}[1]{ \pmb{(} {#1} \pmb{|} }
\newrobustcmd{\Braket}[1]{{\pmb{(}  {#1}  \pmb{)}} }

\newrobustcmd{\Tr}[1]{\underset{#1}{\mathsf{Tr}}}   
\newrobustcmd{\tr}{\mathsf{Tr}}  
\newrobustcmd{\nohat}[1]{#1}   
\newrobustcmd{\sdagger}{{\dagger}}  

\newrobustcmd{\vecg}[1]{{\bm #1}}   
\renewrobustcmd{\vec}[1]{{\mathbf{#1}}}

\renewrobustcmd{\S}{\hl{\text{S}}}  
\renewrobustcmd{\i}{\hl{\text{i}}} 
\renewrobustcmd{\a}{\text{r}} 

\newrobustcmd{\T}{\hl{\mathcal{T}}} 
\renewrobustcmd{\L}{\hl{\text{L}}}  
\newrobustcmd{\R}{\hl{\text{R}}}  
\newrobustcmd{\g}{\hl{\text{g}}}  
\renewrobustcmd{\b}{\hl{\text{b}}}  
\newrobustcmd{\curve}{\hl{\mathscr{C}}}

\newrobustcmd{\W}{\hl{\bar{W}}} 
\newrobustcmd{\F}{\hl{F}} 
\newrobustcmd{\M}{\hl{M}} 

\newrobustcmd{\A}{\hl{\text{A}}} 
\newrobustcmd{\B}{\hl{\text{B}}} 
\newrobustcmd{\C}{\hl{\text{C}}} 
\newrobustcmd{\D}{\hl{\text{D}}} 

\newrobustcmd{\Ref}[1]{Ref.~\onlinecite{#1}}   
\newrobustcmd{\Refs}[1]{Refs.~\onlinecite{#1}}   
\newrobustcmd{\Tab}[1]{Table~\ref{#1}}   
\newrobustcmd{\tab}[1]{\ref{#1}}   
\newrobustcmd{\Fig}[1]{Fig.~\ref{#1}}   
\newrobustcmd{\Eq}[1]{Eq.~(\ref{#1})}   
\newrobustcmd{\Eqb}[1]{Equation~(\ref{#1})}   
\newrobustcmd{\Eqs}[1]{Eqs.~(\ref{#1})}   
\newrobustcmd{\eq}[1]{(\ref{#1})}   
\newrobustcmd{\App}[1]{App.~\ref{#1}}
\newrobustcmd{\app}[1]{\ref{#1}}   
\newrobustcmd{\Sec}[1]{Sec.~\ref{#1}}   
\renewrobustcmd{\sec}[1]{\ref{#1}}   

\robustify{\sum}
\robustify{\int}
\robustify{\nonumber}

\begin{document}

\title{
	Attractive and driven interaction in quantum dots:
	\\
	mechanisms for geometric pumping
}
\author{B. A. Placke$^{(1)}$}
\author{T. Pluecker$^{(2,3)}$}
\author{J. Splettstoesser$^{(1)}$}
\author{M. R. Wegewijs$^{(2,3,4)}$}

\affiliation{
(1) Department of Microtechnology and Nanoscience (MC2),
Chalmers University of Technology, SE-41298 G{\"o}teborg, Sweden
\\
(2) Institute for Theory of Statistical Physics,
RWTH Aachen, 52056 Aachen,  Germany
\\
(3) JARA-FIT, 52056 Aachen, Germany
\\
(4) Peter Gr{\"u}nberg Institut,
Forschungszentrum J{\"u}lich, 52425 J{\"u}lich,  Germany
}
\date{\today}
\begin{abstract}
We analyze time-dependent transport through a quantum dot with electron-electron interaction
that is statically \emph{tunable} to both repulsive and \emph{attractive} regimes,
or even dynamically \emph{driven}.
Motivated by the recent experimental realization [Hamo et. al, Nature 535, \textbf{395} (2016)] of such a system in a static double quantum dot
we compute the geometric pumping of charge in the limit of weak tunneling, high temperature and slow driving.
We analyze the pumping responses for all pairs of driving parameters
(gate voltage, bias voltage, tunnel coupling, electron-electron interaction).
We show that the responses are analytically related
when these different driving protocols are governed by the same pumping mechanism,
which is characterized by effective driving parameters that differ from the experimental ones.
For static attractive interaction we find a characteristic pumping resonance
despite the 'attractive Coulomb blockade' that hinders stationary transport.
Moreover, we identify a pumping mechanism that is unique to driving of the interaction.
Finally, although a single-dot model with orbital pseudo-spin describes most of the physics of the
mentioned experimental setup,
it is crucial to account for the additional (real-)spin degeneracies of the double dot
and the associated electron-hole symmetry breaking.
This is necessary because the pumping response is more sensitive than DC transport measurements
and detects this difference through pronounced qualitative effects.

\end{abstract}

\maketitle

\section{Introduction\label{sec:intro}}

\begin{figure}
	\vspace{1.0cm}
	\centering{\includegraphics[width=0.9\linewidth]{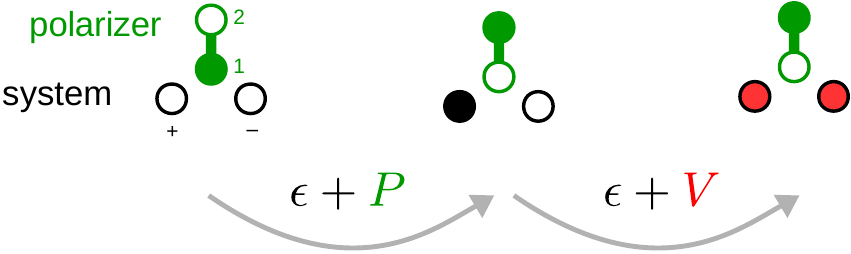}}
	\caption{
		A double-dot system (horizontal, black) with a nearby double-dot polarizer (vertical, green).
		Whereas the system can host either $0,1$ or $2$ electrons with interaction $V$,
		the polarizer contains only one electron that can be excited at an energy cost $P$.
		The system acquires effective attractive electron interaction when the addition of the second electron costs less energy than adding the first.
		This is realized here for $P > V$:
		while the first electron excites the polarizer (on top of the orbital energy $\epsilon$) the second electron only needs to overcome the lower electron repulsion energy.
	}
	\label{fig:balls}
\end{figure}

Recent experimental work has demonstrated the possibility of tuning the interaction between electrons from repulsion to attraction \emph{in situ}.
Following a top-down approach in oxide heterostructures, quantum dots have been realized in which the interaction shows a sharp repulsion-attraction crossover~\cite{Cheng15,Cheng16,Prawiroatmodjo17a} as the electron density is varied electrostatically\footnote
	{See also remarks at Fig. 11 of the supplement of \Ref{Thierschmann17a}}.
The responsible mechanism~\cite{Cheng16}
is of high interest since it is relevant to long standing issues surrounding superconductivity and magnetism in these materials and to related questions for high-$T_\textbf{c}$ superconductors~\cite{Richter13a}.
Importantly, the resulting electron pairing has been shown to occur also \emph{without} superconductivity
('preformed' electron pairs).
Other work has followed a bottom-up approach
which has the advantage that one can start from a conceptually simple mechanism
in which the tuning is well-understood.
Indeed, in \Ref{Hamo16a} the excitonic pairing mechanism~\cite{Little64} has been implemented in a carbon-nanotube double quantum dot. As sketched in \Fig{fig:balls}, an attractive nearest-neighbor interaction $U<0$
can be  generated in this system with the help of a polarizable  ``medium'' consisting of just one electron in an auxiliary nearby double quantum dot (called ``polarizer'').

In general, the tuneability of the interaction opens up new possibilities for
quantum transport through such unconventionally correlated systems,
either in the form of quantum dots~\cite{Cheng15,Cheng16,Prawiroatmodjo17a} or ballistic one-dimensional wires~\cite{Tomczyk16a}.
Early theoretical work pointed out interesting signatures of strong attractive interaction in the stationary transport through a single-level dot~\cite{Koch05c,Koch07a} with possible interesting applications~\cite{CostiZlatic10,Andergassen11b} due to a charge-Kondo effect~\cite{Taraphder91}.
Transport measurements on top-down realizations have indeed demonstrated several of these effects~\cite{Cheng15,Cheng16,Prawiroatmodjo17a}.
Much of the possibilities extend to the double-dot in the bottom-up system of \Ref{Hamo16a}
since it is formally similar to a single quantum dot with a pseudo-spin instead of a real spin.

The present paper sets out to explore the signatures of attractive interaction
in quantum dots probed by \emph{slow driving} of two parameters.
In general, this time-dependent driving leads to an additional contribution to charge transfer called pumping
which is a more sensitive experimental probe than stationary DC transport measurements.
Moreover, we focus on the simpler setting of weak-coupling, high-temperature transport.
The theoretical and experimental works cited above focused on higher-order effects that rely on moderate to strong tunnel coupling.
However, it was demonstrated in \Ref{Prawiroatmodjo17a}
that the measured signatures of attractive interaction in the high-temperature regime
are dominated by first-order effects
with an interesting crossover to the second-order dominated low-temperature regime\footnote
{See discussion of measurements and theory in Fig. 3 of \Ref{Prawiroatmodjo17a}.}.
Pumping effects relying on such first-order processes in quantum dots~\cite{Splettstoesser06,Governale08,Riwar10}
and other strongly interacting open quantum systems~\cite{Sinitsyn07EPL,Ren10}
have been analyzed in great detail,
addressing charge, spin and heat transport.
For example, \emph{qualitative} features of the pumping-response probe level degeneracies~\cite{Reckermann10a,Calvo12a}
-- in contrast to stationary DC transport--
which are different for a single-level quantum dot and for the double-dot of \Ref{Hamo16a} due to the latter's additional degrees of freedom.
Still, such pumping measurements impose only mild experimental requirements:
the driving only needs to be sufficiently fast to generate a small effect
that can be extracted experimentally by using lock-in techniques
and by exploiting its geometric nature\footnote
	{See for example App. C of \Ref{Pluecker17a} for a detailed discussion.}.
Apart from this, the driving can be slow in the sense that many electrons are transferred through the system per driving cycle.

As we show in this paper, static \emph{attractive} interaction introduces intriguing possibilities for 
a new mechanism of pumping
using first order tunnel processes which seems not to have been investigated.
In general, to achieve pumping, one might think that it is required to
have the coupling as one of the driving parameters
to ``clock'' electrons through the system.
However, this is not necessary~\cite{Brouwer98,Altshuler99}:
even with fixed coupling driving any two parameters will do in principle. In particular,  the most natural control parameters of a single-level quantum dot,
the level position (through the gate voltage) and the transport window (through the bias voltage) already result in pumping effects~\cite{Reckermann10a,Calvo12a,Pluecker17a}.
For this a nonzero static electron interaction is necessary and \emph{repulsive} interaction was shown to induce pumping~\cite{Reckermann10a,Calvo12a,Pluecker17a}
similar to earlier observations in other systems~\cite{Sinitsyn07EPL}, cf. also~\cite{Ren10}.
It is thus a natural question whether static \emph{attractive} interaction also enables such pumping for fixed coupling.

Moreover, studies of electron pumping have so far paid little attention to
time-dependent \emph{driving of the interaction} $U$ itself, arguably due to a lack of experimental motivation in electronic systems.
%
The above mentioned experimental breakthroughs
now provide a strong impetus to reconsider even basic pumping effects in the presence of freely tuneable and negative electron-electron interaction.
In particular, pumping resonances associated \emph{uniquely} with driving $U$ are of interest
since their observation provides a strong indication that one has control over the interaction
and thereby gains access to the mechanism that generates $U$.

The resulting variety of pairs of driving parameters of a quantum dot defines several experimental driving protocols.
A key result of the paper is that we map out which possible pumping mechanisms
govern the pumping responses for all these protocols.
In particular, we indeed identify mechanisms that are unique to driving the interaction,
i.e, they cannot be realized otherwise.
Because of our interest in driving the interaction, we focus in particular on the double-dot system of \Ref{Hamo16a} for which the mechanism behind the tuneability of interaction is particularly simple.
However, we will also study the single-dot system in detail as it is interesting in itself and provides a very useful guide to that more complicated double-dot problem.

The outline of the paper is as follows:
In \Sec{sec:model} we describe the single- and double-dot model.
For the latter system we review the generation of attractive interaction by the excitonic mechanism
identifying which experimental parameters can drive the interaction.
In \Sec{sec:method} we set up a master equation, transport current formulas,
and an adiabatic-response approach which are used to compute the pumping response.
We make explicit use of the geometric formulation of \Ref{Pluecker17a,Pluecker18a} by expressing the pumped charge in a curvature tensor
and
give explicit
formulas for the single- and double-dot model.
The discussion of \Sec{sec:results} focuses on the pumping response of the single-orbital quantum dot model.
\section{Quantum dot systems with attractive and tuneable interaction\label{sec:model}}

\subsection{Single quantum dot with spin}
The main focus of our study in \Sec{sec:results} is the single quantum dot model
\begin{align}
	H= \epsilon \sum_{\sigma=\pm} N_\sigma + U N_+ N_-
	\label{eq:H}
\end{align}
with the orbital energy controlled by the gate voltage $\epsilon \propto -V_\g$.
Here, $\sigma= \pm$ labels the electron spin and $N_\sigma=d^\dagger_\sigma d_\sigma$ where $d_\sigma^\dagger$ is the electron creation operator.
We are particularly interested in the situation where the interaction $U$ is negative or tuneable.
The coupling to electrodes to the left ($\alpha=\L$) and to the right ($\alpha=\R$) is described by a tunnel Hamiltonian model,
\begin{subequations}\begin{align}
	H^\text{T} &= \sum_{\alpha\sigma k} \sqrt{\Gamma^{\alpha} / \nu^{\alpha}} d_{\sigma}^\dag c_{k\alpha\sigma}
	+\text{h.c.}
	\\
	H^{\alpha} &= 
	\sum_{k \sigma}
	\omega_{\alpha k}  c_{k\alpha\sigma}^\dag c_{k\alpha\sigma}
	,
	\end{align}\label{eq:coupling1}\end{subequations}
assuming energy- and spin-independent tunnel rates $\Gamma^{\alpha}$ with constant DOS $\nu^{\alpha}$ and electron operators $c_{k\alpha\sigma}$ in electrode $\alpha$.
The time-dependent particle current\footnote
	{Since below we consider period-averaged pumping transport, screening currents need not be discussed due to the invariance of charge measurements under recalibration of the meter, see \Ref{Calvo12a,Pluecker17a}.}
is defined to flow out of reservoir $\alpha=\L, \R$:
\begin{align}
	I_{N^{\alpha}}(t)
	:=
	- \tfrac{d}{dt} \braket{ N^{\alpha} } (t),
	\quad 
	N^{\alpha}
	:=\sum_{\sigma k} c_{k\alpha\sigma}^\dag c_{k\alpha\sigma}
	\label{eq:Idef}
	,
\end{align}
where $N^{\alpha}$ is the charge in reservoir $\alpha$.
We assume a symmetrically applied bias,
entering through the electrochemical potentials of the electrodes
\begin{align}
\mu^\L = \mu + \tfrac{1}{2} V_\b
,\quad
\mu^\R = \mu - \tfrac{1}{2} V_\b
,
\end{align}
each of which is in a grand-canonical equilibrium state with temperature $T$.
Positive source-drain bias $V_\b>0$ drives a DC particle current $\L \to \R$.
Pumping is achieved by driving any pair of the full set of parameters $\epsilon$, $U$, $V_\b$, and $\Gamma^\L$ or $\Gamma^\R$, leading to the variety of pumping responses discussed in \Sec{sec:results}.
\begin{figure*}
	\includegraphics[width=0.9\linewidth]{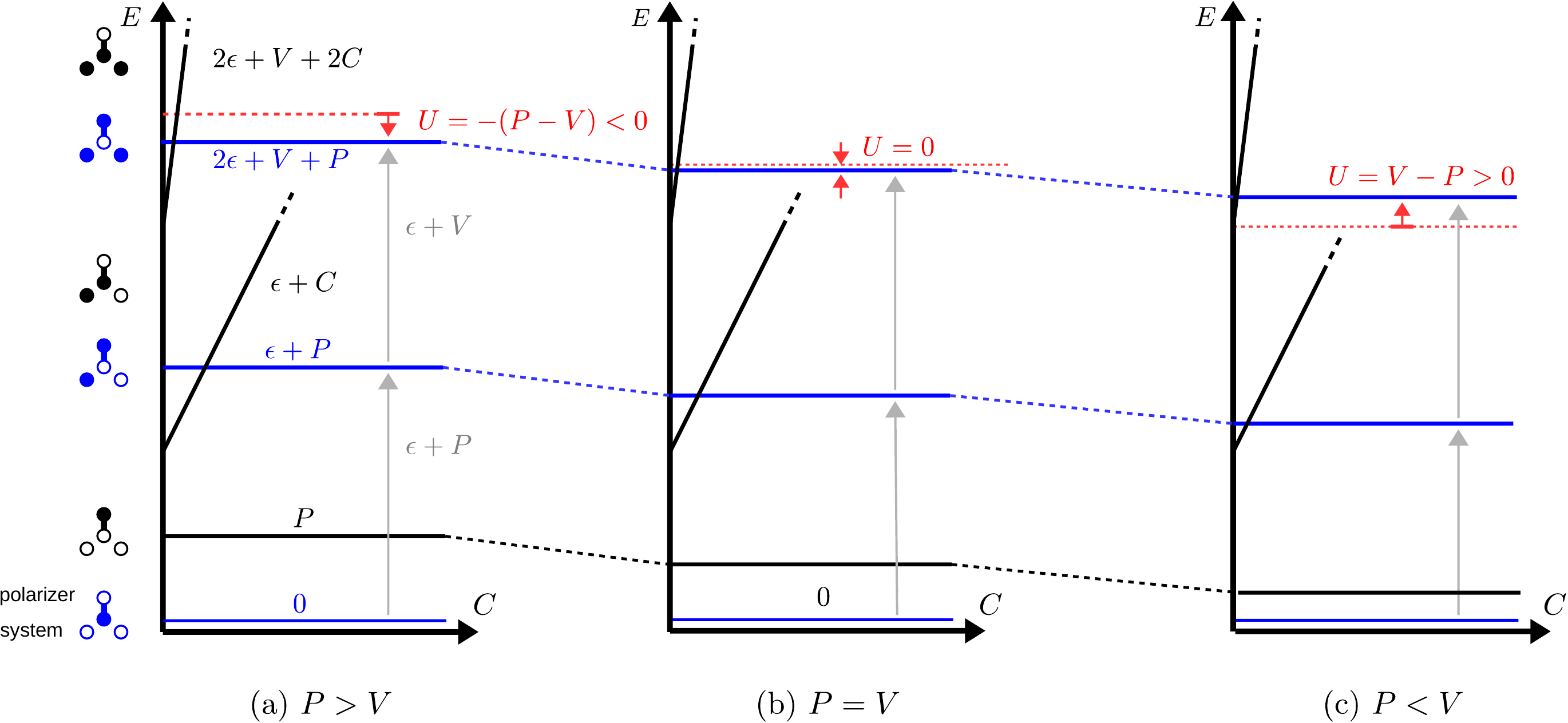}
	\caption{
		Many-body energies of system plus polarizer, i.e., eigenvalues of $H^\text{SP}$ [\Eq{eq:HSP3}] 
		as function of the capacitive-energy gradient $C=W_1 - W_2$
		for different polarization energies $P$ in (a)-(c).
		For each charge multiplet of the system (pairs of blue and black lines)
		there are two configurations of the polarizer as sketched on the left,
		where configurations with one electron in the right dot of the system are not drawn for simplicity.
		The blue configurations are accounted for in the effective Hamiltonian \eq{eq:H} $=$ \Eq{eq:H2}
		and the gray arrows denote the electron-addition energies that were indicated in \Fig{fig:balls}.
		The red dashed line indicates twice the single electron energy $\epsilon+P$.
		Relative to this, the energy of the 2-electron state (uppermost blue line) shows whether the interaction (red arrow) is attractive (a), zero (b), or repulsive (c).
		The sketch shows that by tuning $P$ with respect to $V$ one inverts the effective interaction $U=V-P$
		without a crossing of \emph{energies} as long as $C \gg P$.
		This should not be confused with the crossing of discrete \emph{electrochemical potentials} which does take place in (b) where
		$\epsilon+P=\epsilon+V$.
		For the present work we assume $C$ to take values at the horizontal position of the gray arrows.
	}
	\label{fig:polarizer}
\end{figure*}

\subsection{Double dot with tuneable interaction}

\paragraph{Orbital pseudo-spin.}
The experimental setup in \Ref{Hamo16a},
consisting of a double dot influenced by a polarizer,
can be described by a very similar model.
Let us first focus on the double dot, only (the system).
We assume that this double dot  has a (infinite) dominant intradot repulsion,
such that each system dot is constrained to be at most singly occupied,
$N_\sigma=N_{\sigma \uparrow}+N_{\sigma\downarrow} \leq 1$.
Here, we label the two dots $\sigma=\pm$ and denote their occupations by
\begin{align}
	N_\sigma=\sum_{\tau} d^\dagger_{\sigma\tau} d_{\sigma\tau}
	\label{eq:Nsigma2}
	.
\end{align}
This expression includes a sum over the real spin $\tau=\uparrow,\downarrow$
and denotes the electron operators of dot $\sigma$ by $d_{\sigma\tau}$.
The infinite intradot repulsion implies the constraint\footnote{This breaks the electron-hole symmetry in the double-dot system.}
$N_\sigma^2=N_\sigma + 2 N_{\sigma\uparrow}N_{\sigma\downarrow}=N_\sigma$. 
With this understood, the same model \eq{eq:H} that describes the single dot also describes the double dot, assuming as in the experiment that there is no interdot tunneling (only interdot capacitive coupling).
Thus, the only difference to the single dot is that the charge operators are replaced by \Eq{eq:Nsigma2}, and the coupling Hamiltonian \eq{eq:coupling1} requires a corresponding adjustment (see below).
It is important to note that in the mapping between the two models
the \emph{orbital} index (\emph{not} the real spin) of the double dot
plays the role of the \emph{spin} in the single dot,
which are therefore both labeled by $\sigma=\pm$.

\paragraph{Excitonic mechanism.}

Without the polarizer, the interdot interaction described by \Eq{eq:H} is repulsive, $U>0$.
We now review how due to the presence of the polarizer a new tuneable effective interaction $U$ is obtained,
following the supplement of \Ref{Hamo16a}.
To this end, we start from a model of the system plus polarizer as in \Fig{fig:balls}:
\begin{subequations}\begin{align}
	H^\text{SP} 
	& 
	=
	\varepsilon (N_{+} +N_{-}) + V N_{+} N_{-}
	\label{eq:HSP1a}
	\\
	&
	- \tfrac{1}{2} P (N_1 -N_2)
	\label{eq:HSP1b}
	\\&
	+
	W_1(N_{+} +N_{-}) N_1
	+
	W_2(N_{+} +N_{-}) N_2
	\label{eq:HSP1c}
	,
\end{align}\label{eq:HSP1}\end{subequations}
The term \eq{eq:HSP1a} describes the double dot with
occupations $N_\pm$ given by \Eq{eq:Nsigma2}
and a 'bare' interdot repulsion $V>0$.
The next term \eq{eq:HSP1b} describes the polarizer dots with occupations $N_1$ and $N_2$,
and \Eq{eq:HSP1c} describes the repulsion between electrons on the system and the polarizer.
The two dots of the polarizer together contain just one electron, $N_1+N_2=1$.
Although they are coupled by weak tunneling (relative to their detuning $P$)
the effect of this coupling on the polarizer spectrum is not relevant for the present discussion and it can be left out.
Moreover, the polarizer's energy difference is tuned to $P>0$ such that its electron resides in dot 1 near the system when the latter is empty ($N_{+}=N_{-}=0$).
Since dot 1 (2) of the polarizer is closest to (furthest from) the system
we assume different repulsive Coulomb energies $W_1 > W_2>0$.

In the absence of system electrons, $P$ is the energy change when the polarizer flips from $(N_1,N_2)=(1,0)$ to $(0,1)$.
However, with electrons present on the system
the repulsive interactions $W_1$ and $W_2$ modify this change in energy.
To see this clearly, we rewrite \Eq{eq:HSP1} using $N_1+N_2=1$ as
\begin{subequations}\begin{align}
	H^\text{SP} 
	& =
	[ \varepsilon+\tfrac{1}{2}(W_1+W_2) ] (N_{+}+N_{-})
	+ V N_{+} N_{-}
	\\&
	+ \tfrac{1}{2} [- P+(W_1-W_2)(N_{+}+N_{-}) ] (N_1 -N_2)
	\label{eq:inverted}
	.
\end{align}\label{eq:HSP2}\end{subequations}
We see that once the spatial gradient of the interaction across the polarizer, $W_1 - W_2:=C$, exceeds the potential gradient $P$ of the isolated polarizer
the following happens:
after adding the first electron to the system the polarization energy is effectively inverted [\Eq{eq:inverted}],
thereby attracting the next electron to the system.
To eliminate the polarizer degrees of freedom
we note that for $N_{+}+N_{-}=0,1,2$ the lowest energy state has $N_1-N_2=1,-1,-1$, respectively,
as indicated in blue in \Fig{fig:polarizer}.
This can be summarized as
$N_1-N_2 = 1-2(N_{+}+N_{-}) + 2N_{+} N_{-}$.
Imposing this nonlinear constraint on \Eq{eq:HSP2}
together with $N_\sigma^2=N_\sigma$
gives an effective model for the system only
(ignoring a c-number $-P/2$)
\begin{align}
	H
	&
	=
	(\varepsilon-P-W_2) (N_{+}+N_{-})
	+ (V-P) N_{+} N_{-}
	\label{eq:H2}
	.
\end{align}
Thus, we have obtained an effective model of the form \eq{eq:H} with charge operator \eq{eq:Nsigma2},
but now with a renormalized interaction and orbital energy
\begin{align}
	U:=V-P,
	\quad
	\epsilon := \varepsilon-P+W_2
	,
	\label{eq:Ueps}
\end{align}
due to the presence of the polarizer.

In \Fig{fig:polarizer} we illustrate this mechanism in terms of many-body energies of system plus polarizer.
These are the eigenvalues of $H^\text{SP}$ which we write as
\begin{subequations}\begin{align}
	H^\text{SP}
	& 
	=
	\epsilon (N_{+}+N_{-}) + V N_{+} N_{-}
	\\&
	+ \tfrac{1}{2} [- P+C(N_{+}+N_{-}) ] (N_1 -N_2)
\end{align}\label{eq:HSP3}\end{subequations}
The essence of the mechanism as we sketched in \Fig{fig:balls}
is then understood by just considering the blue states in \Fig{fig:polarizer}.
When the first electron enters the system, the lowest energy state is reached when the polarizer is flipped.
This implies that the second electron does not need to pay the energy $P$ and enters more easily than the first one. This effective energy gain $-P$ counteracts the repulsive interaction $V$ with the other system electron and is responsible for the tuneable interaction.

\Fig{fig:polarizer} makes clear that the elimination of the polarizer is valid if the capacitive-energy gradient is large, $C:=W_1-W_2 \gg P$.
In this case
there is a broad regime in which $P$ can be varied in order to tune the effective interaction $U$ to either sign.
The experiment in \Ref{Hamo16a} demonstrated that this regime of attractive $U<0$
can indeed be achieved when the polarizer is brought close enough to the system,
the latter being realized in a planar geometry.
Note however, that $C$ does not contribute\footnote
{When deriving \Eq{eq:H2} the large gradient $W_1-W_2$ cancels out in the contribution to the interaction terms $\propto N_{+} N_{-}$, even though it does modify the effective potential terms $\propto N_{+}+N_{-}$ via
	$\epsilon = \varepsilon-P+(-C+ W_1+W_2)/2 = \varepsilon-P+W_2$.}
to the expression for $U$.

\Fig{fig:polarizer} furthermore highlights that the inversion of $U$ does not entail an \emph{energy-level} crossing in the full model of double-dot plus polarizer, even though addition energies ($\epsilon+P$ and $\epsilon+V$) do cross. Therefore, the effective low-energy description \eq{eq:H} remains valid in the presence of time-dependent driving when no transitions are induced into states that were eliminated.
The additional condition for the driving frequency is $\Omega \ll P$. This is already implied by the slow driving condition $\Omega \ll \Gamma$ when we require all states on the system plus polarizer to be quantized $T \ll \Gamma \ll P, V$.

Finally, we note that the same mechanism 
in principle can be used to achieve negative $U$ in the \emph{single} dot.
Indeed using the model \eq{eq:H} with $N_\sigma = \sum_{\tau} d_{\sigma\tau}^{\dagger} d_{\sigma\tau}
\to d_\sigma^\dagger d_\sigma$
the above steps show that
a finite 'bare' \emph{intra}dot repulsion term $V' N_{+}N_{+}$,
is renormalized to $U=V'-P$
under the same conditions ($C\gg P$).
Experimentally, the polarization energies $P$ attained so far in the bottom-up approach of \Ref{Hamo16a} suffice to invert the weaker interdot interaction energy scale $V$ in carbon nanotube  double dots, but further progress is required to achieve the larger intradot scale $V'$ in these systems.
In top-down fabricated quantum dots~\cite{Cheng15,Cheng16,Prawiroatmodjo17a}
the effective interaction of a single dot can already be made negative
using a different mechanism.

\paragraph{Coupling to electrodes and transport quantities.}

Although contacting the double dot may be challenging in the original setup of \Ref{Hamo16a}, one may envisage similar structures, for example implementing the double dot in two parallel nanotubes in close proximity, each tube containing one quantum dot.
Regardless of the details, a relevant tunneling model extending \Eq{eq:coupling1} is
\begin{subequations}\begin{align}
	H^\text{T} &= \sum_{\alpha\sigma k} \sqrt{\Gamma^{\alpha} / \nu^{\alpha}}
	\, \sum_{\tau} d_{\sigma\tau}^\dag c_{k\alpha\tau}
	+\text{h.c.}
	,
	\\
	H^{\alpha} &= 
	\sum_{k\sigma}
	\omega_{\alpha k} \sum_{\tau} c_{k\alpha\tau}^\dag c_{k\alpha\tau}
	,
\end{align}\label{eq:coupling}\end{subequations}
again with energy- and real spin- ($\tau$) independent tunnel rates.
\begin{figure}
	\includegraphics[width=0.99\linewidth]{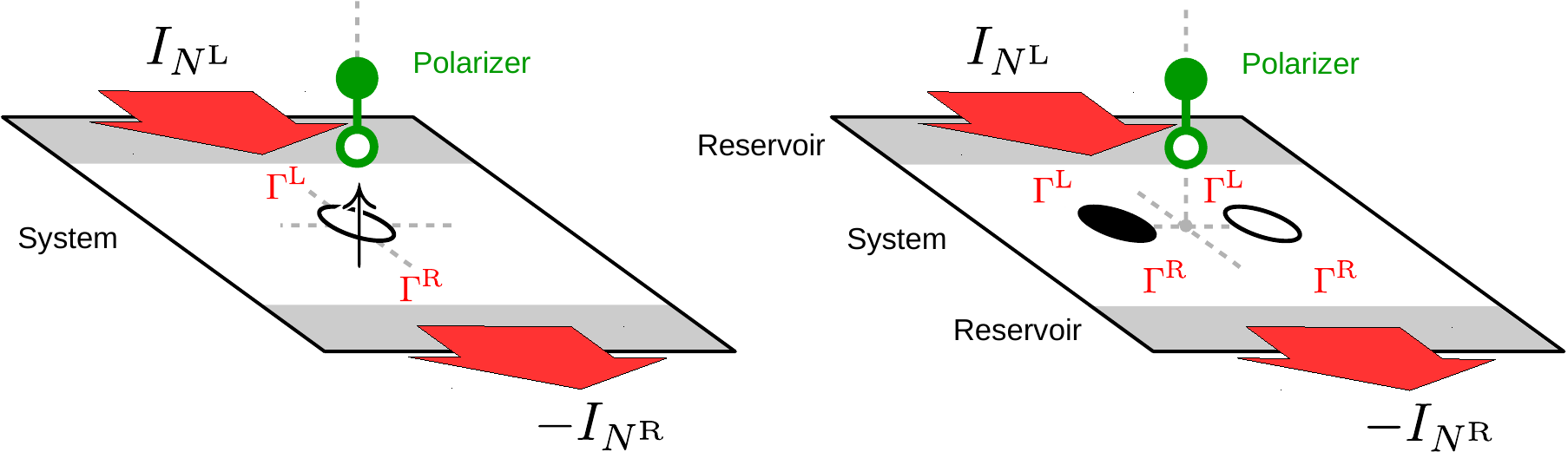}
	\caption{
		Left panel: Single dot with spin $\sigma=+$ (indicated by $\uparrow$)
		which is connected to electrodes $\alpha=\L$ and $\R$ by tunnel junctions.
		Right panel: Double dot with dot occupation of dot $\sigma=+$ indicated by the black filling, connected to a common set of electrodes on either side.
		The two dots are only coupled by electrostatic interaction, not by tunneling.
		These are \emph{schematic} transport setups --experimental details may differ-- the key point being that a contacted planar quantum dot system can be approached by the polarizer transverse to it to modify the electron interactions.
		Carbon-nanotube systems used in \Ref{Hamo16a} are particularly suited since the electronic states are exposed on the tube surfaces.
	}
	
	\label{fig:transport}
\end{figure}
see \Fig{fig:transport} for the considered schematic setup.
For simplicity, these rates are additionally assumed\footnote
	{Lifting this simplifying assumption requires a full account of orbital (pseudo-spin) polarization effects
		which is interesting but beyond the scope of the present study.}
to be the same for each of the two dots:
$\Gamma^{\alpha}$ is $\sigma$-independent.
The electron operators of the dot $\sigma$ (reservoir $\alpha$) are denoted by $d_{\sigma\tau}$ ($c_{k\alpha\tau}$)
where $\tau$ is the real spin.
Importantly, \Eq{eq:Idef} still  when the electrode charge operator is replaced by $N^\alpha =\sum_k c_{k\alpha\tau}^\dag c_{k\alpha\tau}$.
The current $I_{N^\L}$ ($I_{N^\R}$) now denotes the total current out of the left (right) measured electrode, see \App{app:spin}.
Note that the analogy between orbital index $\sigma$ in the double dot and spin index $\sigma$ is not preserved\footnote
	{In general this leads to level renormalization effects, even in the leading order coupling considered here, which are well-known~\cite{Koenig01} to cause observable precession effects in quantum-dot spintronics~\cite{Koenig03,Braun04set,Misiorny13}.
	In the present study these are not relevant due to our assumption of equal couplings to the shared reservoirs.
	}
by the coupling,
compare \Eq{eq:coupling1} with \eq{eq:coupling}.
Below we determine the resulting difference.

\paragraph{Driving the effective single-dot parameters.}
Finally, we address how the parameters in the effective model \eq{eq:H2}
can be driven directly through the gate voltages applied to the system double dot and separately to the polarizer double dot.

(i)
Driving the polarizer's energy $P$ affects both the effective level $\epsilon$ and $U$ [\Eq{eq:Ueps}]. To drive $U$ independently, one thus needs to compensate the side effect of $P$ on $\epsilon$ by driving the gates on the double dot.
As shown in \Fig{fig:polarizer}, $U$ can be driven between positive and negative values without having a crossing of energy levels of the system \emph{plus} the polarizer (which would otherwise invalidate our effective description of just the system).
By slowly driving the parameter $P$ one does not excite the states that are integrated out.

(ii) Although driving of the tunnel coupling strengths can in principle be done by modulating appropriate gate voltages, the \emph{independent} driving of $\Gamma^\L$ or $\Gamma^\R$ seems, however, more challenging.
Because of its conceptual simplicity and qualitatively different impact we will nevertheless analyze this in some detail in \Sec{sec:results}.

(iii)
Finally, driving the spatial separation between polarizer and the system
is equivalent to driving the gates controlling the system double dot.
In particular, modulating the distance would change the Coulomb repulsion energies $W_1$ and $W_2$ in \Eq{eq:HSP1}.
Since $C=W_1-W_2$ is required to be large for the effective description to hold, the quantities $W_1$ and $W_2$ never appear in the interaction $U$ in the effective description as we noted above.
\section{Transport theory of pumping\label{sec:method}}

\subsection{Master equation}

We consider the regime where the coupling to electrodes is weak and temperature is still relatively high, $\Gamma^{\alpha} \ll T$.
In this case, the transport in our model can be described with the help of
a master equation for the probabilities $\rho_N$ of having $N$ electrons on the quantum dot system
and an accompanying current formula,
all to first order in the tunnel coupling strength\footnote{
	In the following, we always assume that $U/T$ is large enough such that even in regions in which the tunneling rates are exponentially suppressed, the second order rates can still be neglected compared to them. This will be particularly relevant for attractive interaction.
}.
These form a closed set of equations making reference to neither the (pseudo-)spin $\sigma$ (in both models)
nor the real spin $\tau$ (in the double-dot model).
Physically speaking, this expresses that information about these quantities is inaccessible.
In \App{app:nospin} and \app{app:spin} we show that this implies that the relevant part of the density operator $\rho$ lies in a linear subspace,
\begin{align}
\Ket{\rho}= \rho_0 	\Ket{0} + \rho_1	\Ket{1}  + \rho_2 	\Ket{2}
.
\label{eq:subspace}
\end{align}
spanned by a basis of three operators denoted $\Ket{0}$, $\Ket{1}$ and $\Ket{2}$
which represent definite \emph{charge} states.
The master equation reads
\begin{align}
\tfrac{d}{dt} \Ket{\rho(t)} = W \Ket{\rho(t)}
\label{eq:master}
,
\end{align}
with $W=\sum_{\alpha=\L,\R} W^\alpha$,
and the current formula for transport quantities $N^\alpha$ is
\begin{align}
I_{N^\alpha}(t)=
\Bra{N} W^\alpha \Ket{\rho(t)}
\label{eq:current}
.
\end{align}
Here and below we use supervector notation where
$\Ket{B}=\hat{B}$ and $\Bra{A}\bullet=\tr \hat{A}^\dag \bullet$, where $\bullet$ denotes any argument such that $\Braket{A|B}=\tr A^\dag B$.
The rates $W^\alpha$ describe the system coupled to one electrode $\alpha$ \emph{only}.
For the single-level model we obtain
\begin{align}
W^\alpha
=
\begin{bmatrix}
-2W_{10}^\alpha & W_{01}^\alpha                & 0               \\
2W_{10}^\alpha  & -W_{01}^\alpha-W_{21}^\alpha & 2W_{12}^\alpha   \\
0              & W_{21}^\alpha                & - 2W_{12}^\alpha
\end{bmatrix} 
\label{eq:Walpha1}
,
\end{align}
where the individual rates are expressed using the Fermi function $f(x)=(e^x+1)^{-1}$ and $n=0,1$
\begin{subequations}\begin{align}
	W_{1,2n}^\alpha & = \Gamma^\alpha f \Big((-1)^n {(\epsilon+nU-\mu^\alpha)/}{T} \Big)
	,
	\\
	W_{2n,1}^\alpha &= \Gamma^\alpha f \Big((-1)^{n+1} {(\epsilon+nU-\mu^\alpha)/}{T} \Big)
	\label{eq:rates}
	.
\end{align}\end{subequations}
Importantly, the double-dot model is described by the same equations when the density operator
is expressed by an expansion \eq{eq:subspace} in a corresponding basis, see \App{app:spin}.
 The only difference with the single dot resides in the degeneracy factors in the first two columns of the rate matrix:
\begin{align}
W^\alpha
=
\begin{bmatrix}
-4W_{10}^\alpha & W_{01}^\alpha                 & 0               \\
4W_{10}^\alpha  & -W_{01}^\alpha-2W_{21}^\alpha & 2W_{12}^\alpha   \\
0              & 2W_{21}^\alpha                 & - 2W_{12}^\alpha
\end{bmatrix}
\label{eq:Walpha2}
.
\end{align}
The difference in degeneracy factors, in contrast to the explicit spin, is accessible via pumping spectroscopy.

\subsection{Adiabatic-response}

\paragraph{Driving parameters.}
The previous section established that all parameters of a double-dot system can be driven in time through applied voltages.
The natural regime for time-dependent spectroscopy
is the limit of slow driving $\dot{R} \sim \Omega |\Delta R| \ll \Gamma$
in which the transport current acquires an additional pumping contribution.
The driving parameters
\begin{align}
	\vec{R} =  \begin{bmatrix}
	{\displaystyle 
		\frac{\epsilon-\mu}{T},
		\frac{V_\b}{T},
		\frac{U}{T},
		\frac{\Gamma^\R}{\Gamma^\L},
		\bar{\Gamma} 
	}		
	\end{bmatrix}
	\label{eq:parameters}
\end{align}
affect the system through the rate matrices $W^\alpha$ [\Eq{eq:Walpha2}, resp. \eq{eq:Walpha1}].
All parameters are dimensionless\footnote
	{For compactness of notation, we will occasionally drop the normalization denominators as well as the constant $\mu$. Whenever we consider a driving parameter, we however always intend the respective component of \eq{eq:parameters}.}
and contribute to pumping, except for the last one, $\bar{\Gamma}:=\sqrt{\Gamma^\L \Gamma^\R}$.

\paragraph{Pumping response -- geometric curvature}
To determine the measurable pumping effect
we employ the adiabatic-response approach to compute the time-dependent solution for $\rho(t)$ and the resulting pumped charge~\cite{Splettstoesser06} in the limit of slow driving.
In particular, we use the geometric formulation of \Refs{Calvo12a,Pluecker17a,Pluecker18a},
which allows for a clear comparison with other approaches (such as FCS pumping~\cite{Sinitsyn07EPL,Sinitsyn09} and Kato-projections~\cite{Avron11,Avron12}, cf. also \Ref{Nakajima15}).
The present approach is, however, quite straightforward.
We first determine the density operator $\Ket{\rho^\i}$ in terms of the kernel $W=\sum_{\alpha} W^{\alpha}$ from the \emph{stationary} master equation \eq{eq:master} for \emph{fixed} parameters, $0=W \Ket{\rho^\i}$. Inserted into the current formula \eq{eq:current}
this gives a nongeometric instantaneous transport of charge
\begin{align}
	\Delta N^{\alpha,\i} = \int_0^\T dt \Bra{N} W^\alpha \Ket{\rho^\i}
	\label{eq:DeltaNi}
\end{align}
which is not discussed further (since it can be experimentally subtracted).
Next, we determine the adiabatic-response $\rho^\a = W^{-1} \Ket{\frac{d}{dt} \rho^\i}$, where $W^{-1}$ is the pseudo inverse. This is the leading-order nonadiabatic correction,
i.e., the contribution linear in the driving velocity $dR(t)/dt$.
It enters the additional geometric \emph{pumping} contribution to the transfered charge, of interest here, caused by the nonadiabatic "lag" of the system:
it can be written as an integral of a geometric curvature
over the surface bounded by the driving cycle $\curve$ traversed in the plane of the two driven parameters $(R_k,R_l)$:
\begin{align}
	\Delta N^\alpha
	 = \int dS \F_{R_k,R_l}^\alpha
	.
	\label{eq:DeltaN}
\end{align}
The pumping curvature is unlike\footnote
	{Although the pumped charge $\Delta N^\alpha$ can be expressed as a proper geometric phase it is \emph{not} simply equal to a Berry phase.
	This geometric phase reflects the invariance of the measurement transported charge under parametrically time-dependent gauge transformations of charge-\emph{observable}:
	pumping is geometric because the charge meter can be \emph{physically} recalibrated or gauged~\cite{Pluecker17a,Pluecker18a}.}
well-known adiabatic Berry-type curvatures that are often discussed. It instead reads
\begin{subequations}\begin{align}
	\F_{R_k,R_l}^\alpha
	& = \Bra{\delR \Phi^\alpha } \times \Ket{ \delR \rho^\i}_{kl}
	\\
	& := \Braket{ \nabla_k \Phi^\alpha |  \nabla_l \rho^\i}
	-
	\Braket{ \nabla_l \Phi^\alpha |  \nabla_k \rho^\i}
\end{align}\label{eq:curvature}\end{subequations}
where $\delR_k := {\partial}/{\partial R_k}$.
It combines the response of the states \emph{and} the response of the transported \emph{observable} (charge) that is measured externally.
Similar responses were first discussed for nonlinear dissipative systems by Ning and Haken and by Landsberg,
see the reviews \Ref{Ning92,Landsberg93,Sinitsyn09}.
Here $\Bra{\Phi^\alpha}
= \Bra{N} W^\alpha W^{-1}$
is a charge-response covector~\cite{Calvo12a} characterizing the nonadiabatic effect\footnote
	{At this point the geometric mean of the tunneling couplings $\bar{\Gamma}$ --the only parameter with a dimension-- cancels out in the pumping curvature [cf. \Eq{eq:Fzero}].}
(called 'adiabatic-response') on the external observable that is transported through the system in a nonequilibrium stationary-state.
An important consequence of \eq{eq:curvature} becomes visible already at this stage, and motivates our parametrization
\eq{eq:parameters} of the tunnel rates by their ratio and geometric mean $\bar{\Gamma}$.
The geometric mean cancels out in the ratio $W^\alpha W^{-1}$
since both $W^\alpha$ and $W$ are proportional to $\bar{\Gamma}$:
\begin{align}
	\Gamma^\L = \bar{\Gamma} \sqrt{ \frac{\Gamma^\L}{\Gamma^\R} }
	,\quad
	\Gamma^\R = \bar{\Gamma} \sqrt{ \frac{\Gamma^\R}{\Gamma^\L} }
	.
	\label{eq:Gamma-par}
\end{align}
The curvature only depends on the coupling asymmetry.

\paragraph{Driving protocols for geometric pumping.}

Selecting a pair of parameters $(R_k,R_l)$ from the list \eq{eq:parameters} to be modulated defines an experimental \emph{driving protocol}
for which the measured response is given by the pumping formula \eq{eq:DeltaN}.
The prime quantity of interested is thus the pumping curvature \eq{eq:curvature} because it contains the full information about the pumped charge for any driving curve $\curve$.
Experimentally, the curvature can be extracted by measuring the pumped charge from small driving cycles only, a method that we call \emph{geometric pumping spectroscopy}~\cite{Reckermann10a,Calvo12a}
(extending the well-known nonlinear dI/dV spectroscopy).
In this limit,
the pumped charge equals the curvature $\F^\alpha[R^{*}]$ at the working point (denoted $R^{*}$)  $\times$ the  driving-parameter area $\pi |\Delta R|^2$ (for a circle of radius $\Delta R$).
As we will illustrate in section \ref{sec:monotonous}, studying the complete profile of the curvature in the  driving parameter plane --its nodes and sign changes-- allows one to directly infer when and how this monotonic increase with $\Delta R$ of the experimental pumping signal $\Delta N^\alpha$ breaks down.

\subsection{Explicit curvature formulas}

\paragraph{Curvature for the single dot.}

For the single-dot model, the curvature $\F^\alpha$ can be computed most easily by noting\cite{Pluecker17a} that the matrix $W^\alpha$ has three eigenvalues,
one of which governs the decay of an excess charge on the quantum dot~\cite{Schulenborg16a}.
This eigenvalue can be written as $-w^\alpha$ where
\begin{subequations}\begin{align}
	w^\alpha &= \sum_{N=0,2} W^\alpha_{1N}
	\\
	& = \Gamma^\alpha \Big[ f^\alpha((\epsilon-\mu^\alpha)/T)+ f^\alpha(-(\epsilon+U-\mu^\alpha)/T)\Big]
	,
	\label{eq:charge-rel}
\end{align}\end{subequations}
is the \emph{charge relaxation rate}.
It determines how fast the charge state $N=1$ is reached due to the coupling to a specific electrode $\alpha=\L$ or $\R$,
irrespective of the initial state of the dot ($N=0$ or $2$).
The pumping curvature \eq{eq:curvature} simplifies to
\begin{align}
	\F^\alpha_{R_k,R_l} = \left\{
	\Big(\delR \frac{w^\alpha}{\sum_{\alpha} w^\alpha} \Big)
	\times
	\delR   \braket{N} \right\}_{kl}
	\label{eq:Falpha}
	,
\end{align}
where $\braket{N}:=\Braket{N|\rho^\i}=\tr \hat{N} \rho^\i$
is the charge on the quantum dot in the parametric stationary state.
Total charge conservation is expressed by
$\sum_\alpha \Bra{N}  W^\alpha = \Bra{N} W$
and implies together with $W \Ket{\rho^\a}=0$,
that we can antisymmetrize in $\alpha$,
$\F:=(\F^\R - \F^\L)/2=\F^\R$
and obtain\footnote
	{The subtraction of a constant $\braket{N} \to \braket{N}-1$ under the gradient in \Eq{eq:F} is motivated by the symmetric role of the $N=0$ and $N=2$ state in \Eq{eq:occupation} which becomes crucial later on, cf. \Eq{eq:motivate} below.}
\begin{subequations}\begin{align}
	\F_{R_k,R_l} := \left\{
	\tfrac{1}{2}
	\Big(\delR \frac{w^\R - w^\L}{w^\R + w^\L} \Big)
	\times
	\delR   \Big( \braket{N}-1 \Big) \right\}_{kl}
	\label{eq:F}
	.
\end{align}
Equations \eq{eq:Falpha} and \eq{eq:F} are the key formulas\footnote
	{\Eq{eq:F} was correctly derived in \Ref{Pluecker17a} [Eq. (D12) and (D14a)], but unfortunately the final result (D19) was written incorrectly. Also, the curvature was not studied for attractive or driven interaction $U$ which is of interest here.}
that allow the origin of any nonzero value of the curvature to be clearly understood
just by plotting the two scalar quantities under the gradients~\cite{Calvo12a}.
Namely, the pumping response is determined by the parametric charge polarization taken relative to $N=1$
\begin{align}
	\braket{N}-1
	& =
	\frac{\sum_{\alpha} ( W^\alpha_{10}- W^\alpha_{12} ) }
	{\sum_{\alpha} \sum_{N=0,2} W^\alpha_{1N}}	
	\label{eq:occupation}
\end{align}
and the asymmetry of the charge-relaxation rates
\begin{align}
	\frac{w^\R - w^\L}{w^\R + w^\L}
	& =
	\frac{\sum_{N=0,2} ( W^\R_{1N}- W^\L_{1N} ) }
	{\sum_{\alpha} \sum_{N=0,2} W^\alpha_{1N}}
	\label{eq:decayratio}
\end{align}\label{eq:Fall}\end{subequations}
Note that in both factors the \emph{magnitude} of these rates is irrelevant.
Thus, even when transport currents are small, it is possible to pump charge,
although one must keep in mind the slow-driving condition
that the driving frequency $\Omega$ must be kept small relative to these rates.
That the factor \eq{eq:occupation} ignores spatial asymmetry ($\L$ vs. $\R$)
whereas the factor \eq{eq:decayratio} ignores charge asymmetry ($N=0$ vs. $2$),
correlates with their very different sensitivity to the bias and and gate voltage
which will be crucial below.

In a way, the ratio \eq{eq:decayratio} quantifies how the parameters modulate the 'effective coupling' to the external electrodes. Importantly, without interaction ($U=0$) the relaxation rates \eq{eq:charge-rel} reduce to $w^\alpha = \Gamma^\alpha$
and all dependence on parameters other than the 'bare' couplings cancels out.
We also note that this factor may seem to be only quantitatively important.
For example, for repulsive interaction and fixed coupling
it was observed~\cite{Reckermann10a} that the geometric pumping spectroscopy can be qualitatively understood by finding the crossings of resonance lines in parameter space where the \emph{occupations} of the quantum-dot states change, as captured by the factor $\braket{N}-1$ in \Eq{eq:F}.
We will see that this intuitive rule is in a way fortuitous :
we find that for fixed \emph{attractive} interaction there are pumping mechanisms which \emph{cannot} be understood -- even qualitatively-- this way
and require explicit consideration of the factor \eq{eq:decayratio}.

\paragraph{Effective parameters and pumping mechanisms.}

\begin{figure*}
	\includegraphics[width=0.7\linewidth]{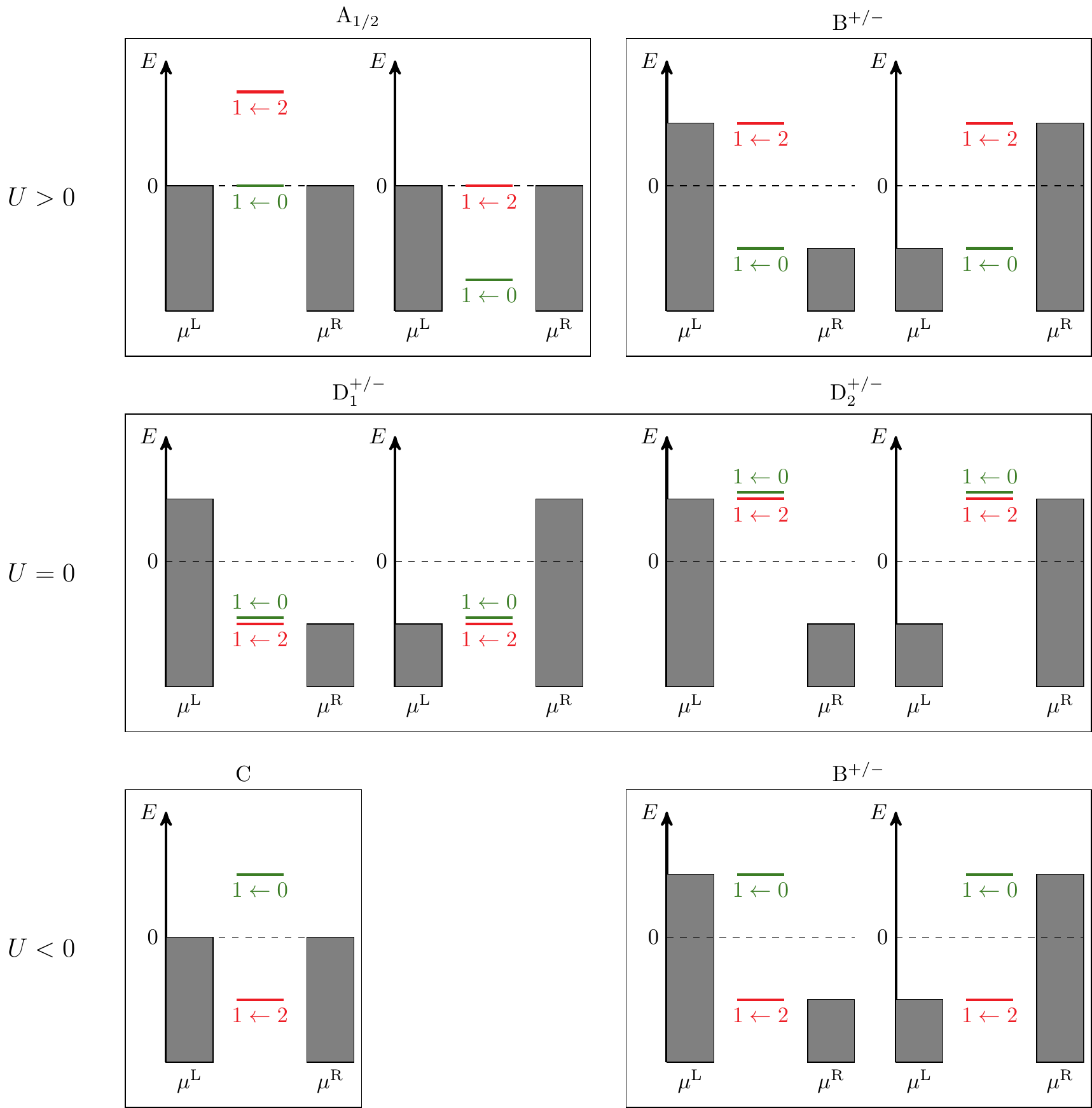}
	\qquad	\qquad
	\caption{
		Electrochemical potential configurations for the \emph{mechanisms} of pumping discussed in the paper.
		Every configuration defines two linearly independent, effective parameters that are zero in the shown configuration.
		A mechanism can thus be specified by the sketched conditions on 
		$\epsilon$, $V_\b$ and $U$,
		where $\mu^\L = \mu + V_\b/2$ and $\mu^\R = \mu - V_\b/2$ with constant $\mu$.
		A mechanism is well-separated from other mechanisms when these conditions differ by more than the scale of thermal broadening  (not indicated).
		By contrast, driving the coupling $\Gamma^\R/\Gamma^\L$ is never equivalent to driving any non-coupling parameter and thus can by definition not access any of the mechanisms associated with the shown configurations.
		We stress that in all cases, the shown configurations are necessary, but not sufficient conditions
		for pumping.
		As an example, $\B^{\pm}$ does not lead to pumping for $U<0$ while $\C$ does.
	}
	\label{fig:levels}
\end{figure*}

While the parameters \eq{eq:parameters} defining the driving protocols are dictated by experimental considerations, the form \eq{eq:F} of the curvature as a combination of transition rates $W^\alpha_{N'N}$ actually suggests that different \emph{effective parameters} govern the response. 
Except for $\Gamma^\R/\Gamma^\L$,  all parameters enter the transition rates via the arguments of Fermi functions. Naively, one then expects a pumping response only in regions, where some Fermi functions are not constant (gradient nonzero). Their arguments are then \emph{effective parameters} modulated around the well known resonance conditions:
\begin{subequations}\begin{align}
	\epsilon - \mu^{\R/\L} &= 0
	\\
	\epsilon + U - \mu^{\R/\L} &= 0.	
\end{align}
\label{eq:resonance-conditions}\end{subequations}
Interestingly, we will find that this simple picture breaks down in the case of attractive interaction. To then find the two effective parameters, which are always needed, is the main subject of \Sec{sec:results}.
We call any such combination of two effective parameters a \emph{pumping mechanism}. Each mechanism corresponds to a configuration in Fig.~\ref{fig:levels}, the value of the coupling being irrelevant.

Since the effective parameters (to be found) are a combination of the experimentally
accessible parameters \eq{eq:parameters}, one \textit{mechanism} can relate the pumping response of different driving protocols to each other. 
For example, we will show [\Eq{eq:relation-A2}] that close to the working point
associated with the mechanism that we label $\A_2$,
\begin{align}
\F_{U,V_\b}(\epsilon)
\approx
\M_{\A_2}[\epsilon+U-\mu,V_\b]
\approx
\F_{\epsilon,V_\b}(U)
.
\label{eq:relation-example}
\end{align}
Here, $M_{\A_2}$ is the curvature that one would obtain from \eq{eq:curvature}, if the effective parameters of mechanism $\A_2$ (indicated in the square bracket) would be chosen as driving parameters. This relates driving of $U$ and $V_\b$ at fixed $\epsilon$ directly with driving of 
$\epsilon$ and $V_\b$ at fixed $U$ in the vicinity $\A_2$ (see Fig. \ref{fig:levels}).
We stress that this is not a linearization of the curvature around the working 
point, but describes its full dependence in the vicinity.

Importantly, close to another working point the relation between the same two curvature components --and thus, two experiments-- may be completely different 
or even absent.
Since in this paper we also allow the interaction $U$ to be one of the driving parameters, it is a key question whether this entails a new \emph{mechanism} of charge pumping or whether it can always be considered as being equivalent to
driving of another parameter as the gate voltage in \Eq{eq:relation-example}.

\paragraph{Explicit formula for the double dot.}

For the double-dot model whose rate matrix $W^\alpha$ is given by \Eq{eq:Walpha2}
the simple trick of \Ref{Pluecker17a} fails because more than one eigenmode plays a role
due to the breaking of electron-hole symmetry (infinite intradot repulsion)
Also, the eigenmodes are no longer simply related to the charge covector $\Bra{N}$.
Absorbing all degeneracy factors into rates $\bar{W}_{i j}$ indicated by an overbar we derive in \App{app:spin} the general result
\begin{align}
	\F^\alpha_{R_k,R_l} =
	\left\{
	\nabla
	\begin{bmatrix}
		\Phi_0  & 	\Phi_2
	\end{bmatrix}
	\times \nabla
	\begin{bmatrix}
		\rho_0^\i  \\ \rho_2^\i
	\end{bmatrix} \right\}_{kl}
	\label{eq:Falpha-result}
\end{align}
\begin{widetext}
expressed in the independent components of the instantaneous stationary state and the response-covector
\begin{subequations}\begin{align}
	\begin{bmatrix}
		\rho_0^\i  \\ 	\rho_2^\i
	\end{bmatrix}	
	& =	
	\frac{1}{ \W_{10} \W_{12} + \W_{10} {\W_{21}} + {\W_{01}} \W_{12} }
	\begin{bmatrix}
		\W_{12} \W_{01}
		\\
		\W_{10} \W_{21}
	\end{bmatrix}
	\label{eq:vector}
	\\
	\begin{bmatrix}
		\Phi_0  \\ 	\Phi_2
	\end{bmatrix}
	& =
	\frac{1}{ \W_{10} \W_{12} + \W_{10} {\W_{21}} + {\W_{01}} \W_{12} }
	\begin{bmatrix}
		- (\W_{10}^{\alpha} + {\W_{01}^{\alpha} - \W_{21}^{\alpha}}) \W_{12} - (\W_{10}^{\alpha} + \W_{12}^{\alpha} ) {\W_{21}}
		\\
		\phantom{-}
		( \W_{21}^{\alpha} + \W_{12}^{\alpha} -\W_{01}^{\alpha}) \W_{10} + (\W_{10}^{\alpha} + \W_{12}^{\alpha} ) {\W_{01}}
	\end{bmatrix}
	.
	\label{eq:covector}
\end{align}\label{eq:Fall2}\end{subequations}\end{widetext}
Further simplifications can be made by evaluating the gradients,
antisymmetrizing [see \App{app:spin}],
and finally using that the rates without degeneracy factors sum to constants, $W_{ij}^\alpha+W_{ji}^\alpha=\Gamma^\alpha$.
For the double dot the result remains unwieldy.
The formula \eq{eq:Fall2} also applies to the single-level model when the corresponding rates \eq{eq:Walpha1} are substituted: only then it simplifies to the much simpler result \eq{eq:Falpha}.

\begingroup
\begin{table*}
	\caption{\label{tab:mechanisms}
		Driving protocols with static couplings $\Gamma^\R/\Gamma^\L$
		and the accessed pumping mechanisms (sketched in \Fig{fig:levels}).
		Blank cells in certain columns mean that the mechanism of the respective column is not accessible in the driving protocol of a given row.
		The signs and factors 2 and 1/2 indicate relations between curvature components of the type \eq{eq:relation-example}
		that are discussed in the main text.
	}
	\begin{ruledtabular}
	{\renewcommand{\arraystretch}{1.2} 
	\begin{tabular}{|c|c|c|c|c|c|c|c|}
		\hline
		\textbf{Driven}                       & \textbf{Constant} &        &         			&                          		&                   			&                        	&  \\\hline
		\multirow{3}{*}{(i): $(V_\g,V_\b )$ } & $U>0$             & $\A_1$ & $\phantom{-}\A_2$  & $\phantom{-\tfrac{1}{2}}\B^{\pm}$&                   			&                        	&  \\
		                                      & $U=0$             &        &         			&                          		&                   			&                        	&  \\
		                                      & $U<0$             &        &         			&                          		& $\phantom{\tfrac{1}{2}}\C$    &                        	&  \\ \hline
		\multirow{3}{*}{(ii): $(U,V_\g)$ }    & $V_\b >0$         &        &         			& $\phantom{-\tfrac{1}{2}}\B^{+}$  &                   			& $-2\D_1^{+}$           	& $\phantom{-} 2\D_2^{+}$  \\
		                                      & $V_\b = 0$        &        &         			&                         			&                   			&                         	&  \\
		                                      & $V_\b < 0$        &        &         			& $-\phantom{\tfrac{1}{2}}\B^{-}$  &                   			& $\phantom{-}2\D_1^{-}$ 	& $-2\D_2^{-}$ \\ \hline
		\multirow{3}{*}{(iii): $(U,V_\b )$ }  & $V_\g > 0$        &        & $-\A_2$ 			& $-\tfrac{1}{2} \B^{\pm}$ 		&                   			& $\phantom{-2}\D_1^{\pm}$  &  \\
		                                      & $V_\g = 0$        &        &         			&                          		& $\tfrac{1}{2} \C$ 			& $\phantom{-2}\D_1^{\pm}$  & $\phantom{-2}\D_2^{\pm}$ \\
		                                      & $V_\g < 0$        &        &         			&                          		& $\tfrac{1}{2} \C$ 			&                        	& $\phantom{-2}\D_2^{\pm}$ \\ \hline
	\end{tabular}
	}
	\end{ruledtabular}
\end{table*}\endgroup

\section{Pumping response -- Single dot\label{sec:results}}

We now turn to the main results of the paper,
focusing on the role of the tunable interaction as a \emph{static} parameter which can be \emph{negative} or as a parameter that is \emph{driven},
while using the familiar case of static repulsive interaction $U$ as a reference.
To distinguish new effects
we work out a map containing all possible situations and analyze them carefully.
In \Fig{fig:levels} we sketch the electrochemical potential diagrams for all pumping mechanisms as introduced in section \ref{sec:method}.
Which of these mechanisms is accessed in a given driving protocol is summarized in \Tab{tab:mechanisms}.

From \Tab{tab:mechanisms} one immediately sees that there are \textit{new} mechanisms both for driven (mechanism $\D$) as well as for constant attractive interaction (mechanism $\C$), which are not accessible using any driving protocol with static, repulsive interaction $U$.
Mechanism $\D$ is of interest since the experimental detection of its pumped-charge signature provides a strong indication that one has independent dynamical control over the interaction in the engineered structure.
Furthermore, the table shows that in other situations pumping by driving the interaction can be due to the same mechanism as when dealing with the static interaction (eg. mechanism $\B$).
Finally, since some of these mechanisms ($\B$) are only available for asymmetric tunnel coupling
we plot in the following all our results for generic values of the ratio $\Gamma^\R/\Gamma^\L \neq 1$.

\subsection{Coupling strength as one pumping parameter\label{sec:coupling}}

We first consider driving protocols in which charge pumping is achieved
by choosing the coupling strength as one of the pumping parameters.\footnote
	{As driving parameter, the tunneling coupling is doubly restricted:
	the time-dependent values of the coupling should always remain less than temperature $T$ (weak coupling limit),
	while exceeding the driving frequency $\Omega=2\pi/\T$ (slow driving).}
We discuss this class of driving protocols separately since the coupling strength is the only parameter that enters the transition rates linearly (compared to all other parameters entering via Fermi functions). We will here see that this leads to some fundamental differences in the pumping features.

As a first example of this, driving \emph{both} couplings does not lead to \emph{any} pumping response for \emph{any} value of the other parameters,
\begin{align}
		\F_{\Gamma^\L,\Gamma^\R} = 0  \quad\text{always}
		.
		\label{eq:Fzero}
\end{align}
The reason is that the geometric mean of the couplings $\bar{\Gamma}=\sqrt{\Gamma^\L \Gamma^\R}$ [\Eq{eq:Gamma-par}] cancels out in the pumping curvature \eq{eq:curvature},
even though it does modify\footnote
	{As a result, symmetric modulation of the couplings offers an additional way to experimentally extract the pumping response, see for a discussion App. B of \Ref{Pluecker17a}.}
the instantaneous response \eq{eq:DeltaNi} (not discussed).
Therefore driving both $\Gamma^\L$ and $\Gamma^\R$ amounts to varying only a single effective coupling parameter $\Gamma^\R/\Gamma^\L$ and thus no pumping is
possible, regardless of the bias voltage (both equilibrium and nonequilibrium)
and the interaction $U$ (both attractive and repulsive).
Therefore, in the following we modulate one coupling strength and one further independent parameter to achieve pumping.

\begin{figure*}
	\subfigure[~Curvature $\F_{\epsilon,\Gamma^\R} \cdot \Gamma^\L$ vs. coupling and gate \emph{driving} parameters for $U>0$ (left) and $U<0$ (right). ]{
		\includegraphics[width=0.4\linewidth]{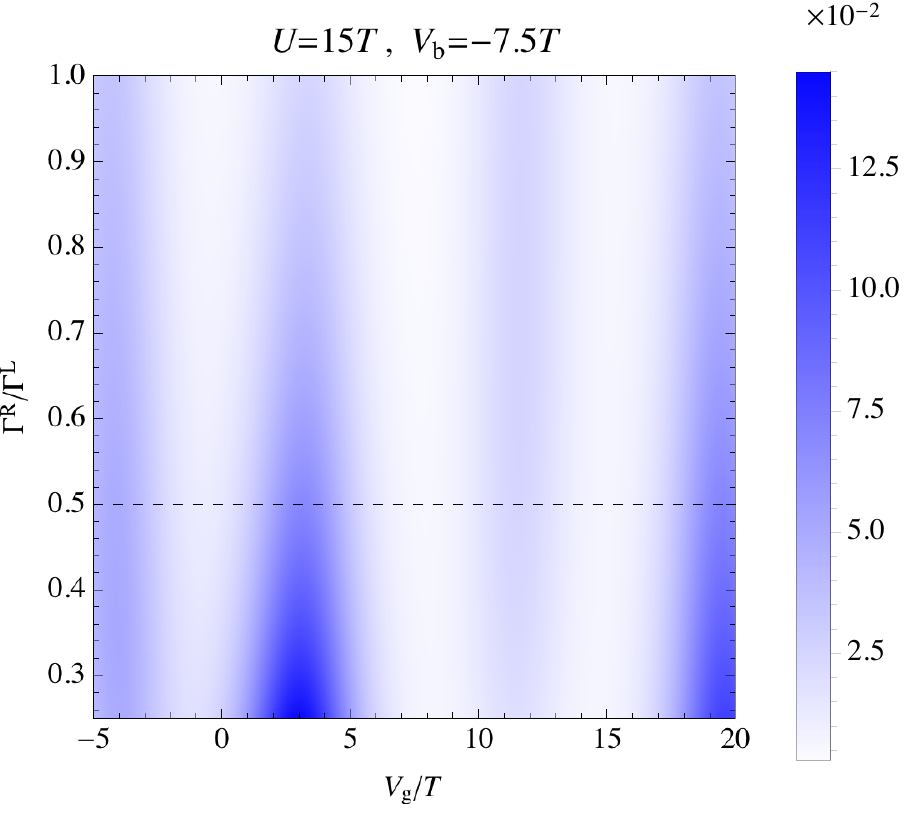}
		\includegraphics[width=0.39\linewidth]{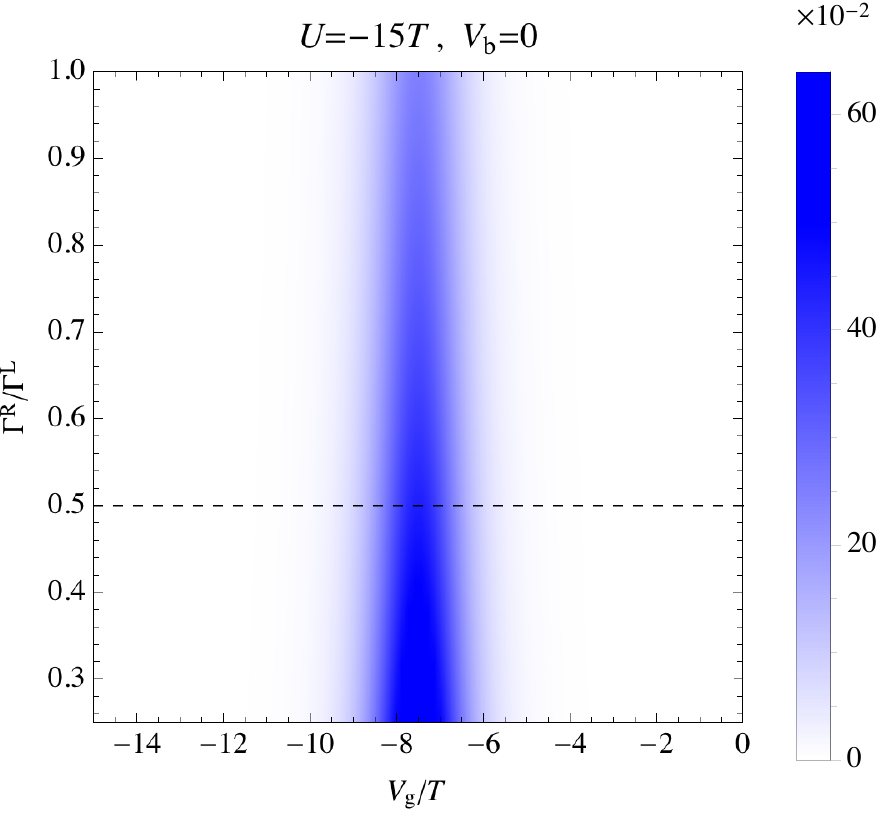}
		\label{fig:coup-gate-curv}
	}\\
	\subfigure[~Curvature $\F_{V_\b,\Gamma^\R} \cdot \Gamma^\L$ vs. coupling and bias \emph{driving} parameters for $U>0$ (left) and $U<0$ (right).]{
		\includegraphics[width=0.4\linewidth]{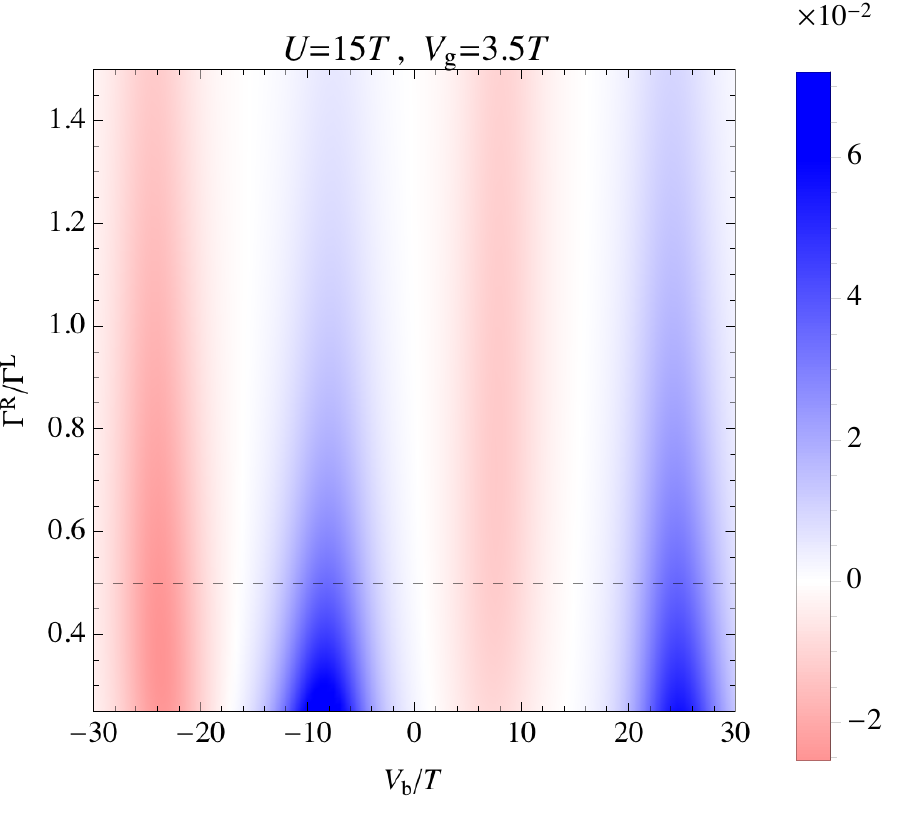}
		\includegraphics[width=0.4\linewidth]{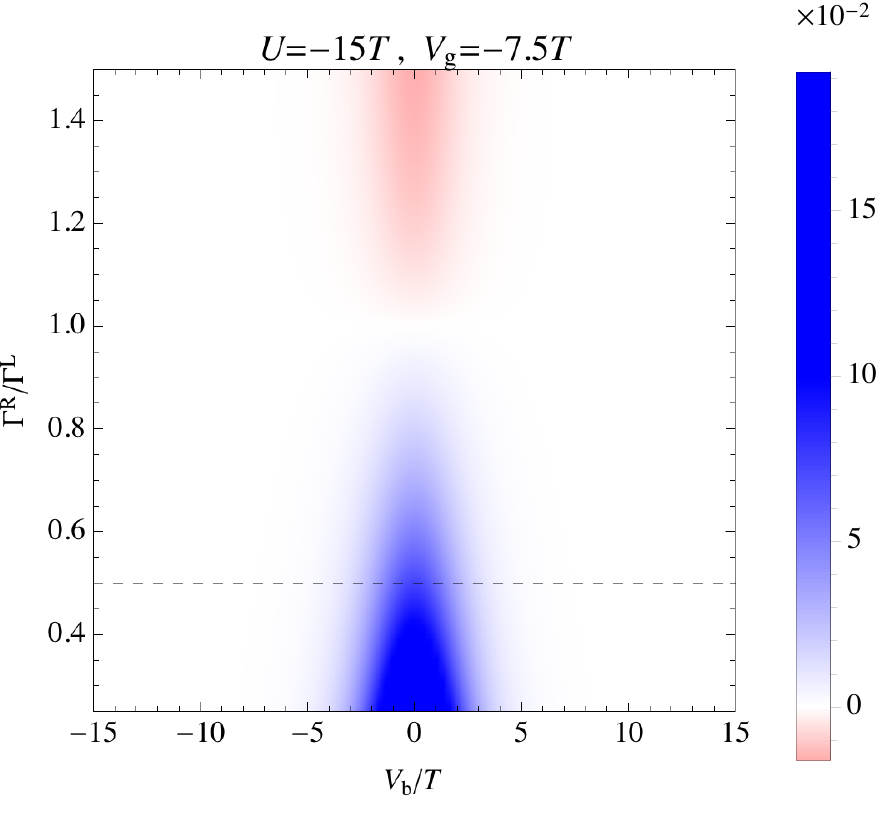}
		\label{fig:coup-bias-curv}
	}\\
	\subfigure[~Curvature $\F_{U,\Gamma^\R} \cdot \Gamma^\L$ vs. coupling and \emph{interaction driving} parameter.]{
		\includegraphics[width=0.4\linewidth]{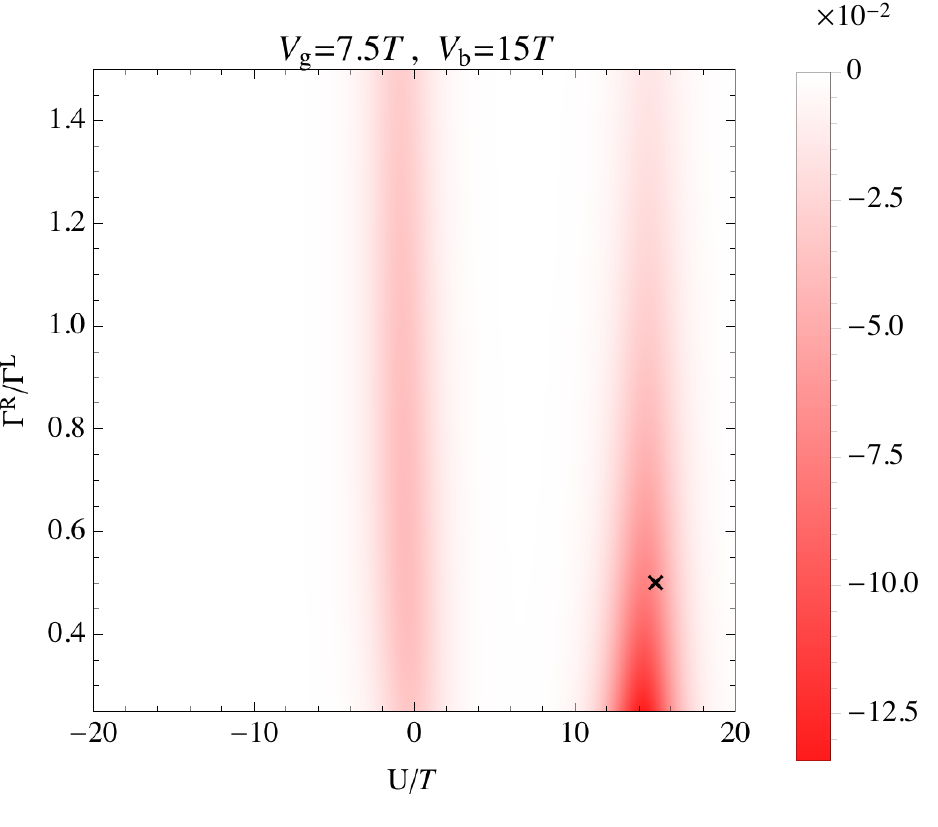}
		\includegraphics[width=0.39\linewidth]{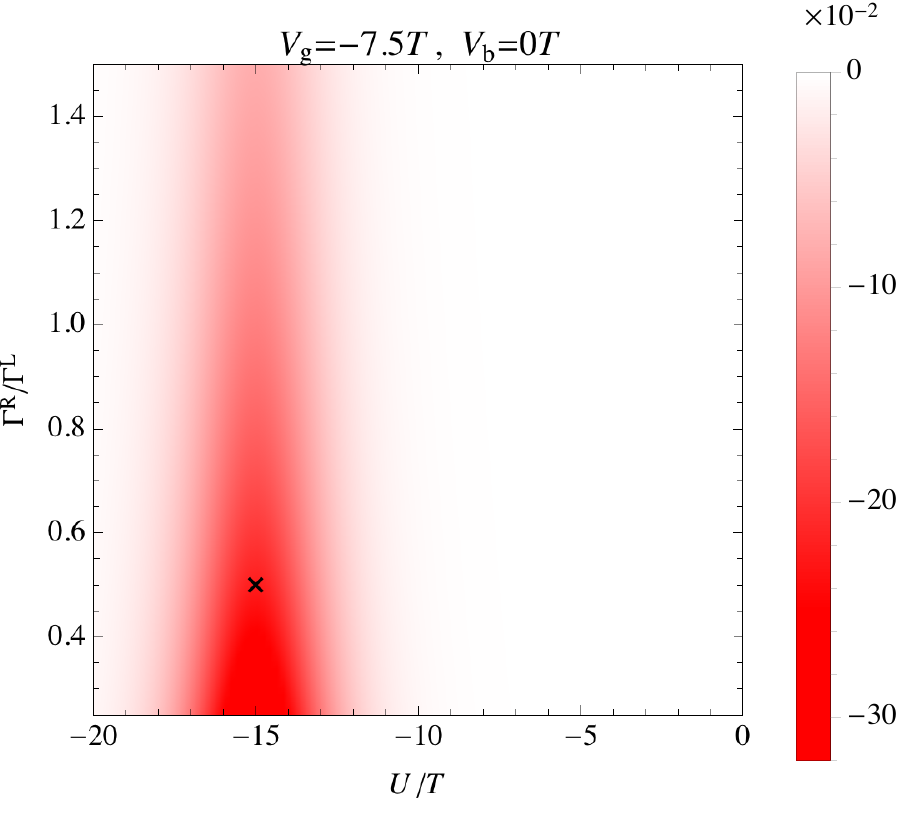}
		\label{fig:coup-int-curv}
	}\\
	\caption{
		Pumping curvatures as function of the \emph{driven} parameters,
		the coupling $\Gamma^\R$ and one other parameter ($\epsilon$, $V_\b$ or $U$).
		All curvatures and parameters are dimensionless.
		The dependencies on other static parameters are plotted in \Fig{fig:coup},
	taking $\Gamma^\R/\Gamma^\L = \tfrac{1}{2}$ (indicated by dashed lines in (a) and (b))
		or taking $\Gamma^\R/\Gamma^\L = \tfrac{1}{2}$ and $U=15T$ (indicated by the crosses in (c)).
		\label{fig:coup-curv}
	}
\end{figure*}

\begin{figure}
	\centering{\includegraphics[width=0.99\linewidth]{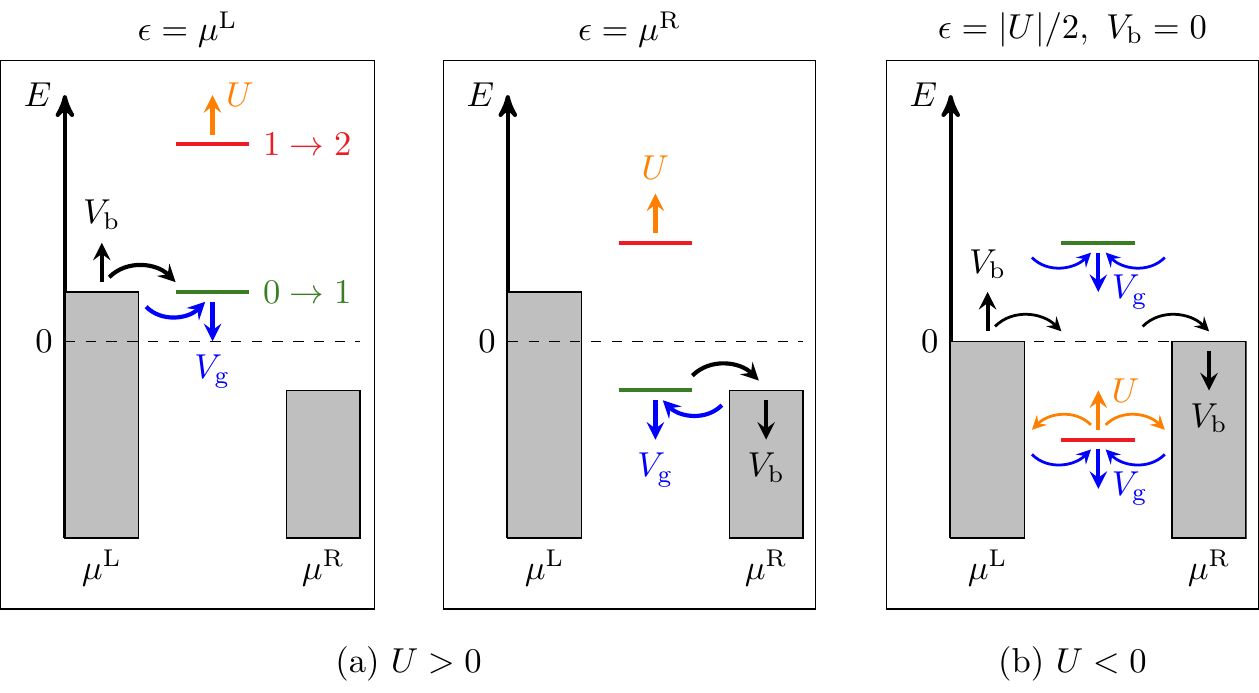}}
	\caption{
		Effect of changing the gate, bias or interaction parameter.
		The vertical colored arrows show the direction in which parameters of the same color will drag the indicated respective electrochemical potential. The bent arrows indicate the resulting changes of occupation. 
		(a) Repulsive interaction:
		In the left configuration of (a), increasing the bias and gate voltage
		both will tend to \emph{fill} the dot, in the right configuration
		the bias voltage tends to \emph{deplete} the dot.
		This is responsible for the sign-change in the pumping response 
		shown in \Fig{fig:coup-curv} and \Fig{fig:coup} when driving $V_\b$.
		(b) Attractive interaction:
		Equivalent sketch for pumping configuration for attractive interaction.
		In both (a) and (b) changing the interaction parameter has qualitatively the same effect as
		changing $\epsilon \propto - V_\g$ or no effect at all.
		\label{fig:bias-effect}
	}
\end{figure}

\subsubsection{Repulsive interaction $U>0$.}

In \Fig{fig:coup-gate-curv} we show the curvature as function of the \emph{driven} parameters,
the coupling $\Gamma^\R/\Gamma^\L$ and the experimental gate voltage $V_\g := - \epsilon$ (incorporating the gate lever-arm factor).
This graphic representation is the only one that allows to obtain the pumped charge just by drawing the driving cycle at a working point and then computing the flux of $\F_{\epsilon,\Gamma^\R}$ through the covered area.

The vertical stripes in the plot are a consequence of the fact, that the coupling asymmetry enters the 
transition rates linearly and is therefore always a possible effective parameter. 
In addition it is however necessary to have a second effective parameter.
The four lines in the figure correspond precisely to one of the resonance conditions \eq{eq:resonance-conditions}
and thus indicate the effective parameters.

This is verified in \Fig{fig:coup-gate-B} where for a generic fixed value of $\Gamma^\R/\Gamma^\L$ we plot the curvature as a function of the driven parameter $\epsilon$ and the additional static parameter $V_\b$. 
In contrast to the ``natural" way of plotting the curvature as function of the
driven parameters (\Fig{fig:coup-curv}), here it is easy to spot the familiar 
resonance conditions and identify the mechanism at work:
the lines of nonzero curvature (blue) coincide with the lines where the
occupation changes, as one would measure by a stationary DC spectroscopy ($dI
/dV_\b$ Coulomb diamonds).
Related to this, the curvature in \Fig{fig:coup-gate-curv} has the same sign for
all $V_\g$ working points, reflecting that the gate voltage always has the same
effect on the occupations, no matter what the other parameters are:
making $V_\g$ more positive always attracts charge to the dot, see 
\Fig{fig:bias-effect}(a).

Next, we show in \Fig{fig:coup-bias-curv}
the curvature when driving the bias voltage $V_\b$ (instead of $\epsilon$) together with the coupling $\Gamma^\R/\Gamma^\L$.
In this case the vertical lines have alternating signs (blue, red).
As explained in \Fig{fig:bias-effect}(a),
the sign changes reflect that the qualitative effect of the bias voltage $V_\b$ in comparison with the gate voltage $V_g$ depends on the transition energy configuration which depends on other non-driven parameters.

\subsubsection{Attractive interaction $U<0$.}
\begin{figure*}
	\subfigure[~Curvature $\F_{\epsilon,\Gamma^\R} \cdot \Gamma^\L$ vs. driven gate parameter $\epsilon$ and static bias parameter $V_\b$.]{
		\includegraphics[width=0.4\linewidth]{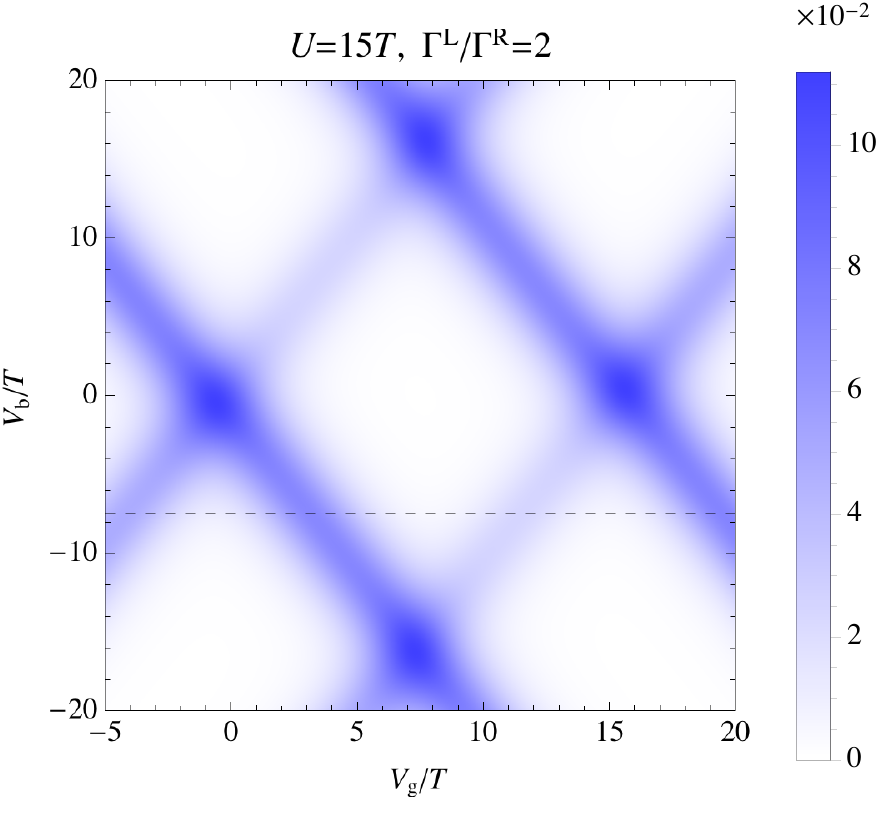} 
		\includegraphics[width=0.4\linewidth]{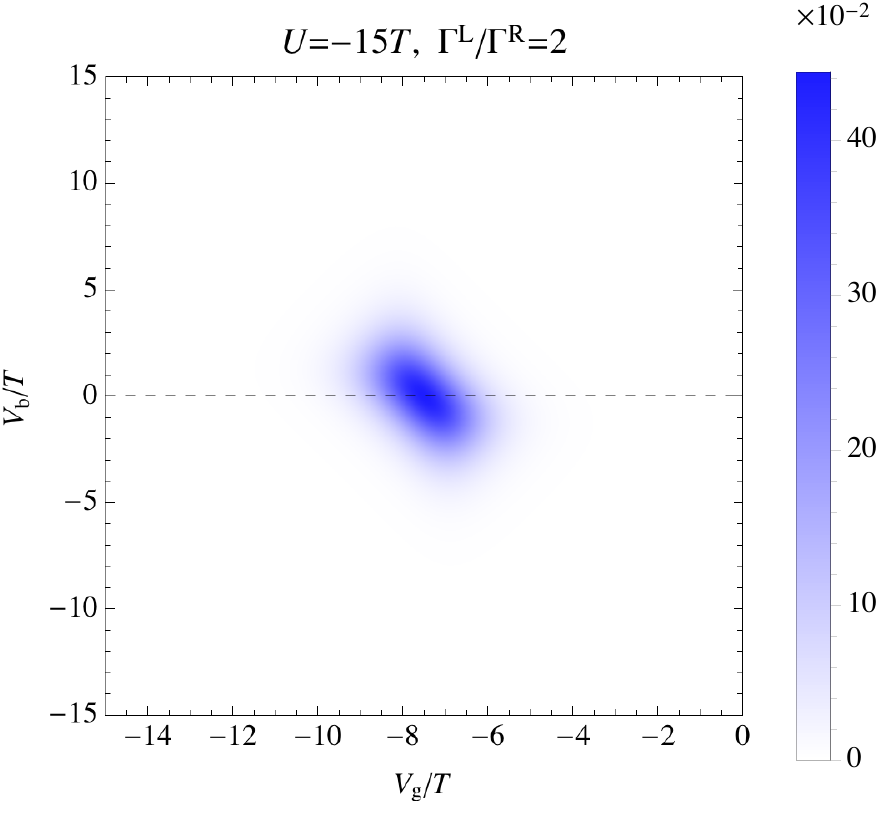}
		\label{fig:coup-gate-B}
	}
	\subfigure[~Curvature $\F_{V_\b,\Gamma^\R} \cdot \Gamma^\L $ vs. driven bias parameter $V_\b$ and static gate parameter $\epsilon$.]{
		\includegraphics[width=0.4\linewidth]{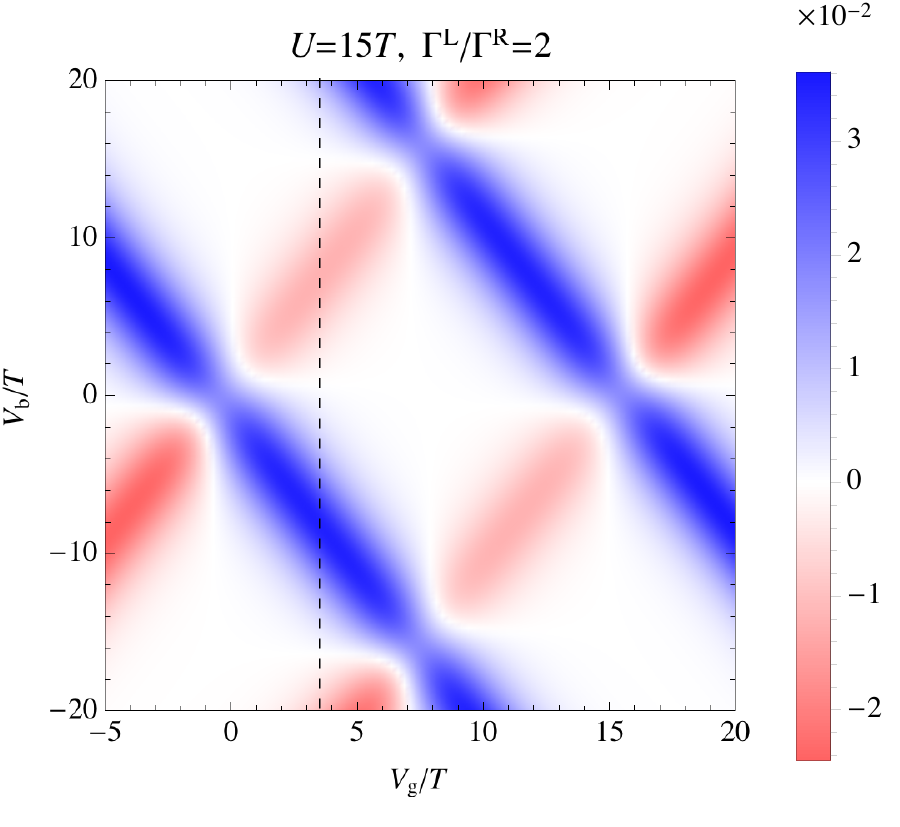}
		\includegraphics[width=0.39\linewidth]{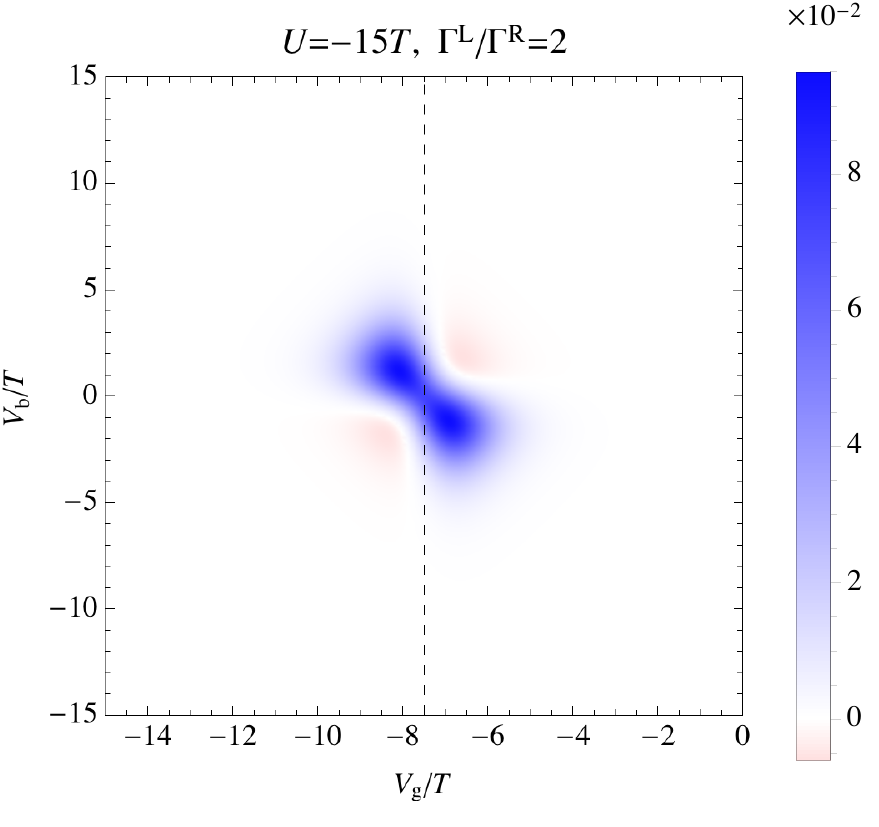}
		\label{fig:coup-bias-B}
	}
	\subfigure[~Curvature $\F_{U,\Gamma^\R} \cdot \Gamma^\L$ vs. static parameters $\epsilon$ and $V_\b$.]{
		\includegraphics[width=0.4\linewidth]{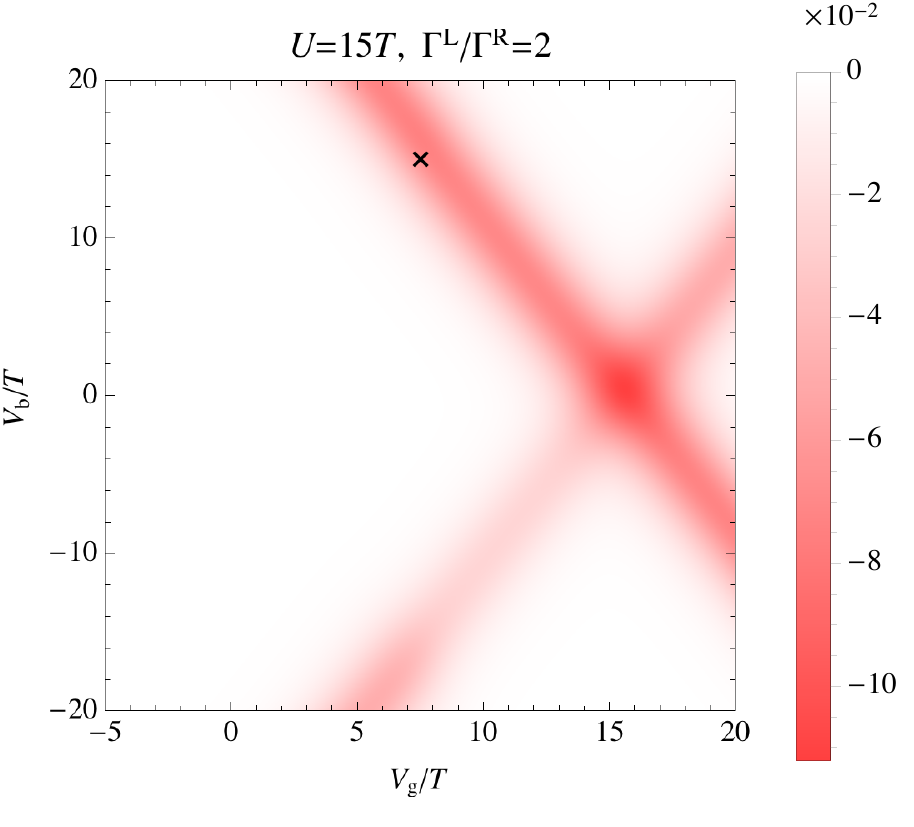}
		\includegraphics[width=0.4\linewidth]{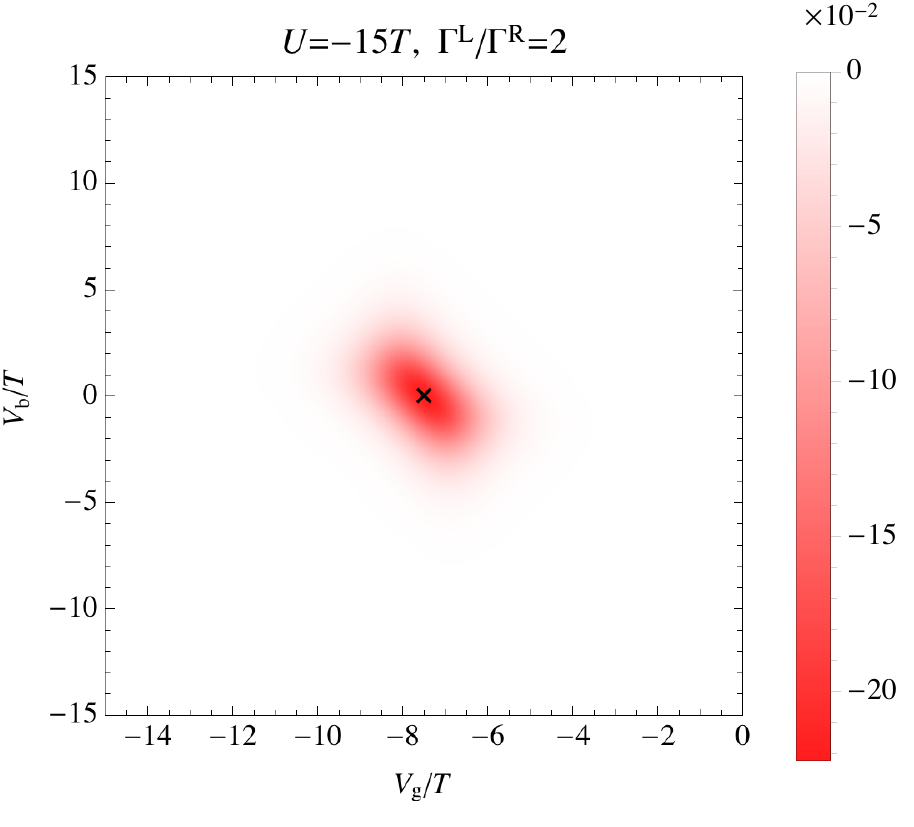}
		\label{fig:coup-int-B}
	}
	\caption{
		Pumping curvatures of \Fig{fig:coup-curv} as function of gate and bias voltage,
		i.e., these curvatures are plotted as function of one \emph{driven} and one \emph{static} parameter,
		having fixed the driven $\Gamma^\R$ as indicated in \Fig{fig:coup-curv}.
		\label{fig:coup}}
\end{figure*}

In the right panels of \Fig{fig:coup-curv}(a)-(b) we show the corresponding results for attractive interaction (same strength but opposite sign, $U=-|U|$).
In contrast to the case $U>0$, for the first two driving protocols the response is nonzero only at a \emph{single}, thermally-broadened vertical line, $\epsilon-|U|/2=\mu$.
For this line to appear at all, a second condition must be satisfied\footnote
	{For all plots for $U<0$ in the right panels of \Fig{fig:coup-curv} we choose the static parameters such that both conditions can be satisfied somewhere in the driving plane.
	For other static parameters, the curvature in the right panels is zero throughout the entire plane (not shown), in contrast to the $U>0$ cases on the left which generically show some response.},
$V_\b=0$.
These conditions correspond to the configuration labeled $\C$ in \Fig{fig:levels}.
Notably, neither of them is contained in \Eq{eq:resonance-conditions}.
The reason for this is, that attractive interaction suppresses all rates
in the region around the symmetry point\footnote
	{Recall that we suppose the temperature to be large enough to ensure that the exponentially suppressed first order rates are still larger than their second order tunneling correction. In the figures, we nonetheless chose relatively large values of $|U|/T$ for a clear comparison with the figures of the repulsive case. However, the discussed effects dominate as long as $|U| > T$.}.
This renders the resonance conditions \eq{eq:resonance-conditions} irrelevant and gives rise to new effective parameters as will become understandable from a later discussion (see paragraph \ref{sec:C-resonance}).
This is further underlined by the right panels of \Fig{fig:coup} where we plot each curvature
as function of another non-driven parameter (instead of $\Gamma^\R/\Gamma^\L$).
In each case the $U<0$ response reduces to a single thermally broadened point defined by the above two conditions, in contrast to the results for $U>0$ in the left panels of \Fig{fig:coup}.

Finally, the sign changes in the curvature when driving $\Gamma^\R/\Gamma^\L$ and bias are a qualitative difference in the pumping response when compared to driving $\Gamma^\R/\Gamma^\L$ and the gate voltage.
The reason is, that for attractive $U < 0$, bias and gate driving cannot be mapped into each other, see \Fig{fig:bias-effect} (b).

\subsubsection{Driving the interaction $U$}
Finally, we consider driving $\Gamma^\R/\Gamma^\L$ together with the interaction $U$, which can be driven around both repulsive ($U>0$) and attractive ($U<0$) values,
see \Fig{fig:coup-int-curv}.
In these cases, \emph{whenever} there is a response, the driving of $U$ can be understood as effective driving of $\epsilon = - V_\g$,
meaning that no new mechanisms are accessed by driving $U$ in addition to $\Gamma^\R/\Gamma^\L$.
Qualitatively, this may be rationalized in terms of the levels sketched in 
\Fig{fig:bias-effect}.

More quantitatively, 
for $U>0$ the response is nonzero at the single line defined by
a condition $\mu^\alpha= \epsilon+U$
for either $\alpha=\L$ or $\R$.
[\Eq{eq:resonance-conditions}]
Close to each resonance line $\mu^\alpha= \epsilon+U$:
\begin{align}
	\F_{U,\Gamma^\R}(\epsilon)
	&
	\approx
	\M_{\alpha}[\epsilon+U-\mu^\alpha,\Gamma^\R/\Gamma^\L]
	\approx
	\F_{\epsilon,\Gamma^\R}(U)
	\label{eq:equal-coup-int-gate2}
	,
\end{align}
where $\M_\alpha$ is the curvature due to driving of the effective parameters
$\epsilon + U - \mu^\alpha$ and $\Gamma^\R/\Gamma^\L$.
The configuration corresponding to this single condition is not listed in \Fig{fig:levels}
nor in \Tab{tab:mechanisms}.
The different sign in \Fig{fig:coup-int-B} relative to \Fig{fig:coup-gate-B} is merely because we plot versus $V_\g = -\epsilon$.
Note however, that \Fig{fig:coup-int-B} shows no response to $U$-driving at the other two lines $\mu^\alpha= \epsilon$,
whereas $V_\g$-driving clearly has an effect there, see \Fig{fig:coup-gate-B}.
This is clear from the transition energies sketched in \Fig{fig:bias-effect}(a) and the fact that the effective parameter $\epsilon -\mu^\alpha$ is independent of $U$.

For $U<0$, there is again only a single resonance line at $\epsilon-\tfrac{1}{2}|U|=\mu$,
which, moreover, only appears if the additional condition $V_\b=0$ is satisfied.
This corresponds to the configuration labeled $\C$ in \Fig{fig:levels}.
For $|U| \gg T$, the response around this line obeys
\begin{align}
	\F_{U,\Gamma^\R}(\epsilon)
	&
	\approx
	\M_{\C'}[\epsilon+\tfrac{1}{2}U-\mu, \Gamma^\R/\Gamma^\L]
	\approx
	\tfrac{1}{2} \F_{\epsilon,\Gamma^\R}(U)
	\label{eq:equal-coup-int-gate}
	.
\end{align}
This relation reflects that the shared effective parameter that is driven is now
$\epsilon + \tfrac{1}{2}U - \mu$.
Here $\C'$ indicates that the working point is the one labeled $\C$ in \Fig{fig:levels},
whereas the prime denotes that the coupling is the second driving parameters
(rather than the bias $V_\b$, as discussed later in \Eq{eq:relation-C}).
We stress that \Eq{eq:equal-coup-int-gate2} and \eq{eq:equal-coup-int-gate} are two \emph{different} relations (governed by two different mechanisms) between the \emph{same} pair of curvature components.

\subsubsection{Summary.} 
Although for repulsive interaction $U>0$ driving the coupling $\Gamma^\R$ is indeed a simple way to achieve pumping,
for attractive $U<0$ the possibilities are limited by the effect of the inverted Coulomb gap.
This also applies when the second driving parameter is the interaction $U$ itself:
whenever this leads to an effect, it can be understood as an effective gate-driving which is subject to the same limitations.
Driving $U$ is nevertheless interesting since it selectively picks out a transition in the \emph{many-body} spectrum of the dot ($1 \leftrightarrow 2$), which the other considered parameter drives cannot do.
\subsection{Driving two parameters for static coupling\label{sec:nocoupling}}

We now turn to driving protocols in which the coupling ratio $\Gamma^\R/\Gamma^\L$ is fixed.
In all these cases, the pumping is localized in thermally broadened regions around points
(rather than around lines as for coupling driving, see Fig. \ref{fig:coup-curv}).
This is interesting for the purpose of geometric pumping spectroscopy~\cite{Reckermann10a,Calvo12a,Pluecker17a}.

\subsubsection{Repulsive interaction $U>0$}

\begin{figure*}
	\subfigure[~Curvature $\F_{\epsilon,V}$ vs. gate and bias \emph{driving} parameters.]{
		\includegraphics[width=0.4\linewidth]{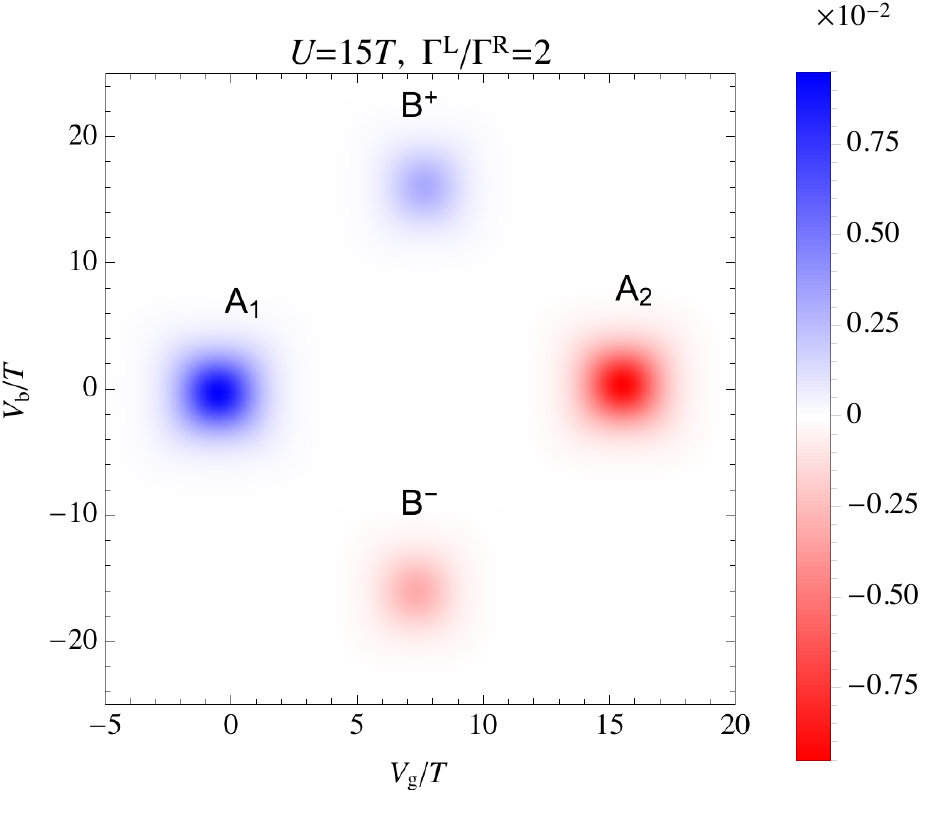}
		\includegraphics[width=0.39\linewidth]{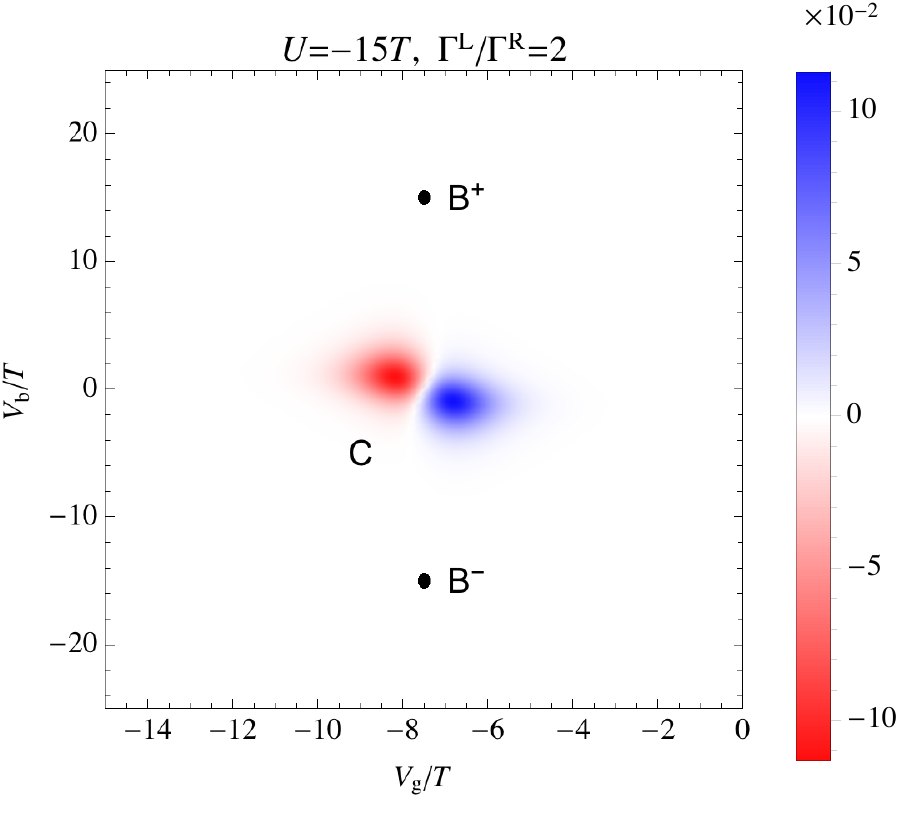}
		\label{fig:gate-bias-B}
	}\\
	\subfigure[~Occupation number $\braket{N}$ given by \Eq{eq:occupation}.
	]{
		\includegraphics[width=0.4\linewidth]{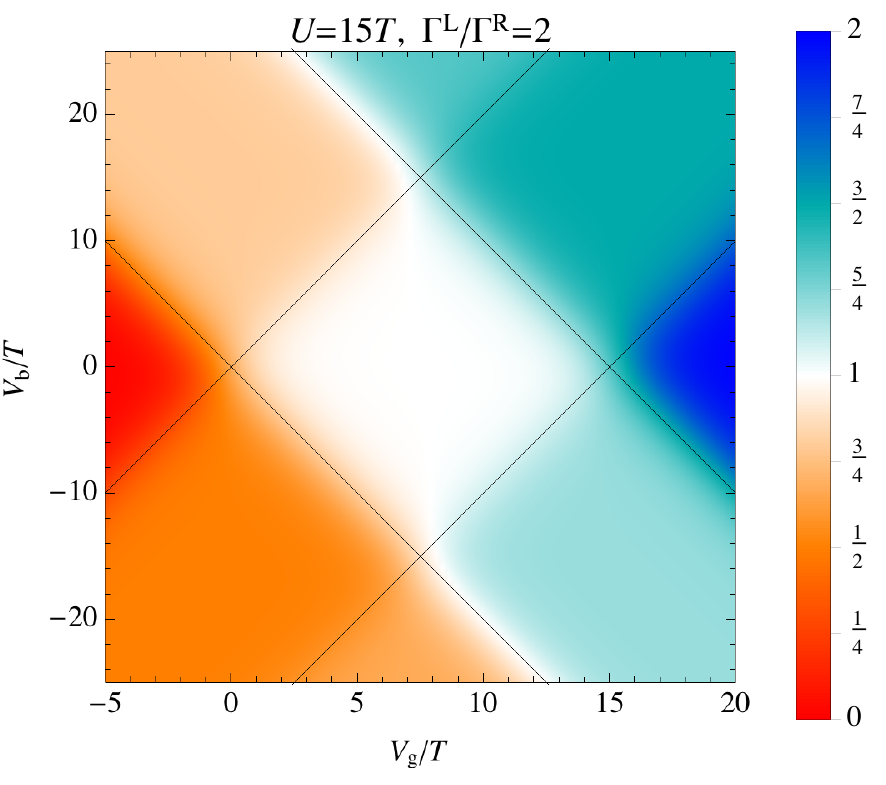}
		\includegraphics[width=0.4\linewidth]{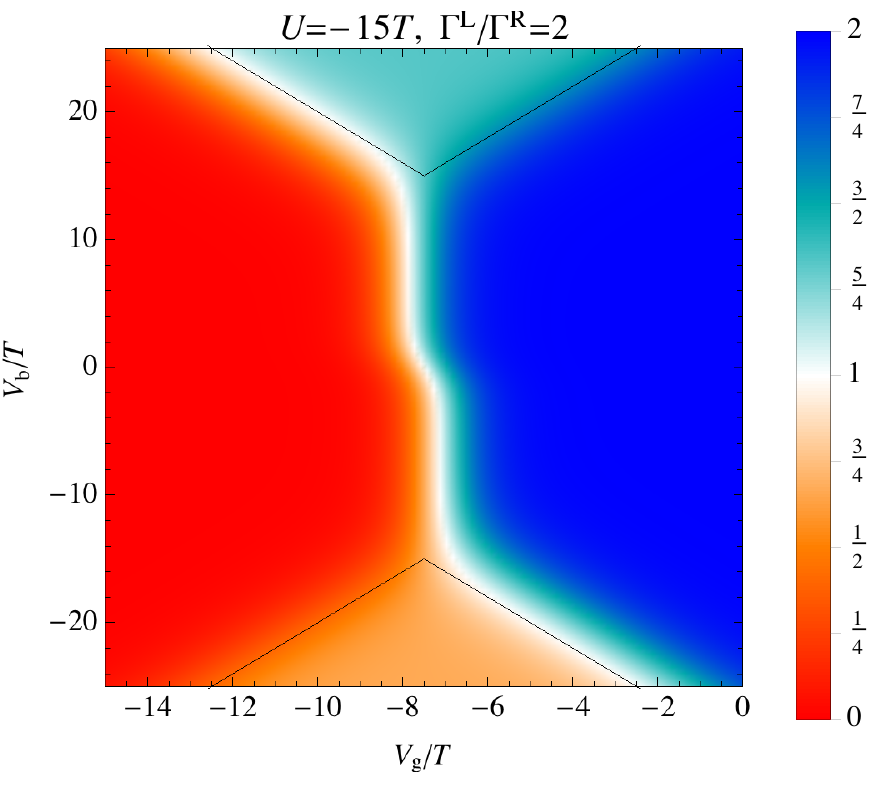}
		\label{fig:gate-bias-nExp}
	}\\
	\subfigure[~Ratio of charge-relaxation rates $w^\R/(w^\R + w^\L)$ given by \Eq{eq:decayratio}.
	]{
		\includegraphics[width=0.4\linewidth]{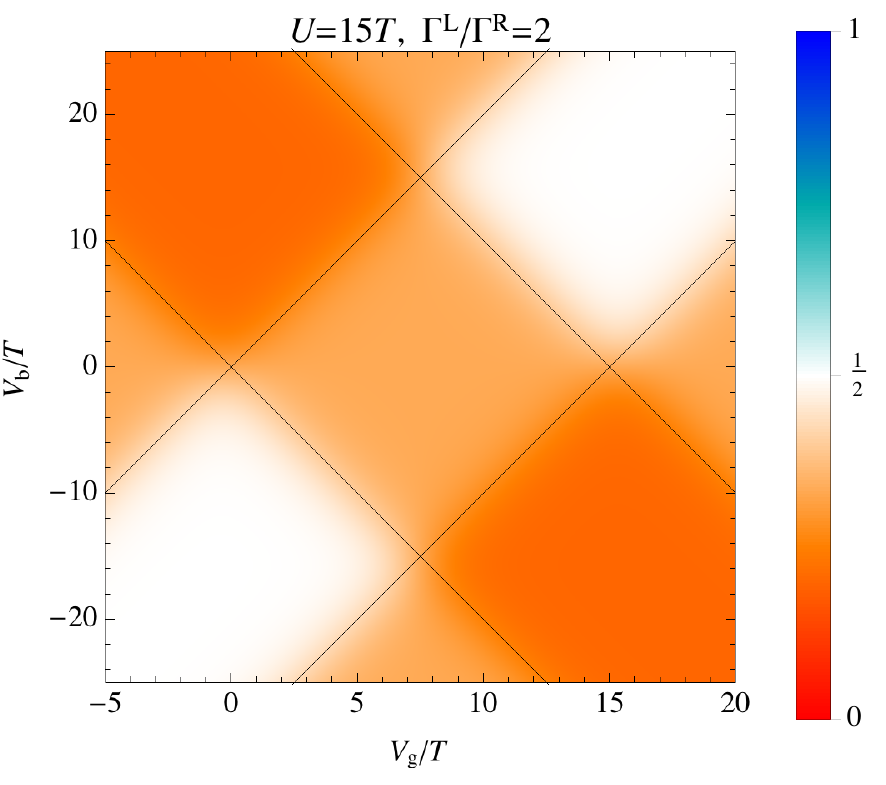} 
		\includegraphics[width=0.4\linewidth]{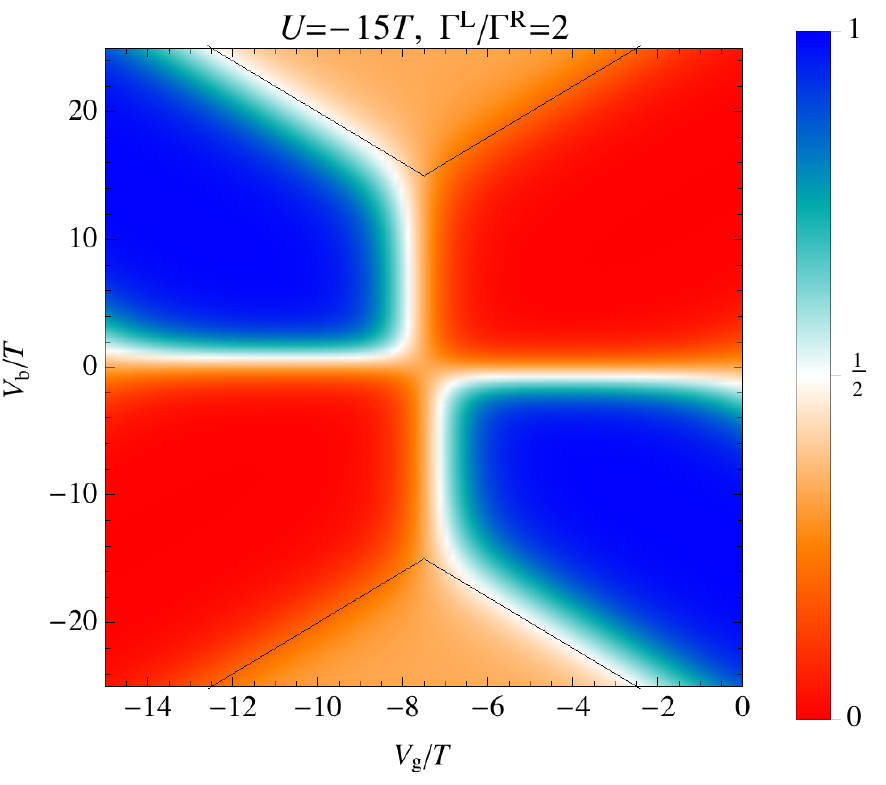} 
		\label{fig:gate-bias-decayRatio}
	}\\
	\caption{
		Pumping response for gate and bias driving for static $U>0$ (left) and $U<0$ (right).
	}
	\label{fig:gate-bias}
\end{figure*}

In \Fig{fig:gate-bias}(a) we show for reference the curvature when driving gate ($\epsilon$) and bias voltage ($V_\b$).
The response in the driving parameter plane is now restricted to thermally broadened crossing points of the edges of the Coulomb diamonds.

This has been related to the requirement of varying (at least) two independent parameters to achieve pumping,
in particular two parameters that change the occupations~\cite{Reckermann10a,Calvo12a,Pluecker17a}.
Indeed, using \Eq{eq:Fall} we can separate the charge response into its two factors which are plotted separately in \Fig{fig:gate-bias}(b)-(c).
Whenever both of these quantities depend on the same single effective parameter (as happens at the edges between crossing points), the gradients in the crossproduct are parallel and the pumping curvature is zero.
In the present case of fixed coupling and repulsive interaction, the two gradients can be nonparallel only  only at the crossing of two resonance lines \eq{eq:resonance-conditions} where two effective parameters emerge.
This is where the occupations change, confirming the above intuitive explanation in this case.

The pumping response points come in pairs with opposite sign. However, around each resonance point, the curvature has a definite sign ('monopolar' character) which has been related to the change of the ground state degeneracy in Refs.~\cite{Reckermann10a,Calvo12a,Pluecker17a}

\subsubsection{Attractive interaction $U<0$}
The corresponding results for attractive interaction $U<0$ are shown in the right panels of \Fig{fig:gate-bias}(a).
The curvature shows only a single, thermally-broadened resonance when the two conditions
$V_\b=0$ and $\epsilon-|U|/2 = \mu$ are satisfied.
This resonance is thus due to the $\C$-mechanism. It has an internal node where the curvature changes sign ('dipolar' character) in the driving-parameter plane.

Importantly, the response in the right panel of \Fig{fig:gate-bias}(a) cannot be understood --even qualitatively-- based on the changes in the occupations of the quantum dot [conditions \Eq{eq:resonance-conditions}]
plotted in \Fig{fig:gate-bias}(b).
The charge is plotted in \Fig{fig:gate-bias}(b) and changes only along a vertical line $\epsilon-|U|/2 = \mu$ with a kink that is discussed below.
However, there is no \emph{crossing} of resonance lines (in the occupation) here.
Furthermore, when at much larger bias $|V_\b| \geq U$ there are such crossing lines where the charge changes, then the pumping response is absent.
Thus, the $\B$ mechanism sketched for attractive $U<0$ in \Fig{fig:levels} does not lead to a pumping response in the single-dot model [cf. \Sec{sec:results-real}].
Thus, the observations that $\C$ arises at all and $\B$ is missing
are surprising in view of the success of the intuitive explanation for the $U>0$ case.
However, the origin of the pattern becomes clear from the following analysis of the two factors plotted in \Fig{fig:gate-bias}(b)-(c), of which the gradients need to be calculated in order to obtain the pumping curvature [\Eq{eq:Fall}].

\paragraph{Presence of a single $C$ resonance for $|V_\b| \ll -U$. \label{sec:C-resonance}}
Whereas $\braket{N}$ changes only at the vertical line \Fig{fig:gate-bias}(b),
the factor containing the charge relaxation rates $w^\alpha$ additionally changes at a horizontal line $V_\b\approx0$ in \Fig{fig:gate-bias}(c).
Both factors have a fundamentally different dependence on gate and bias voltage compared to repulsive case (left panels in the same figure).
Close to the $\C$-point the two gradients are thus orthogonal, leading to a resonance restricted by the thermal energy in both the $\epsilon$ and $V$ direction.
These
features of the two factors are intimately tied to the strong attractive interaction on the quantum dot as we explain in the following.
In simple terms, the attractive gap stabilizes charge states $N=0,2$ on the quantum dot.
In the weak coupling, high temperature regime that we consider,
a transition between the $N=0$ and $N=2$ states is induced already by two sequential, first order processes
both of which are suppressed.
What matters for $\braket{N}$ entering the curvature formula \eq{eq:Fall} is only the balance between these two competing charge transitions, \emph{irrespective} of which electrode induces them, which occurs when
\begin{align}
	\sum_{\alpha} W^\alpha_{10} = \sum_{\alpha} W^\alpha_{12}
	\label{eq:balance-charge}
	,
\end{align}
and charge state 0 (2) is stable when the right (left) side dominates.
Because the attractive interaction with $-U=|U| \gg T$ suppresses all rates that appear in the condition \eq{eq:balance-charge} up to sizeable bias $|V_\b| \lesssim U$ and gate $0 \lesssim \epsilon \lesssim |U|$, the balance is determined by the tails of the reservoir distribution functions. The line at which $\braket{N}$ changes is thus given by the condition
\begin{align}
 	\epsilon - \tfrac{1}{2} |U| - \mu =
 	 T
 	 \tfrac{1}{2}
 	 \ln \frac{\Gamma^\L e^{V_\b/T} + \Gamma^\R e^{-V_\b/T}}
 	          {\Gamma^\R e^{V_\b/T} + \Gamma^\L e^{-V_\b/T}}
 	 \label{eq:balance}
	 .
\end{align}
This condition is only weakly bias dependent: the left hand side depends on $\mu$ only;
the right-hand side introduces a kink shifting the vertical line's position to
$\epsilon = \pm T \ln \Gamma^\L/\Gamma^\R$ for $U \gg \pm V_\b \gg T$.

In contrast, the balance of charge relaxation rates, 
the other factor in the curvature formula \eq{eq:Fall}, strongly depends on the bias.
The charge relaxation rate $w^\alpha$, given by \Eq{eq:charge-rel}, quantifies how fast the state $N=1$ can be reached due to a transition induced by a specific reservoir $\alpha$,
\emph{irrespective} of the initial state of the dot (0 or 2).
In this case there is thus a balance when
\begin{align}
	\sum_{N=0,2} W^\L_{1N} = \sum_{N=0,2} W^\R_{1N}
	.
	\label{eq:motivate}
\end{align}
This yields a further condition: up to sizeable bias $|V_\b| \lesssim U$ and gate $0 \lesssim \epsilon \lesssim |U|$,
this implies 
\begin{align} \label{eq:balance_b}
	V_\b =  \pm T \ln \frac{\Gamma^\L}{\Gamma^\R}
\end{align}
for $\epsilon - \tfrac{1}{2}|U| - \mu \gg T$
and $\epsilon - \tfrac{1}{2}|U| - \mu \ll  -T$, respectively.
\Eq{eq:balance} and \Eq{eq:balance_b} explain why the naive conditions \eq{eq:resonance-conditions}
do not define the effective pumping parameters,
which in this case are $\epsilon - \tfrac{1}{2} |U| - \mu$ and $V_\b$.

\begin{figure*}
	\subfigure[~Curvature $\F_{U,V_\b}$ vs. interaction and bias \emph{driving} parameters.
	]{
		\includegraphics[width=0.4\linewidth]{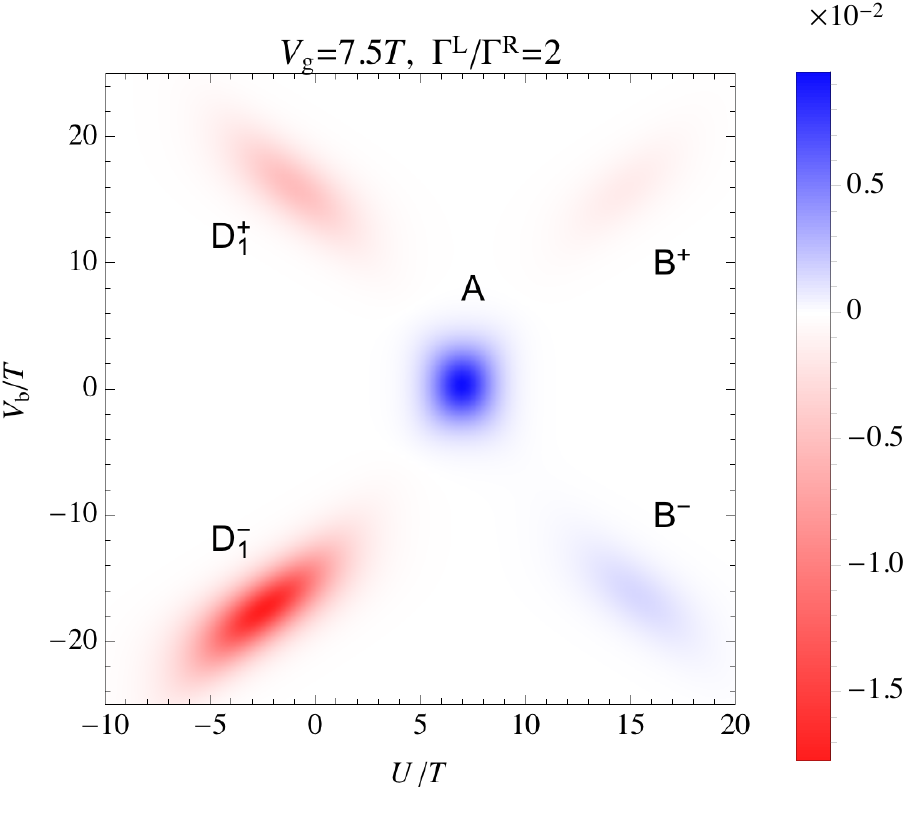}
		\includegraphics[width=0.39\linewidth]{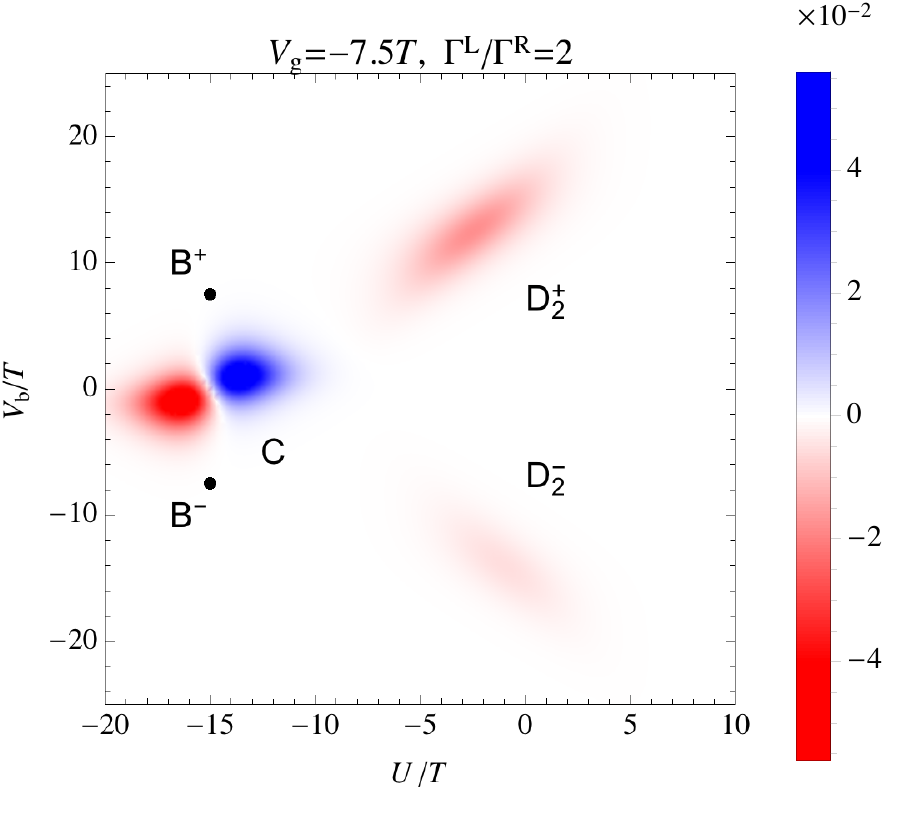}
		\label{fig:bias-int-B}
	}\\
	\subfigure[~Occupation number $\braket{N}$ given by \Eq{eq:occupation}.]{
		\includegraphics[width=0.4\linewidth]{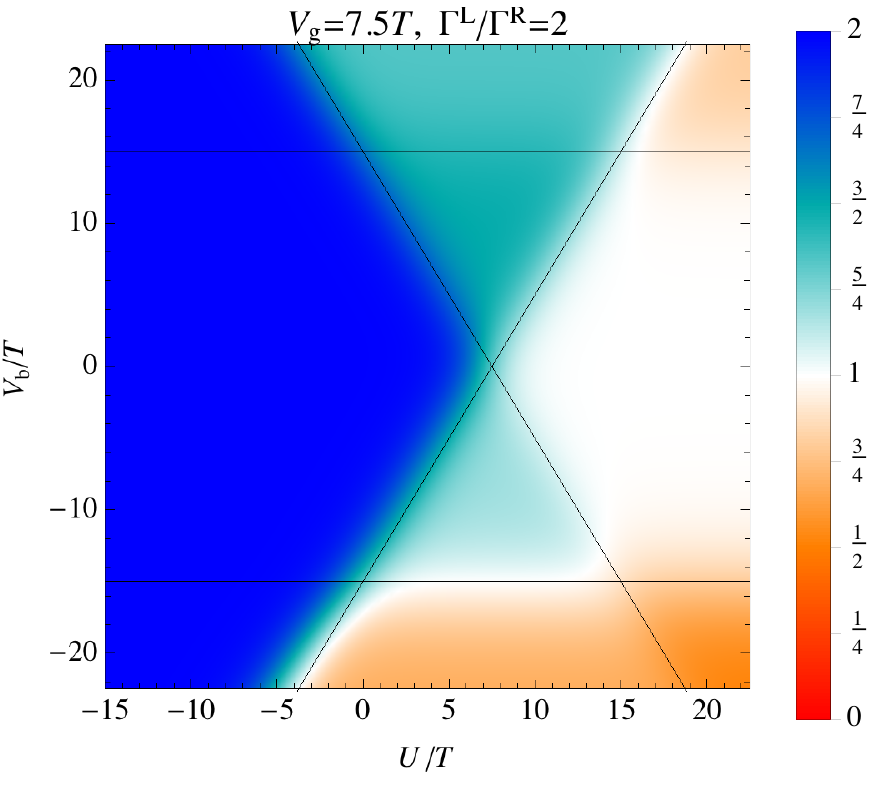}
		\includegraphics[width=0.4\linewidth]{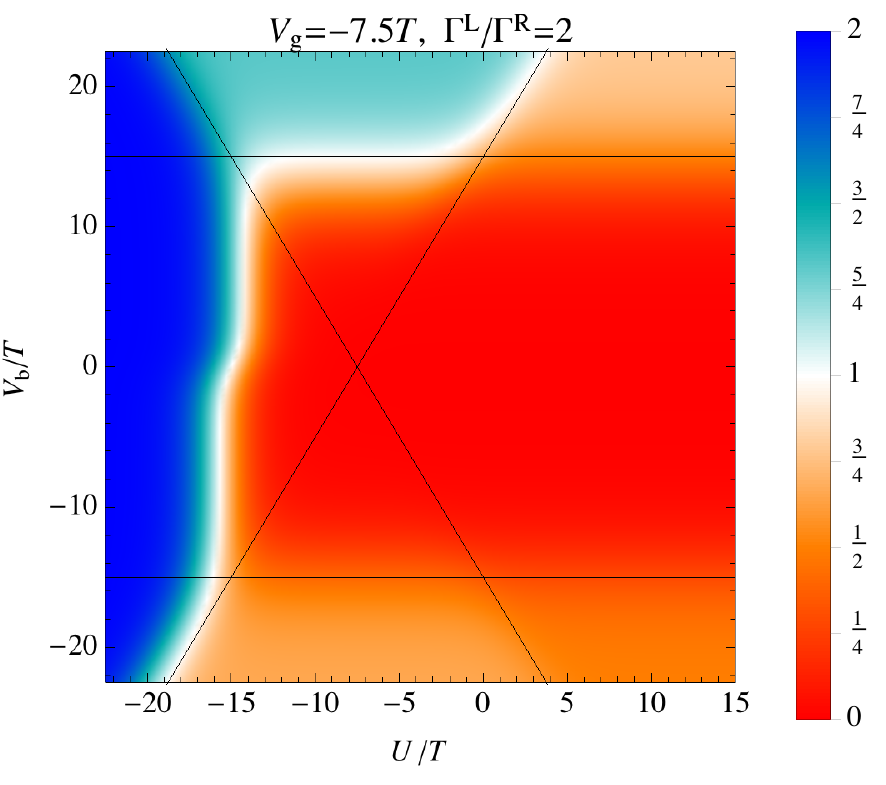}
		\label{fig:bias-int-nExp}
	}\\
	\subfigure[~Ratio of charge-relaxation rates $w^\R/(w^\R + w^\L)$ given by \Eq{eq:decayratio}.]{
		\includegraphics[width=0.4\linewidth]{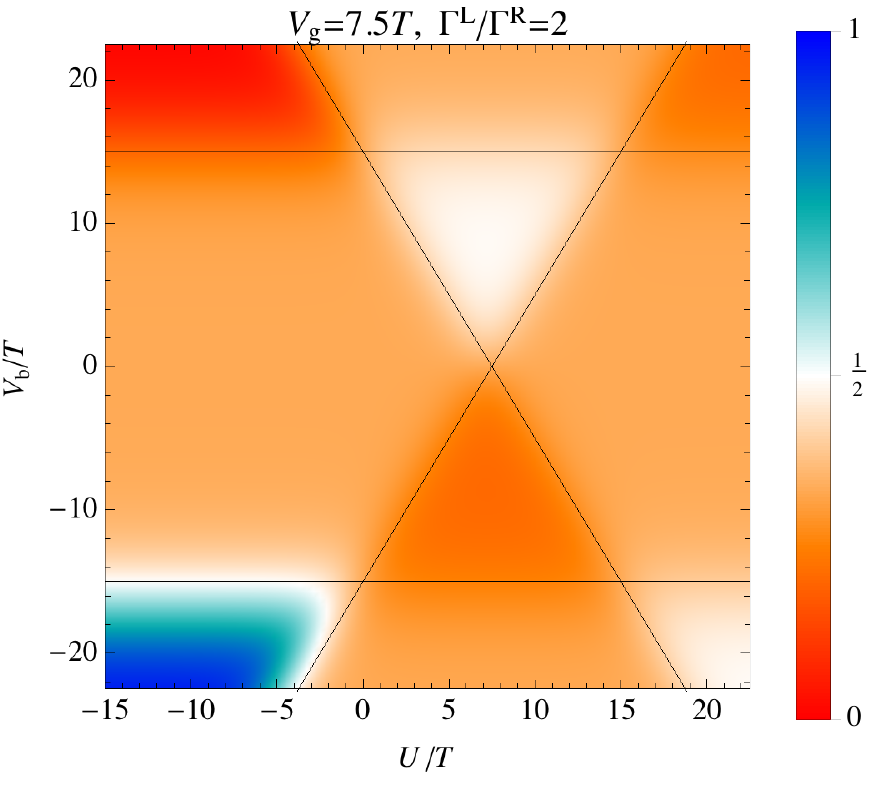}
		\includegraphics[width=0.4\linewidth]{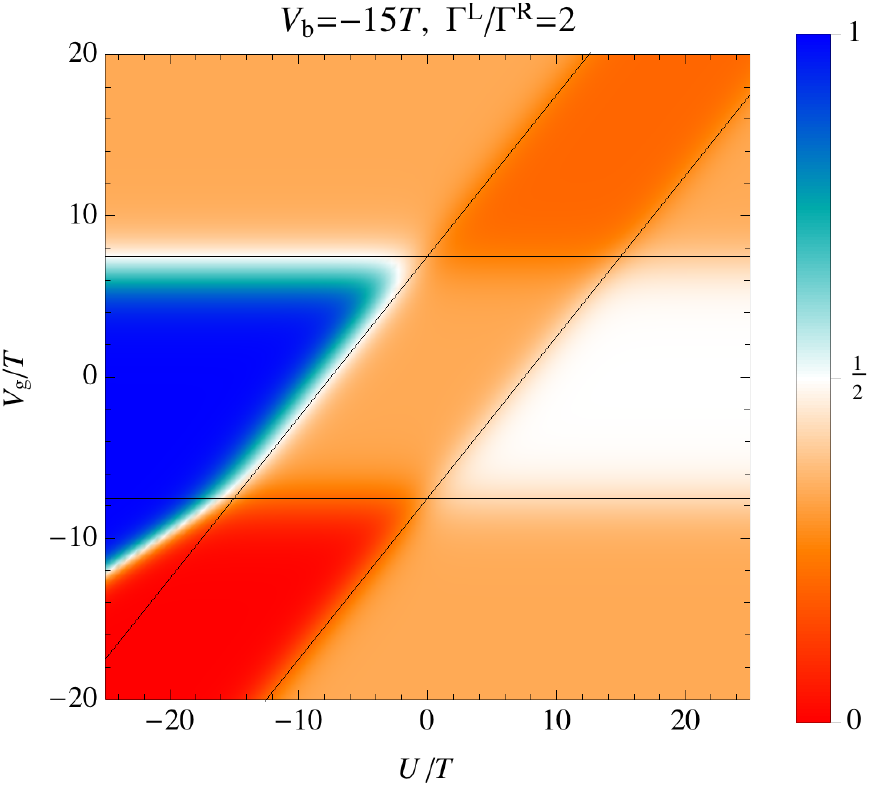}
		\label{fig:bias-int-decayRatio}
	}\\
	\caption{
		Pumping response for interaction and bias driving for static $\epsilon < \mu$ (left) and $\epsilon > \mu$ (right). We note that even for small gate voltages, $|\epsilon - \mu| \ll T$,  (not shown)
			the resonances $\A,\B,\C$ shown in the two upper panels merge to a nonvanishing pumping response in the vicinity of  $U=0$ and $V_\b =0$.		
	}
	\label{fig:bias-int}
\end{figure*}

As a unique feature of the $\C$-resonance is that its curvature profile is 'dipolar'. 
This can now be understood as caused by the competition between two
suppressed processes,
involving the $0 \rightarrow 1$ or the $2 \rightarrow 1$ transition.
 However, for
negative interaction these transitions only become active together around the point marked $\C$.
We either get a positive or negative pumped charge to the left or right of this point
when one of the processes dominates.
Which one dominates depends on both the asymmetry in the couplings ($\Gamma^\L$ vs $\Gamma^\R$)
and in the chemical potential differences ($\epsilon-\mu$ vs. $\epsilon-|U|-\mu$).
In order to fully analyze the shape, we use that for $-U\gg T$ the curvature is well-described by\footnote{This expression also shows explicitly that the curvature indeed only depends on the effective parameters $\epsilon-{U}/2-\mu$ and $V_\b$.}
\begin{align}
	\F_{\epsilon,V_\b}
	=
	& {\Gamma^\L \Gamma^\R} \times
	\label{eq:shape}
	\\
	&
	\frac{\Gamma^\R \sinh(\frac{\epsilon+U/2-\mu^\L}{T})
		+ \Gamma^\L  \sinh(\frac{\epsilon+U/2-\mu^\R}{T})}
	{\Big[ \Gamma^\L \cosh(\frac{\epsilon+U/2-\mu^\L}{T})
		+ \Gamma^\R \cosh( \frac{\epsilon+U/2-\mu^\R}{T} )\Big]^3 }
	\notag
	.
\end{align}
The asymmetry of the two-lobed resonance in \Fig{fig:gate-bias}(b) is due the coupling asymmetry and can be quantified by the slope of the nodal curve separating the two lobes:
linearizing the numerator of \Eq{eq:shape}, 
using $\epsilon+U/2 - \mu_\alpha$ with $\alpha=\L, \R$ as variables, shows that the slope of the tangent directly gives the junction asymmetry:
\begin{align}
	\frac{ \partial (\epsilon +U/2- \mu^\L) }
	{ \partial (\epsilon +U/2 - \mu^\R) }
	= - \frac{ \Gamma^\L }{\Gamma^\R}\ .
\end{align}

\paragraph{Absence of other resonances.\label{sec:absentB}}
It remains to explain why the factors in the curvature \eq{eq:Fall}
do \emph{not} lead to any other response
just described,
in particular due to the $\B$-mechanism.
This is surprising since
\emph{both} occupations and the ratio of charge-decay rates [\Fig{fig:gate-bias} (b)-(c)]
show drastic changes around the $\B$-configuration
and, moreover, for $U > 0$ the corresponding $\B$-mechanism in \Fig{fig:levels} does lead to pumping.
However, when $U<0$ the positions of $\epsilon$ and $\epsilon+U$ are interchanged
which causes the two factors in \eq{eq:Fall} to become \emph{equal}
for $\mu^\L > \epsilon > \epsilon - |U| > \mu^\R$
and $V_\b \gg T$
\begin{align}
	\frac{w^\R - w^\L}{w^\R + w^\L}
	 \approx
	\frac{W^\R_{10}- W^\L_{12} }
	{W^\R_{10} + W^\L_{12}}
	\approx \braket{N}
	.
	\label{eq:reason1}
\end{align}
For opposite bias $-V_\b \gg T $ they are \emph{opposite}: 
for $\mu^\R > \epsilon > \epsilon - |U| > \mu^\L$ we find
\begin{align}
	\frac{w^\R - w^\L}{w^\R + w^\L}
	\approx
	\frac{W^\R_{12} - W^\L_{10}}
	{W^\R_{12}+ W^\L_{10} }
	\approx - \braket{N}
	.
	\label{eq:reason2}
\end{align}
Thus, the gradients in \Eq{eq:F} are either parallel or antiparallel and the response, given by their cross product \eq{eq:F}, remains zero around the $\B$-points in \Fig{fig:gate-bias}.

\subsubsection{Driving the interaction $U$}

\begin{figure*}
	\subfigure[~Curvature $\F_{U,V_\g}$ vs. interaction and gate-voltage \emph{driving} parameters.]{
		\includegraphics[width=0.4\linewidth]{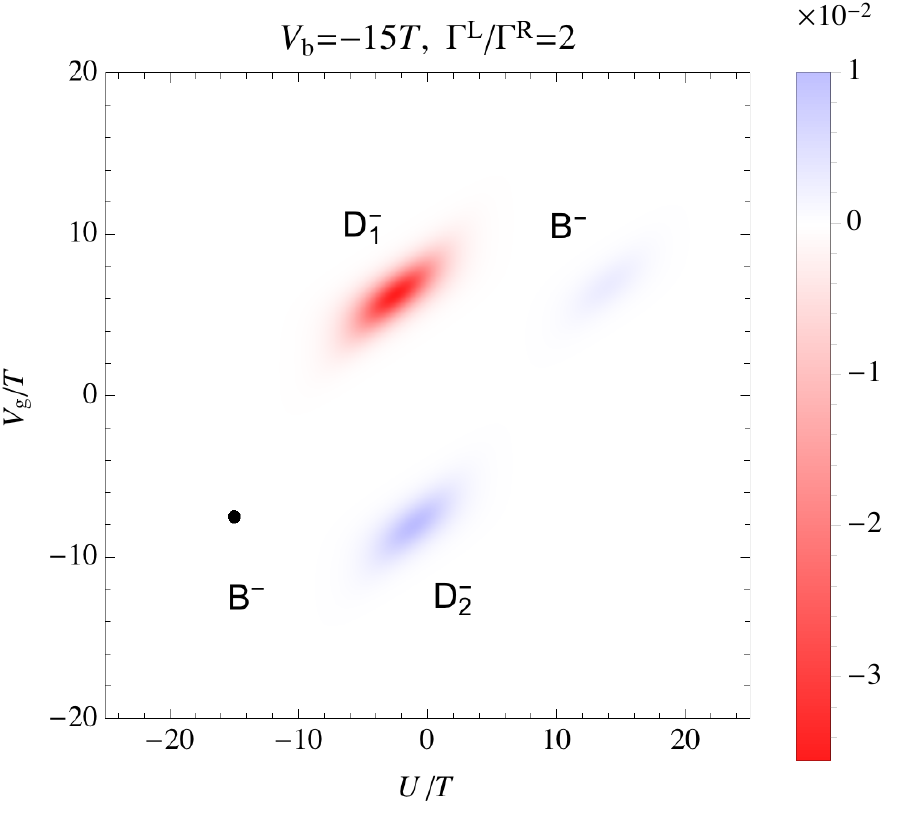}
		\includegraphics[width=0.4\linewidth]{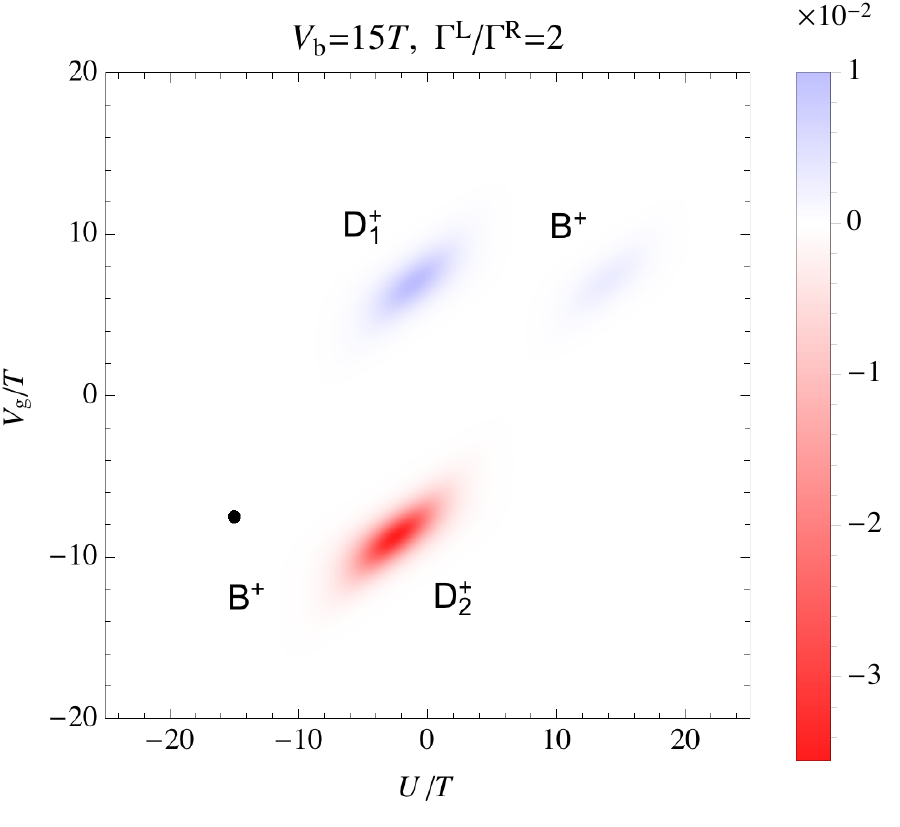}
		\label{fig:gate-int-B}
	}\\
	\subfigure[~Occupation number $\braket{N}$ given by \Eq{eq:occupation}.]{
		\includegraphics[width=0.4\linewidth]{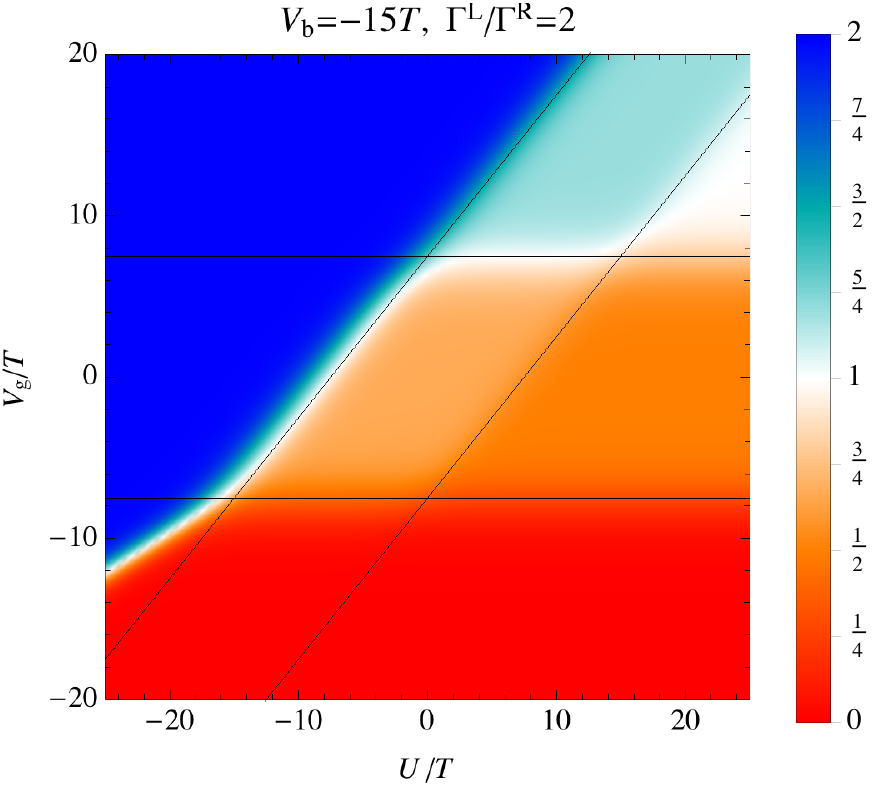}
		\includegraphics[width=0.4\linewidth]{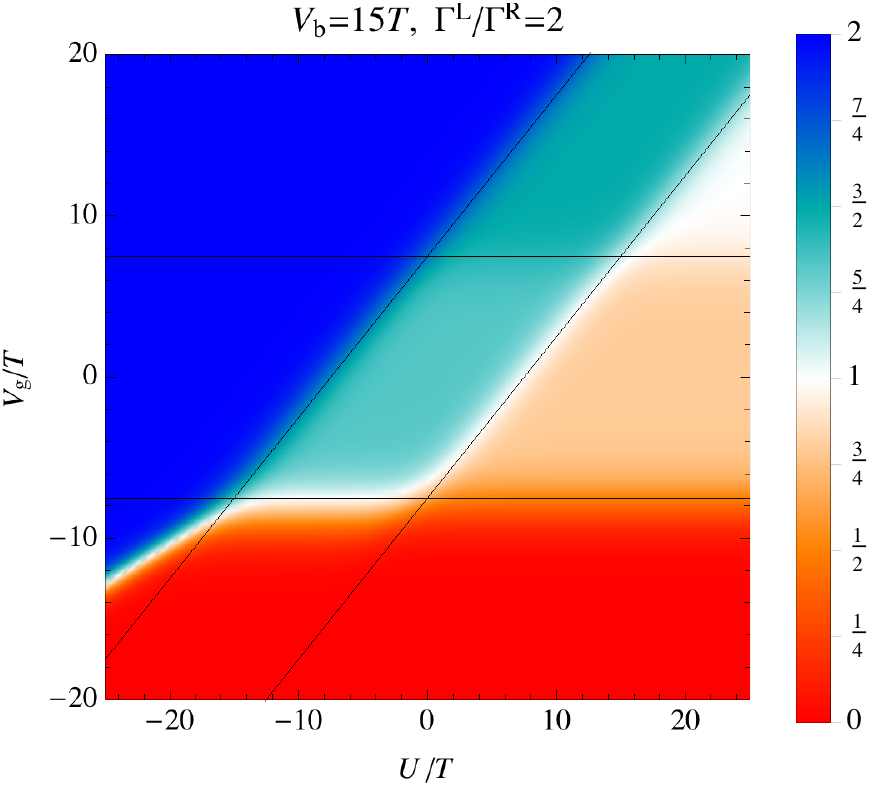}
		\label{fig:gate-int-nExp}
	}\\
	\subfigure[~Ratio of charge-relaxation rates $w^\R/(w^\R + w^\L)$ given by \Eq{eq:decayratio}.]{
		\includegraphics[width=0.4\linewidth]{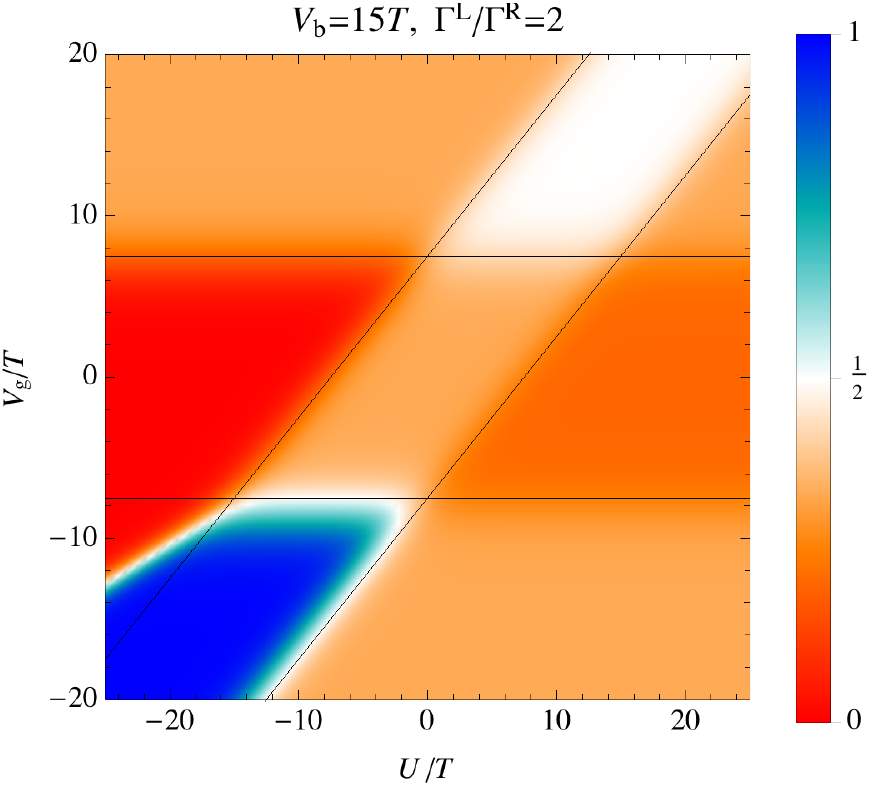}
		\includegraphics[width=0.4\linewidth]{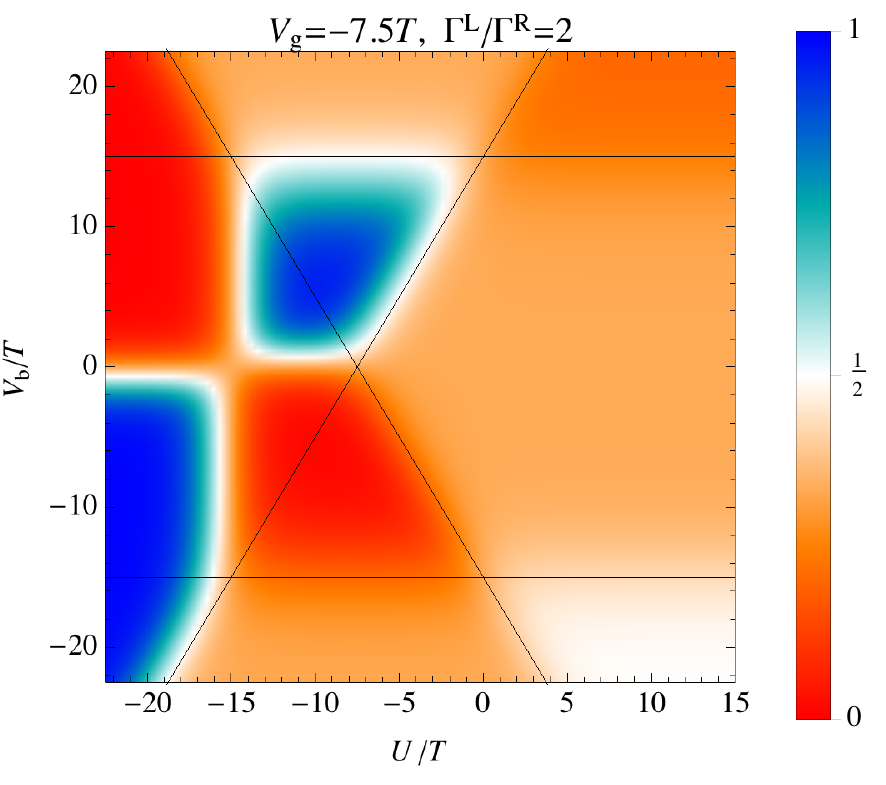}
		\label{fig:gate-int-decayRatio}
	}\\
	\caption{Pumping response for interaction and gate driving
		for static $V_\b > 0$ (left) and $V_\b < 0$ (right).
	}
	\label{fig:gate-int}
\end{figure*}

Finally, we discuss the response when driving the interaction
together with a second parameter, as summarized in Figs.~\ref{fig:bias-int} and \ref{fig:gate-int}.
In addition to a number of features that can be mapped to other driving protocols via the previously discussed mechanisms, we importantly also find a new mechanism that is unique to interaction driving, the mechanism $\D$.
It is operative at working points with zero interaction $U=0$
and either $\mu^\L = \epsilon$ or $\mu^\R = \epsilon$,
i.e., where the $0\to 1$ and $1\to 2$ transitions are degenerate
and both are resonant with either source or drain,
as sketched in \Fig{fig:levels}.
The two effective pumping parameters of the $\D$ mechanism are thus
$\epsilon-\mu^{\alpha}$ and $\epsilon+U-\mu^{\alpha}$ for $\alpha=\L$ or $\R$,
or, equivalently,
$\epsilon-\mu^\alpha$ and $U$.
Only by driving $U$ we can modulate both of them independently.

This pumping is remarkable, since when $U$ is not driven but fixed (together with the couplings),
pumping is not possible for $U=0$.
Observation of the $\D$-resonances it thus a particularly clear indication
that one has gained independent experimental control of the interaction,
even when it is too small to be detected in stationary DC spectroscopy. 

We first consider the pumping curvature as a function driving parameters $U$ and $V_\b$ in \Fig{fig:bias-int}.
The left panels are for static gate voltage $\epsilon - \mu<0$ (such that $\braket{N}=2$ at the origin of the plane);
the right panels are for static gate voltage $\epsilon - \mu>0$ (such that $\braket{N}=0$ at the origin of the plane).
When the static gate voltage $\epsilon$ is reduced to zero, the $\D$-resonances seen in \Fig{fig:bias-int} move towards $V_\b=0$ where their amplitude vanishes (not shown).
We also observe that the qualitative effect of driving $V_\b$ does not depend on the static value of $\epsilon$  or the working-point value of $V_\b$: inverting the sign of either leaves the sign at a $\D$-resonance unaltered, in contrast to the $\B$-resonances.

The new $\D$-mechanism that is specific to driving $U$ also shows up when driving $U$ and $\epsilon$,
see \Fig{fig:gate-int}. We can map all the $\D$ features occurring here to the previous ones:
\begin{subequations}\begin{align}
		\F_{U,V_\b}(\epsilon)
		&
		\approx
		\M_{\D_1^-} [ U,\epsilon-\mu^\L ]
		\approx
		- \tfrac{1}{2} \F_{U, \epsilon}(V_\b)
		,
		\\
		\F_{U,V_\b}(\epsilon)
		&
		\approx
		\M_{\D_2^+} [ U,\epsilon-\mu^\L ]
		\approx
		- \tfrac{1}{2} \F_{U, \epsilon}(V_\b)
		,
		\\
		\F_{U,V_\b}(\epsilon)
		&
		\approx
		\M_{\D_1^+} [ U,\epsilon-\mu^\R ]
		\approx
		\phantom{+}
		\tfrac{1}{2} \F_{U, \epsilon}(V_\b)
		,
		\\
		\F_{U,V_\b}(\epsilon)
		&
		\approx
		\M_{\D_2^-} [ U,\epsilon-\mu^\R ]
		\approx
		\phantom{+}
		\tfrac{1}{2} \F_{U, \epsilon}(V_\b)
		,
\end{align}\end{subequations}
using $\mu^{\L,\R} = \mu \pm V_\b/2$ (see \App{app:effective} for details).
In this case, however, the qualitative effect of driving $\epsilon$ depends
both on the static value of $V_\b$ and the working-point value of $\epsilon$:
inverting the sign of either reverses the sign at a $\D$-resonances.\footnote
{As the static bias is reduced to $|V_\b| \ll T$  all resonances seen in \Fig{fig:gate-int}(a) merge at the working point $U=0$ and $\epsilon=0$ (not shown). Notably, to have nonzero curvature at that point the coupling needs to be asymmetric (otherwise pumping is prohibited by spatial symmetry).}

We now discuss how the remaining features in Figs.~\ref{fig:bias-int} and \ref{fig:gate-int} map to pumping features due to \textit{static, negative or positive} interaction $U$. Let us start with mechanisms $\A$. There is no feature due to mechanism $\A_1$ since $U$ does not enter any of its effective parameters.
Mechanism $\A_2$  can be accessed by driving $U$ and $V_\b$ and it occurs around the point $U = -(\epsilon-\mu) > 0$ and $V_\b=0$. It relates to driving with a static $U$ via
\begin{align}
	\F_{U,V_b}(\epsilon)
	&
	\approx
	\M_{\A_2}[\epsilon+U-\mu,V_\b]
	 \approx
	 \F_{\epsilon,V_b}(U)
	\label{eq:relation-A2}
	.
\end{align}
In contrast to mechanism $\A$ which always involves only one transition energy, at the $\B$-points the large bias voltage $|V_\b| \approx U$
generates nonequilibrium populations of all states, thereby 'coupling' the pumping responses of the $\epsilon$ and $\epsilon+U$ transitions. This is of interest since it allows for pumping with $\epsilon$ and $U$ as independent driving parameters (in contrast to a number of previous cases where we found that $U$ may effectively act as a gate voltage).
We therefore now have a relation  to the static $U$ case both when driving $U$ and $V_\b$,
\begin{align}
	\F_{U,V_\b}(\epsilon)
	&
	\approx
	\M_{\B^{+}} [\epsilon + U-\mu^\L, \epsilon-\mu^\R]
	\approx
	\tfrac{1}{2} \F_{\epsilon,V_\b}(U)
	,
	\notag
	\\
	\F_{U,V_\b}(\epsilon)
	&
	\approx
	\M_{\B^{-}} [\epsilon + U-\mu^\R,\epsilon-\mu^\L]
	\approx
	\tfrac{1}{2} \F_{\epsilon,V_\b}(U)
	\label{eq:relation-B}
	.
\end{align}
as well as when driving $U$ and $\epsilon$,
\begin{align}
	\F_{U,\epsilon}(V_\b)
	&
	\approx
	\M_{\B^{+}} [\epsilon + U-\mu^\L,\epsilon-\mu^\R ]
	\approx
	\phantom{-}
	\F_{\epsilon,V_\b}(U)
	,
	\notag
	\\
	\F_{U,\epsilon}(V_\b)
	&
	\approx
	\M_{\B^{-}} [\epsilon + U-\mu^\R,\epsilon-\mu^\L ]
	\approx
	-
	\F_{\epsilon,V_\b}(U)
	\label{eq:relation-B-again}
	.
\end{align}
Here, the factor of 2 between the two curvatures in \eq{eq:relation-B} stems from
a coupled transformation of parameters, see \App{app:effective} for details. The pumping response due to mechanism $\B$ at attractive interaction is again completely absent, as explained in section~\ref{sec:absentB}.

For driven interaction, mechanism $\C$ can again only be accessed by driving the bias voltage as a second parameter.
It is operative around the working point $U = - 2(\epsilon-\mu) < 0$ and $V_\b=0$ and obeys the relation
\begin{align}
	\F_{U,V_\b} (\epsilon)
	&
	\approx
	\M_{\C}[ \epsilon+\tfrac{1}{2}U-\mu, V_\b ]
	\approx
	\tfrac{1}{2} \F_{\epsilon,V_\b}(U)
	\label{eq:relation-C}
	.
\end{align}
The function $F_{\epsilon,V_\b}(U)$ was explicitly written in \Eq{eq:shape}.

The relations \eq{eq:relation-A2}, \eq{eq:relation-B} and \eq{eq:relation-C} express that around the discussed resonances the two factors that make up the pumping curvature \eq{eq:Fall} locally show the same structure as in the cases discussed earlier on, see \Fig{fig:bias-int}(b)-(c).
In particular, the vertical line with a kink in the plot of $\braket{N}$ and the corresponding pattern in the right panels of \Fig{fig:gate-bias}(b)-(c) for the ratio \eq{eq:decayratio}, can be clearly identified, even though we are plotting as a function of the interaction $U$ and not the gate voltage.

\subsubsection{Summary}

Driving two parameters with fixed coupling
shows a rich set of pumping mechanisms
as compared to protocols in which one coupling is driven.
Some resonances appear at equilibrium working points ($\A$),
where the pumping may dominate the transferred charge,
whereas others arise at strong nonequilibrium ($\B$),
where one is 'pumping with/against the flow' of the instantaneous current, \Eq{eq:DeltaNi}.
We have shown that the pumping mechanism $\C$ is specific to the physics of the '\emph{attractive} Coulomb blockade'.
Moreover, the new pumping mechanism $\D$ is unique to \emph{driven interaction}.
Remarkably, its response arises at working points where the static small $|U| \ll T$ forbids pumping with other parameters ($\epsilon$, $V_\b$).

\subsection{Pumped charge from integrated curvature\label{sec:monotonous}}

In the previous two sections, we discussed the curvature and its qualitative differences between different driving protocols. We now turn to the pumped charge that can be obtained from it
by integrating the curvature over the area of the driving cycle in the plane of driving parameters.
We stress again, that whenever the same mechanism is at work, for corresponding driving cycles of its effective parameters, the pumped charge will always be identical regardless of the actual experimental protocol used to realize it.

\paragraph{Coupling driving.}
\Fig{fig:coup-curv} shows that the curvature has either a constant sign or an alternating signs depending on the second driving parameter.
For  gate voltage $V_\g$ as well as the interaction $U$ being the second parameter, increasing the driving amplitude of both driven parameters will result in a monotonically increasing pumped charge $\Delta N^\R$. 
When increasing only the amplitude of $V_\g$ or $U$ (for fixed $\Gamma^\R$ amplitude) the pumped charge eventually saturates when all resonance-lines are covered by the driving cycle.\footnote
	{To maintain the slow driving condition for large amplitude, the frequency needs to be reduced accordingly, see \Ref{Pluecker17a}.}.
The amplitude for which this happens depends on the other parameters.
In contrast,
as a consequence of the sign changes of the curvature in \Fig{fig:coup-bias-curv}, the dependence of the pumped charge $\Delta N^\R$ on the $V_\b$-driving amplitude is not monotonic:
instead of saturating it may even approach zero depending on the driving cycle.

\paragraph{Gate and bias driving.}
For repulsive interaction [left panel of \Fig{fig:gate-bias-B}], around each resonance point, the curvature has a definite sign ('monopolar' character). Thus the pumped charge initially increases monotonically and already when the driving amplitude of both voltages is large on the thermal scale $T$ the pumped charge saturates at an intermediate plateau. However, since these points come in pairs with opposite sign and thus eventually, the pumped charge decreases again for amplitudes exceeding the \emph{interaction} energy $U$ and finally goes to zero:
\begin{align}
	\int  dS \F_{\epsilon,V_\b} = 0
	,
	\label{eq:zero}
\end{align}
This has been connected to the electron-hole symmetry of the single-dot model\footnote
	{See relation of  (A8a) and (A13a) in \Ref{Calvo12a}.}.
Quantitative relations between the pumped charges of the $\A$ and $\B$ mechanisms where already discussed in detail in \Ref{Calvo12a}.

For attractive interaction, (right panel of \Fig{fig:gate-bias-B}), the feature resulting from the $\C$ mechanism has an very different, 'dipolar' character.
This implies that the pumped charge depends nonmonotonically on the driving amplitude and goes to zero already when the amplitude exceeds the \emph{thermal} energy $T$.
For symmetric coupling the contribution from just one of the lobes of the $\C$ resonance can be obtained exactly using from our explicit result \eq{eq:shape}:
\begin{align}
	\int
	d S \F_{\epsilon,V_\b} = \tfrac{1}{2}
	\label{eq:one-half}
	.
\end{align}
Experimentally, this implies a characteristic feature of a net pumping of 1/2 an electron per cycle for a sufficiently large driving curve that passes through the node of the resonance, tangent to the nodal line.

\paragraph{Interaction driving.}
Finally, the new mechanism $\D$ unique to driving the interaction
has two curvature resonances of the same sign in \Fig{fig:bias-int-B}.
In combination with the other resonances this leads nonmonotonic behavior of the pumped charge
depending on the chosen working point.
In contrast, in \Fig{fig:gate-int-B} the $\D$ resonances resonances with opposite signs and are the sole cause of nonmonotic behavior.\section{Pumping response -- double dot\label{sec:results-real}}

In this final section, we discuss the pumping response for the double-dot model \eq{eq:H2}, which
only differs from the single dot by level-degeneracy factors
[\Eq{eq:Walpha2} replaces \eq{eq:Walpha1}].
Since the orbital index in the double dot plays the role of the spin in the single dot
(both labeled by $\sigma$),
the degeneracy difference is entirely due to the \emph{real spin} of the double dot ($\tau$).
In contrast to stationary transport, where the additional spin degeneracy would only lead to quantitative changes, for the pumping response this leads to \emph{qualitative} changes relative to the single-dot model
and in particular to a much more complicated curvature formula [\Eq{eq:Falpha-result}].
For pumping, replacing $N=\sum_{\sigma } d^\dagger_\sigma d_\sigma \to \sum_{\sigma\tau } d^\dagger_{\sigma\tau} d_{\sigma\tau}$
in \Eq{eq:H} is thus not an innocent operation.

\begin{figure}
	\subfigure[~Coupling-gate driving response $\F_{\epsilon,\Gamma^\R}$.]{
		\includegraphics[width=0.82\linewidth]{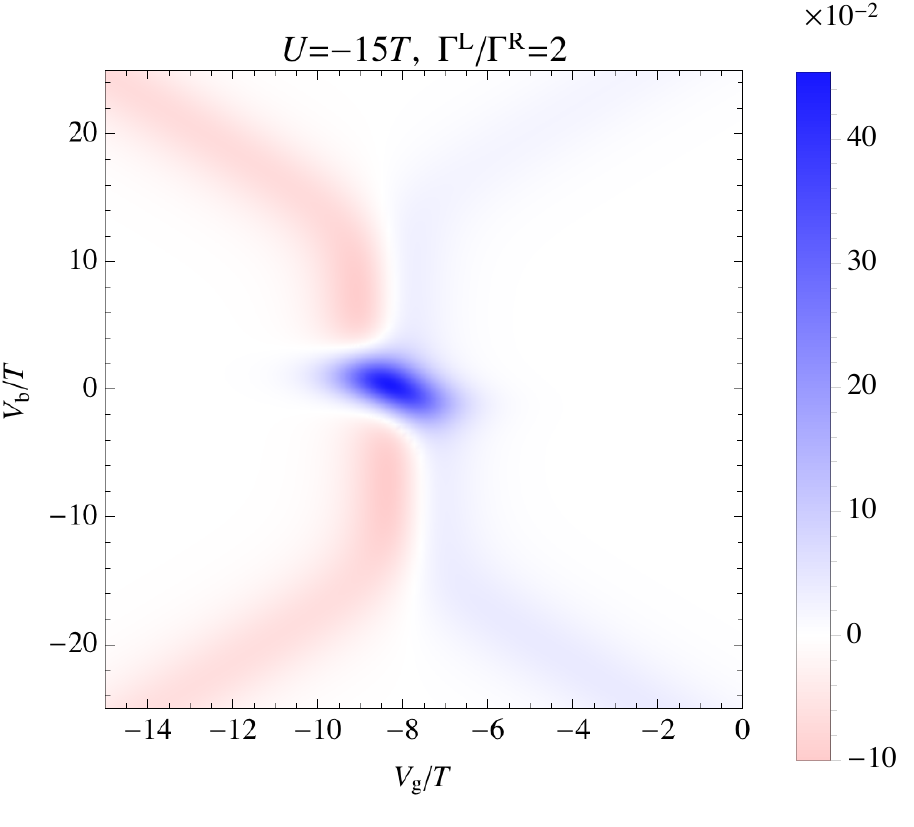}
		\label{fig:F2-coup-gate}
	}
	\\
	\subfigure[~Coupling-bias driving response $\F_{V_\b,\Gamma^\R}$.]{
		\includegraphics[width=0.82\linewidth]{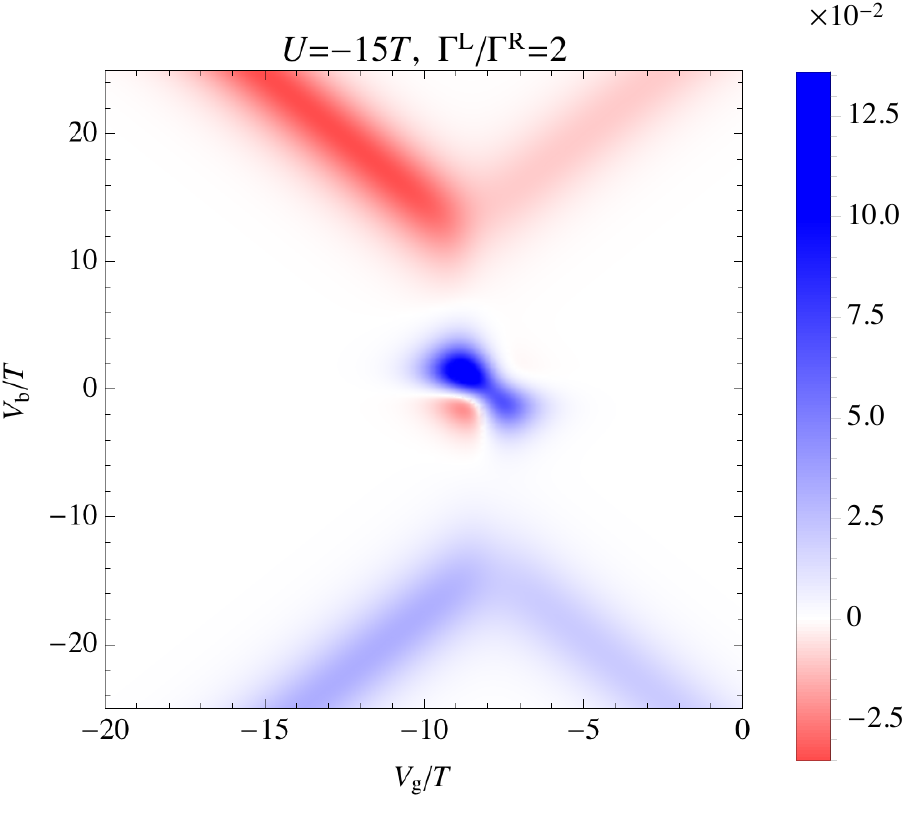}
		\label{fig:F2-coup-bias}
	}
	\\
	\subfigure[~Coupling-interaction driving response $\F_{U,\Gamma^\R}$.]{
		\includegraphics[width=0.82\linewidth]{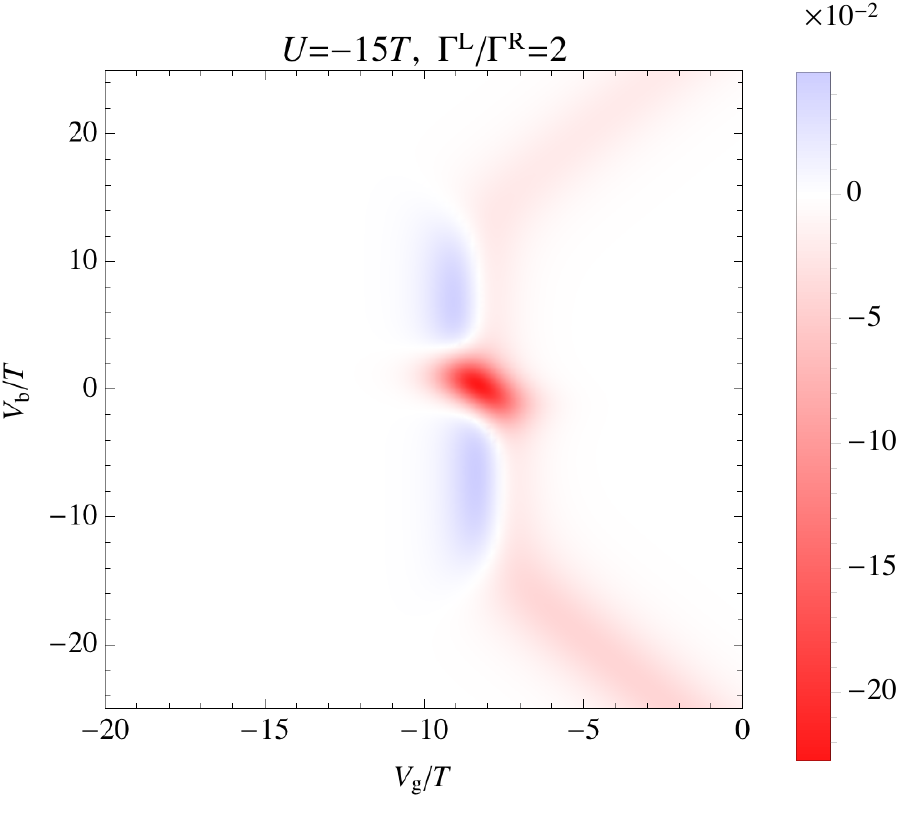}
		\label{fig:F2-coup-int}
	}
	\caption{Pumping curvatures for the double-dot system computed from \Eq{eq:Falpha-result}
		for driving protocols involving one of the couplings, $\Gamma^\R$.
		These correspond to right panels in \Fig{fig:coup} computed for the single-dot model using \Eq{eq:Falpha}.
		\label{fig:F2-drivencoup}
	}
\end{figure}

As before, we start by comparing the results for driving the coupling $\Gamma^\R/\Gamma^\L$ together with one second driving parameter
[\Sec{sec:coupling}].
For repulsive interaction $U>0$, the results (not shown) are qualitatively unaltered relative to the left panels of \Fig{fig:coup}.
Also for attractive interaction $U<0$ similarities persist: a comparison of the panels in \Fig{fig:F2-drivencoup} with the right panels in  \Fig{fig:coup} shows that the same mechanism still dominates the pumping response at low bias.
However, the curvature is also nonzero along lines, at which the occupation of the double dot changes.
The sign of the pumping response at these lines depends on the polarity of the gate voltage ($\epsilon - \mu$ relative to $-|U/2|$)
in \Fig{fig:F2-coup-gate} and \Fig{fig:F2-coup-int} or the bias polarity in \Fig{fig:F2-coup-int}.

Thus, when driving the coupling $\Gamma^\R/\Gamma^\L $ of the double dot, we find that even for attractive interaction, $U<0$, there is a non-vanishing pumping response, whenever the occupation changes. Exceptions to this are the missing lines at $\epsilon-\mu^\alpha$ in Fig.~\ref{fig:F2-drivencoup}, which are not accessible by driving $U$ as before in \Fig{fig:coup-curv}.
All together, this means that some of the intuition that holds for $U>0$ is restored.
The breakdown of this intuition for the attractive single-dot model is thus a result of its electron-hole symmetry. which causes in particular the resonance lines at large bias $V_\b>|U|$ to vanish.

\begin{figure*}
	\subfigure[~Gate-bias driving response $\F_{\epsilon,V_\b}$]{
		\includegraphics[width=0.4\linewidth]{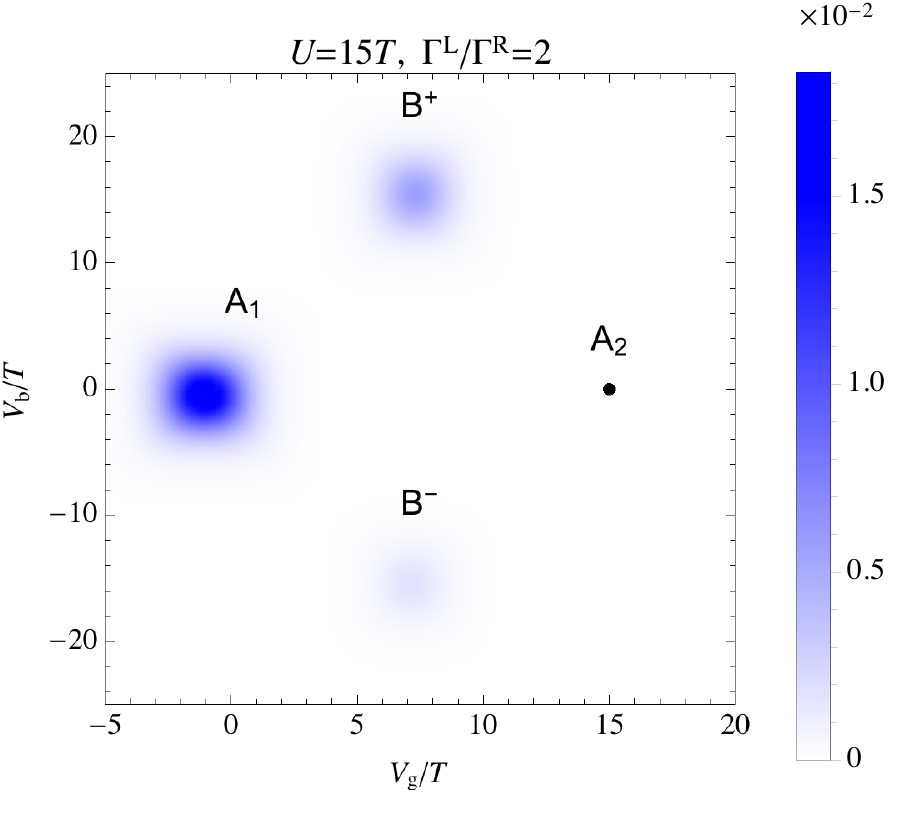}
		\includegraphics[width=0.4\linewidth]{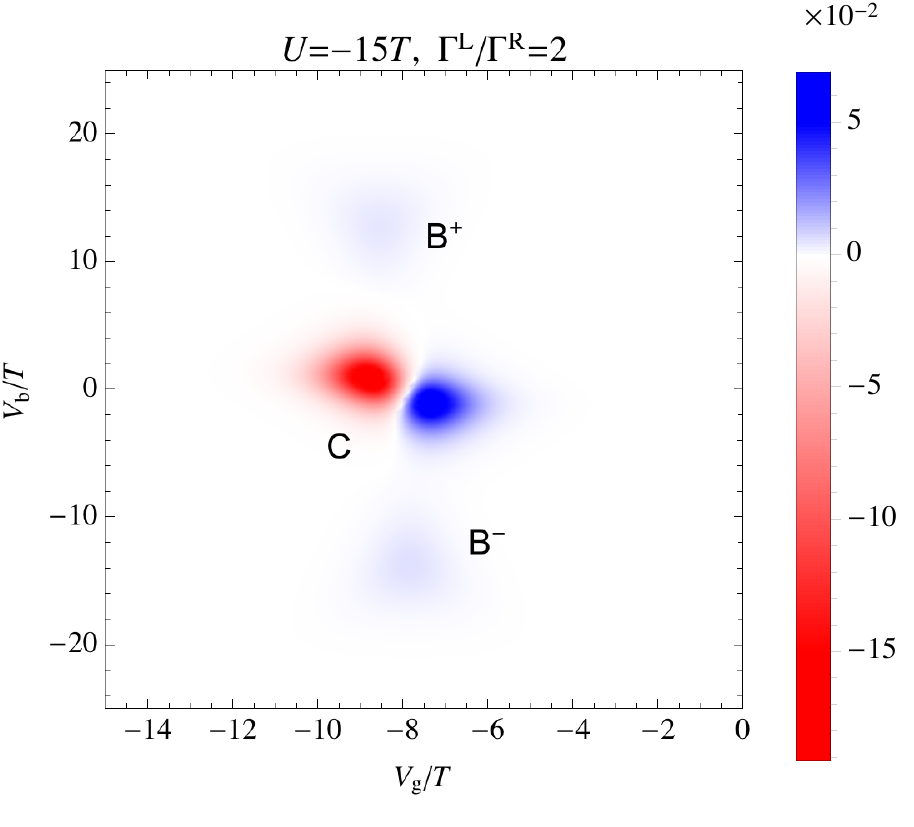}
		\label{fig:F2-gate-bias}
	}
	\subfigure[~Interaction-bias driving response $\F_{U,V_\b}$.]{
		\includegraphics[width=0.4\linewidth]{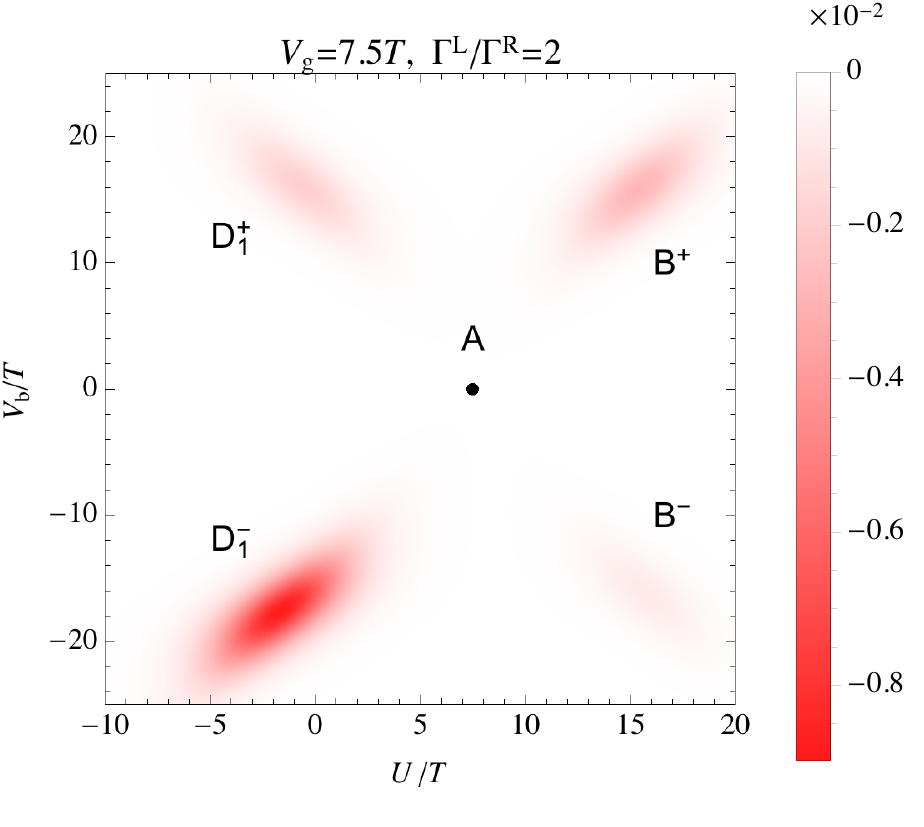}
		\includegraphics[width=0.4\linewidth]{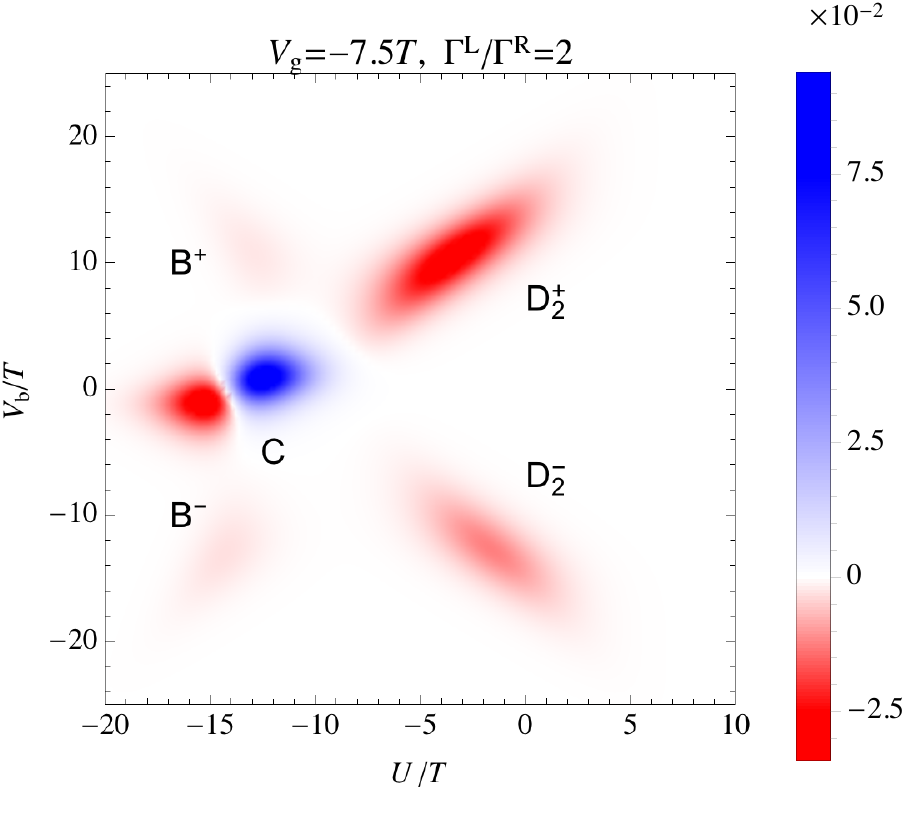}
		\label{fig:F2-int-bias}
	}
	\subfigure[~Interaction-gate driving response $\F_{U,V_\g}$]{
		\includegraphics[width=0.4\linewidth]{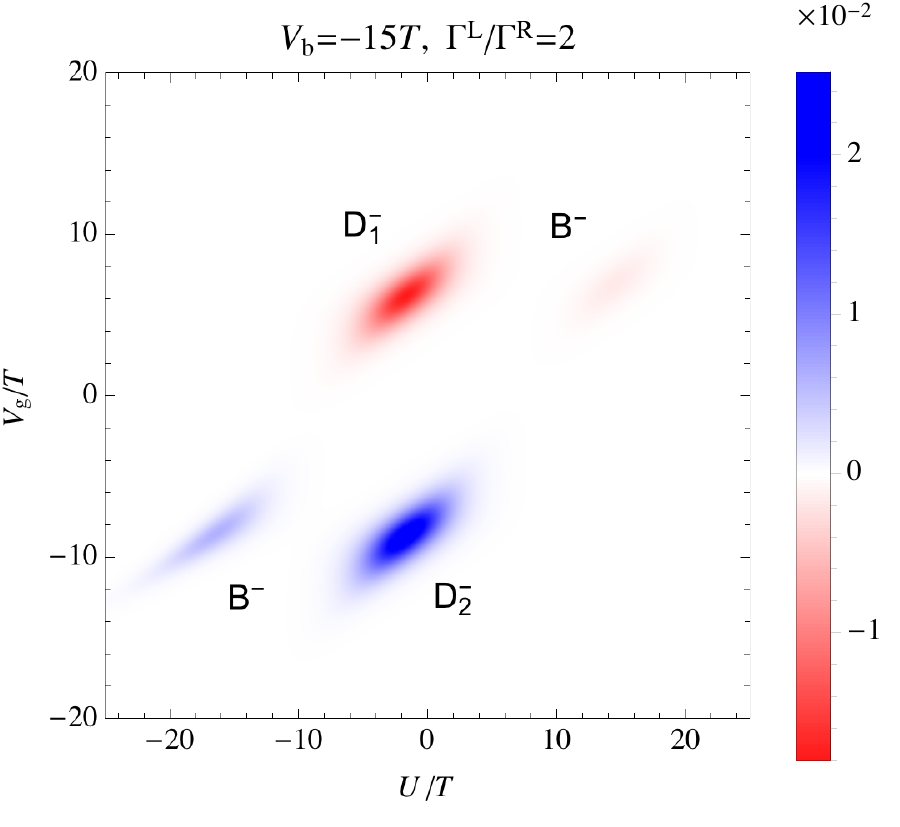}
		\includegraphics[width=0.4\linewidth]{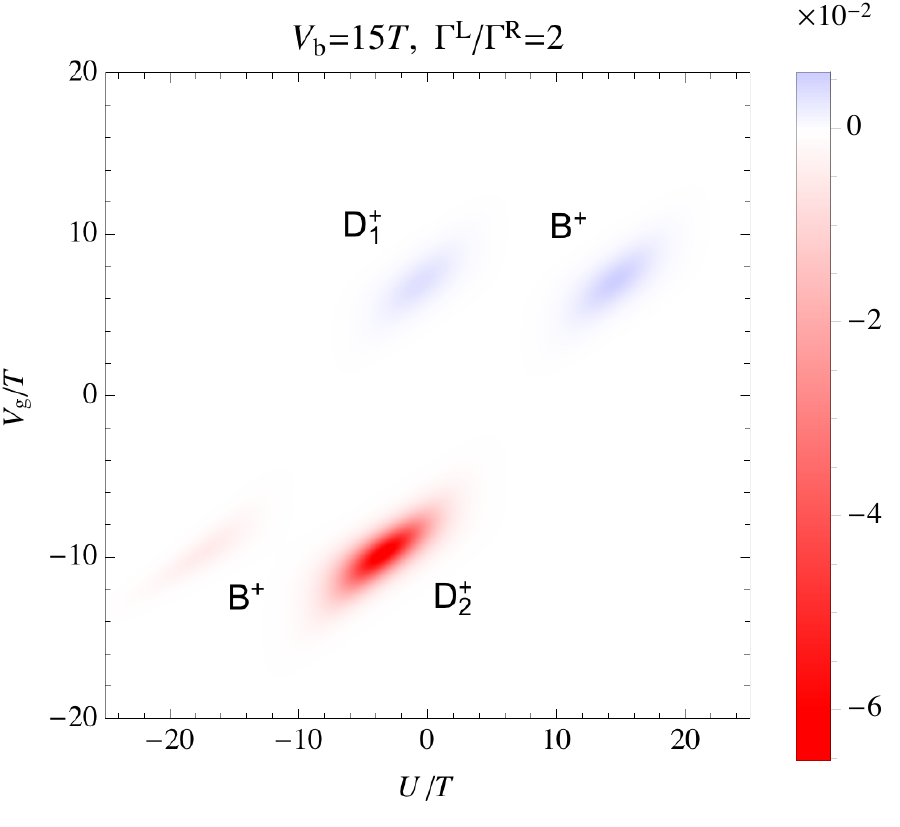}
		\label{fig:F2-int-gate}
	}
	\label{fig:F2-fixedcoup}
	\caption{Pumping curvatures for the double-dot system computed from \Eq{eq:Falpha-result}
		for driving protocols with fixed couplings $\Gamma^\L$ and $\Gamma^\R$.
		These correspond to the top panels of Figs.~\ref{fig:gate-bias}, \ref{fig:bias-int} and \ref{fig:gate-int} for a single-dot model which were computed using \Eq{eq:Falpha}.
	}
\end{figure*}

Next we analyze the impact of the additional spin degeneracy of the double dot
when driving two parameters at fixed couplings [\Sec{sec:nocoupling}].
Comparing results for repulsive interaction $U>0$ in the left panels of \Fig{fig:F2-gate-bias} and \Fig{fig:gate-bias-B} (gate and bias driving), one immediately notes the complete absence of a pumping response due to the $\A_2$ mechanism.
This qualitative difference is due to the equal degeneracy of the $N=1$ and $N=2$ charge states (both 4-fold degenerate): it was noted in \Ref{Reckermann10a} that the zero-bias resonances in the pumping curvature are sensitive to the \emph{change} in the degeneracy of the adjacent ground states. This makes pumping an interesting spectroscopy tool for quantum-dot systems~\cite{Calvo12a,Pluecker17a} independent of the DC stationary transport.
%

Another difference is that although the pumping responses due to the $\B^\pm$-mechanisms at large bias $V_\b \approx \pm U$ are still visible, their curvature values no longer have the same magnitude.
Depending on the coupling asymmetry $\Gamma^\R/\Gamma^\L$, they may even have the same sign as seen in the left panel of \Fig{fig:F2-gate-bias}.
For symmetric coupling $\Gamma^\L = \Gamma^\R$ both features at the $\B^\pm$-resonances survive with the same sign (not shown), in contrast to the single-dot case, where they both vanish due to electron-hole symmetry.

Comparing the results for attractive interaction $U<0$ in the right panels of \Fig{fig:F2-gate-bias} and \Fig{fig:gate-bias-B}, we note that the response due to the $\C$ mechanism  still dominates in the low bias regime, as in the single-dot model. However, the amplitudes of the two lobes now differ (note also the nonsymmetric color scale), even for symmetric coupling $\Gamma^\L = \Gamma^\R$ (not shown).
Qualitatively new is the non-vanishing pumping response due to the $\B$-mechanism. This response was suppressed in the single-dot case, see \Fig{fig:gate-bias-B} and \Eq{eq:reason1}-\eq{eq:reason2} ff.

Similar observations apply when comparing \Fig{fig:F2-int-bias} and \Fig{fig:bias-int-B} (interaction bias driving):
The $\A_2$ mechanism is again missing due to equal degeneracy of the $N=1,2$ state while for the same reason the $\B$-resonances appear\footnote
	{Going from \Fig{fig:F2-gate-bias} to \Fig{fig:F2-int-bias} the $\B$-resonance change sign, in accordance with the relation \eq{eq:relation-B} for the $\B$-mechanism.}, 
even for attractive interaction (right panels). Also as before, the magnitudes of the response due to the $\B$ mechanisms differ and the $\C$-resonance continues to dominate the low bias regime of attractive interaction, but with asymmetric lobes.
Importantly, the new $\D$-resonances --unique to interaction driving-- do not change qualitatively, although one should note the nonsymmetric color scale.

Finally, comparing \Fig{fig:F2-int-gate} and \Fig{fig:gate-int-B} (interaction gate driving), the $\B$-resonances appear also at working points with attractive interaction $U<0$, in contrast to the single-dot model.
 
\emph{Summary.} The real spin in the double dot indeed leads to three measurable qualitative deviations from the simpler Anderson model, all due to the now equal degeneracies for $N=1$ and $2$: 
the $\A_2$ mechanism becomes inoperative for $U>0$, the $\B$-mechanisms become operative even in the attractive interaction regime, and for repulsive interaction the $\B$ mechanism does no longer require nonsymmetric coupling.

\section{Summary\label{sec:summary}}

Motivated by recent experiments\cite{Cheng15,Tomczyk16a,Prawiroatmodjo17a,Hamo16a}
we have analyzed the pumping response of quantum dot systems with fully tuneable parameters,
in particular, in which the electron interaction can be statically tuned or even dynamically driven.
We have mapped out which possible mechanisms govern the pumping response for different experimentally realizable driving protocols.
The geometric formulation of the pumped charge in terms of curvatures
was a crucial tool for the understanding of the pumping mechanisms.

We here highlight two key results arising from our detailed analysis:
(i) \emph{Static attractive interaction} --nowadays accessible~\cite{Cheng15,Tomczyk16a,Hamo16a,Prawiroatmodjo17a}-- results in a novel characteristic pumping response ($\C$ mechanism)
whose characteristics and explanation are completely different from the repulsive case.
(ii) While we can show that \emph{driven interaction} is sometimes equivalent to driving of other parameters (gate or bias voltage), we also found a \textit{unique} pumping response ($\D$ mechanism) that cannot be realized without interaction  driving.
In all cases studied, we quantitatively demonstrated relations between the pumping responses of \emph{different driving protocols}
that are governed by the \emph{same pumping mechanism}.
These analytical relations between different geometric curvature components make precise the nontrivial difference between experimental driving parameters and the \emph{physical, effective} parameters that drive the electron pump.

Experimentally, the resulting pumping responses are observable both in a single quantum dot with real spin~\cite{Cheng15,Tomczyk16a,Prawiroatmodjo17a}
as well as in a double-dot with orbital pseudo-spin~\cite{Hamo16a}.
We, however, also identified pumping responses that are characteristic
of the additional real-spin degree of freedom of the double-dot model
(yielding a broken electron-hole symmetry).
These differences would remain undetected when comparing the stationary transport spectroscopy of the two systems.

Finally, it is noteworthy that 
pumping by the $\C$-mechanism (leading to a response at a two-particle resonance) is \emph{not} suppressed in the weak coupling limit. This is because it relies on tunnel rate asymmetries and \emph{not} on their magnitude.
Indeed, the pumping effects predicted here rely on leading-order tunneling process which were found to play a role at the two-particle resonance of an attractive quantum dot in a recent experiment~\cite{Prawiroatmodjo17a}.
Although corrections to pumping from next-to-leading order processes are of interest,
the mechanisms that we have described seem quite generic
and are expected to remain relevant for stronger tunneling.
Moreover, the effects do not rely on exact electron-hole symmetry, as our analysis of the double-dot case showed.

\acknowledgments

We thank Thibault Baquet  and Jens Schulenborg for helpful discussions. 
T.~P. was supported by the Deutsche Forschungsgemeinschaft (RTG 1995) and J.~S. by the Knut and Alice Wallenberg Foundation and the Swedish VR.

\appendix
\section{Elimination of spin from a single quantum dot}
\label{app:nospin}

Here we derive the master equations for the effective model \eq{eq:H2}.
The key point is to clarify the degeneracy factors that appear in the rate matrix \eq{eq:Walpha1}
due to the presence of the spin $\sigma= \pm$.
This procedure will be extended in \App{app:spin} to deal with the double-dot model.

\subsection{Master equation without spin}
The single quantum dot model has 4 possible states:
0-electron state $\ket{0}$,
four 1-electron states $\ket{\sigma}$ with $\sigma=+,, -$.
and a 2-electron state $\ket{2}=\ket{\uparrow\downarrow}$.
We consider a single reservoir and drop the superscript $\alpha$,
the general result follows by restoring this index and summing the rates over $\alpha$, i.e., we consider $W^\alpha$ in the decomposition of the kernel $W=\sum_{\alpha=\L,\R} W^\alpha$.
We start from the master equation for the occupation probabilities
\begin{subequations}\begin{align}
	\tfrac{d}{dt} \rho_{0}
	&=
	W_{0,0} \rho_{0}
	+
	\sum_{\sigma} W_{0,\sigma} \rho_{\sigma},
	\label{eq:master-0single}
	\\
	\tfrac{d}{dt} \rho_{\sigma}
	&=
	W_{\sigma,0} \rho_{0}
	+
	W_{\sigma,\sigma} \rho_{\sigma}
	+
	W_{\sigma, 2}  \, \rho_{2},
	\label{eq:master-1single}
	\\
	\tfrac{d}{dt} \rho_{2}
	&=
	\phantom{ W_{\sigma,0} \rho_{0}  }
		\sum_{\sigma} W_{ 2,  \sigma} \,  \rho_{\sigma}
	+
	W_{2,2} \,  \rho_{2},
	\label{eq:master-2single}	
\end{align}\end{subequations}
which is derived in the standard way assuming weak coupling and high temperature,
see, e.g., \Ref{Schulenborg16a}.
The diagonal entries are fixed to $W_{i,i} = -\sum_{f\neq i} W_{i,f}$ by trace preservation
where $i=0, \sigma$ or $2$.
In the main text we consider tunneling independent of the spin $\sigma$:
\begin{subequations}\begin{alignat}{2}
	W_{\sigma,0} & = W_{1,0}
	,
	&\quad
	W_{ 2,\sigma} & = W_{2,1}
	,
	\\
	W_{0,\sigma} & = W_{0,1} 
	,
	&
	W_{\sigma,2} &= W_{1,2}
	,
\end{alignat}\end{subequations}
where the right-hand sides are given in \Eq{eq:rates}.
Introducing the probability of single occupation,
\begin{align}
	\rho_{1}  := \sum_{\sigma} \rho_{\sigma}
	\label{eq:partial-prob1}
\end{align}
we integrate out the spin $\sigma$
by taking \Eq{eq:master-0single}, the linear combination $\sum_{\sigma}$ \Eq{eq:master-1single} and \Eq{eq:master-2single}:
\begin{align}
	\frac{d}{dt}
	\hspace{-4pt}
	\begin{bmatrix}
		\rho_0 \\ \rho_1 \\ \rho_2
	\end{bmatrix}
	=
	\begin{bmatrix}
		-2W_{10}^\alpha & W_{01}^\alpha                 & 0               \\
		2W_{10}^\alpha  & -W_{01}^\alpha-W_{21}^\alpha & 2W_{12}^\alpha   \\
		0              & W_{21}^\alpha                 & - 2W_{12}^\alpha
	\end{bmatrix}
	\hspace{-4pt}
	\begin{bmatrix}
		\rho_0 \\ \rho_1 \\ \rho_2
	\end{bmatrix}
\end{align}
where the diagonal entries are again dictated by trace preservation 
$W_{i,i} = -\sum_{f\neq i} W_{i,f}$
where now $i=0,1,2$.
Restoring the $\alpha$ index, this completes our derivation of the rate matrix \eq{eq:Walpha1}.
These degeneracy factors express that
the $N=0 \leftrightarrow 1$ transitions occur with ratio $2:1$  due to the spin degeneracy for $N=1$,
as do the $N=2\leftrightarrow 1$.
The doubling of rates $2W_{1,0}$  ($2 W_{1,2}$) occurs in the outer columns of the matrix because the spin provides two processes for the decay of state 0 (2) electrons.

The fact that the spin can be eliminated by introducing the $N=1$ occupation \eq{eq:partial-prob1} implies that (the relevant part of) the density operator
\begin{align}
	\rho = 
	&
	\sum_{N=0,1,2} \rho_{N} \Ket{N}
	\label{eq:rho-exp}
\end{align}
is confined to a linear subspace spanned by proper quantum states with $N=0,1$ or $2$ electrons:
whereas the 0- and 2-electron states are pure,
\begin{subequations}\begin{align}
		\Ket{0}:=\ket{0}\bra{0}
		,
		\quad
		\Ket{2}& :=	\ket{2}\bra{2}
\end{align}
the 1-electron state is maximally mixed
\begin{align}
		\Ket{1}& :=\tfrac{1}{2}
		\sum_{\sigma=\pm}
		\ket{\sigma}\bra{\sigma}
		.
\end{align}\label{eq:mixtures1}\end{subequations}
This statistical mixing simply expresses that due to the assumed spin-symmetries
the transport measurements are unable to detect the spin $\sigma$.
Each of these basis states is trace normalized, $\Braket{\one|N}=1$,
such that normalization is expressed as $\tr \hat{\rho}=\Braket{\one|\rho}=\sum_{N=0,1,2} \rho_N=1$.

\subsection{Current formula without spin}

The current flowing out of reservoir $\alpha$ into the dot is given by:
\begin{align}
		I_{N^\alpha}
		 = - \tfrac{d}{dt} \braket{N^{\alpha}}(t)
		 = \tr N W^{\alpha} \rho(t)
\label{eq:current-formula1}\end{align}
Where, in the first step, we used that $[N+N^{\alpha},H^{\text{T} \alpha}]=0$
for the coupling Hamiltonian decomposed as $H^\text{T}=\sum_\alpha H^{\text{T} \alpha}$,
see App. A of \Ref{Pluecker17a}.
The signs are chosen to agree with those of the master equation
$\tfrac{d}{dt} \rho(t) = (\sum_{\sigma\alpha} W^{\alpha\sigma} ) \rho(t)$.
An expression $\tr N \bullet$ as it appears in \Eq{eq:current-formula1} can be written in the same way
\begin{subequations}\begin{align}
		\tr N \rho
		=
		\Braket{N|\rho}
		=
		\sum_{n=0,1,2} n \rho_n
	\end{align}\label{eq:trNrho}\end{subequations}
due to the trace-normalization of the basis states $\Braket{N|n}=1$.
This gives
\begin{align}
	I_{N^\alpha}
	&
	= \Bra{N} W^{\alpha} \Ket{\rho}
	\\
	&= \begin{bmatrix}
		0 & 1 & 2
	\end{bmatrix}
	\begin{bmatrix}
		-2W_{10}^\alpha & W_{01}^\alpha                 & 0               \\
		2W_{10}^\alpha  & -W_{01}^\alpha-W_{21}^\alpha & 2W_{12}^\alpha   \\
		0              & W_{21}^\alpha                 & - 2W_{12}^\alpha
	\end{bmatrix}
	\begin{bmatrix}
		\rho_0 \\ \rho_1 \\ \rho_2
	\end{bmatrix}
	\notag
	\\
	&=
	2W_{10}^\alpha \rho_0 + (-W_{01}^\alpha+W_{21}^\alpha)\rho_1 - 2W_{12}^\alpha \rho_2
	\label{eq:current1}
\end{align}
Here the current contributions are enhanced by factors 2 respectively,
due to $\sigma$ degeneracy of the final state.
\section{Elimination of the real spin of the double dot\label{app:spin}}

Closely following \App{app:nospin}, we obtain the master equations and the current formula [\Eq{eq:master} and \Eq{eq:current} with rate matrix \Eq{eq:Walpha2}] for the double quantum dot,
highlighting the additional assumptions relative to \App{app:nospin}
and the role of real spin ($\tau$) degeneracy factors.
These constitute the essential difference to the single dot (not the pseudo spin $\sigma$!).

\subsection{Master equation without real- and pseudospin}

The double dot model \Eq{eq:H2} has 9 possible states:
one 0-electron state $\ket{0}$,
four 1-electron states $\ket{\sigma\tau}$ with one real spin $\tau$ in dot $\sigma$
and four 2-electron states $\ket{\tau\tau'}$ with spin $\tau$ in dot $\sigma=+$ and $\tau'$ in dot $\sigma = -$.
We excluded double occupation of the each dot by the very large (infinite) intradot interaction.
If the intradot interaction is not much larger then the interdot interaction, the experimental setup would just as well be able to invert the sign of the intradot interaction in a single quantum dot, simplifying matters significantly.
We also assumed negligible tunneling between the dots and therefore work with product states
$\ket{\tau\tau'}=\ket{\tau}\otimes\ket{\tau'}$.

As before, first consider a single reservoir, not writing a superscript $\alpha$,
and start from the master equation for the occupations of the 9 states:
\begin{widetext}\begin{subequations}\begin{align}
	\tfrac{d}{dt} \rho_{0}
	&=
	W_{0,0} \rho_{0}
	+
	\sum_{\sigma\tau} W_{0,\sigma\tau} \rho_{\sigma\tau}
	\label{eq:master-0}
	\\
	\tfrac{d}{dt} \rho_{\sigma\tau}
	&=
	W_{\sigma\tau,0} \rho_{0}
	+
	W_{\sigma\tau,\sigma\tau} \rho_{\sigma\tau}
	+
	\delta_{\sigma + }\sum_{\tau'} W_{+\tau, \tau \tau'}  \, \rho_{\tau \tau'}
	+
	\delta_{\sigma - }\sum_{\tau'} W_{-\tau, \tau' \tau}  \, \rho_{\tau'\tau}
	\label{eq:master-1}
	\\
	\tfrac{d}{dt} \rho_{\tau,\tau'}
	&=
	\phantom{ W_{\sigma\tau,0} \rho_{0}  }
		W_{ \tau\tau',  +\tau} \,  \rho_{+\tau}
	+W_{ \tau\tau',  -\tau'} \,  \rho_{-\tau'}
	+
	W_{\tau \tau',  \tau\tau'} \,  \rho_{\tau \tau'}
	\label{eq:master-2}	
\end{align}\end{subequations}\end{widetext}
with $W_{i,i} = -\sum_{f\neq i} W_{i,f}$ for $i=0, \sigma\tau$ or $\tau\tau'$.
The assumptions made in the main text that the tunneling
(i) of each dot $\sigma$ to the left/right side ($\alpha$) is the same
and
(ii) independent of the real spin $\tau$
imply [\Eq{eq:rates}]
\begin{subequations}\begin{alignat}{2}
	W_{\sigma\tau,0} & = W_{1,0}
	,
	& \quad
	W_{ \tau \tau',  + \tau} & = W_{ \tau \tau',  - \tau'}= W_{2,1}
	,
	\\
	 W_{0,\sigma\tau} & = W_{0,1} 
	,
	&
	W_{+\tau, \tau\tau'} &= W_{-\tau', \tau\tau'} =  W_{1,2}
	.
\end{alignat}\end{subequations}
This allows us to integrate out the real spin $\tau$ and the pseudo spin $\sigma$
by introducing partial sums of probabilities
\begin{align}
	\rho_{1}  := \sum_{\sigma} \sum_{\tau} \rho_{\sigma\tau}
	,
	\qquad
	\rho_{2}  := \sum_{\tau\tau'} \rho_{\tau \tau'}
	\label{eq:partial-prob}
	,
\end{align}
and taking the linear combinations
\Eq{eq:master-0}, $\sum_{\sigma\tau}$ \Eq{eq:master-1} and $\sum_{\tau\tau'}$ \Eq{eq:master-2}:
\begin{align}
	\frac{d}{dt}
	\hspace{-4pt}
	\begin{bmatrix}
		\rho_0 \\ \rho_1 \\ \rho_2
	\end{bmatrix}
	=
	\begin{bmatrix}
		-4W_{10}^\alpha & W_{01}^\alpha                 & 0               \\
		4W_{10}^\alpha  & -W_{01}^\alpha-2W_{21}^\alpha & 2W_{12}^\alpha   \\
		0              & 2W_{21}^\alpha                 & - 2W_{12}^\alpha
	\end{bmatrix}
	\hspace{-4pt}
	\begin{bmatrix}
		\rho_0 \\ \rho_1 \\ \rho_2
	\end{bmatrix}
\end{align}
with $W_{i,i} = -\sum_{f\neq i} W_{i,f}$ for  $i=0,1,2$.
Restoring the $\alpha$ index, this completes our derivation of the rate matrix $W^\alpha$ given in \Eq{eq:Walpha2}.

The degeneracy factors express that
the $N=0 \leftrightarrow 1$ transitions occur with ratio $4:1$  due to real and pseudo spin for $N=1$
whereas
the $N=2\leftrightarrow 1$ transitions occur with equal
ratio $2:2$ due having two real spins for $N=2$ and one real spin and one pseudo spin for $N=1$.

Also in this case, the introduction of partial sums of probabilities \eq{eq:partial-prob} implies that the density operator can expanded trace-normalized basis states as in \Eq{eq:rho-exp}
Although the 0-electron state is still pure, now both the 1- and 2-electron states are maximally mixed
\begin{subequations}\begin{align}
	\Ket{1}& :=\tfrac{1}{4}
	\sum_{\sigma=\pm} \sum_{\tau=\uparrow, \downarrow}
	\ket{\sigma \tau}\bra{\sigma \tau}
	,
	\\
	\Ket{2}& :=	\tfrac{1}{4}
	\sum_{\tau,\tau'=\uparrow, \downarrow}
 \ket{\tau \tau'}\bra{\tau \tau'}
	 \label{eq:2}
	 .
\end{align}\label{eq:mixtures}\end{subequations}
The statistical mixing now expresses that due to the assumed spin- and spatial symmetries
the transport measurements are able to detect neither the real spin $\tau$ nor the pseudo spin $\sigma$.
Note that the four 2-electron states are degenerate (the dots are not tunnel-coupled but only capacitively coupled) which rules out any spin-exchange effects. Indeed, \Eq{eq:2} can also be written as a statistical mixture of singlet and triplet states.

\subsection{Current formula without real- and pseudo-spin}

The sum of currents flowing out of reservoir $\alpha$ into both dots $\sigma=\pm$ is
\begin{subequations}\begin{align}
	I_{N^\alpha}
	& = -  \tfrac{d}{dt} \braket{N^{\alpha}}(t)
	\\
	& = \sum_{\sigma}  \tr N^{\sigma} W^{\alpha\sigma} \rho(t)
	 = \tr N W^{\alpha} \rho(t)
	 .
\end{align}\label{eq:current-formula2}\end{subequations}
Where now we used that $[\sum_\sigma N_\sigma+N^{\alpha},H^{\text{T},\alpha}]=0$ when decomposing the coupling as $H^\text{T} =\sum_{\alpha} H^{\text{T} \alpha}$.
We also decomposed $W=\sum_{\alpha\sigma} W^{\alpha\sigma}$ into contributions involving dot $\sigma$ and reservoir $\alpha$.
In the second step, we assumed the coupling strength on the $\alpha$-side to be the same for each dot,
such that $W^{\alpha\sigma}=W^{\alpha}$.
Also in this case \Eq{eq:current1} holds due to the trace-normalization of the basis states $\Braket{N|n}=1$, with the same modified rate matrix:
\begin{align}
	I_{N^\alpha}
	& = \tr N W^{\alpha} \rho
	  = \Bra{N} W^{\alpha} \Ket{\rho}
	\\
	&= \begin{bmatrix}
		0 & 1 & 2
	\end{bmatrix}
	\begin{bmatrix}
		-4W_{10}^\alpha & W_{01}^\alpha                 & 0               \\
		4W_{10}^\alpha  & -W_{01}^\alpha-2W_{21}^\alpha & 2W_{12}^\alpha   \\
		0              & 2W_{21}^\alpha                 & - 2W_{12}^\alpha
	\end{bmatrix}
	\begin{bmatrix}
		\rho_0 \\ \rho_1 \\ \rho_2
	\end{bmatrix}
	\notag
	\\
	&=
	4W_{10}^\alpha \rho_0 + (-W_{01}^\alpha+2W_{21}^\alpha)\rho_1 - 2W_{12}^\alpha \rho_2
\end{align}
Now the current contributions are enhanced by factors 4, 2 and 2 respectively,
due to $\sigma \tau$, $\tau$ and $\sigma$ degeneracy of the final state.\section{Curvature formula\label{app:curvature}}

In this appendix we derive the key result \Eq{eq:Falpha-result} of the main text.
%
The adiabatic-response equations $W\Ket{\rho^\i}=0$ and $W\Ket{\rho^\a}= d/dt\Ket{\rho^\i}$ for both cases [\Eq{eq:master} with rates \eq{eq:Walpha2} or \eq{eq:Walpha1}] can be written in the same form
\begin{align}
	\begin{bmatrix}
		0 \\ 0 \\ 0
	\end{bmatrix}
	&=
	\begin{bmatrix}
		-\W_{10} & \W_{01}                 & 0               \\
		\W_{10}  & -\W_{01}-\W_{21} & \W_{12}   \\
		0              & \W_{21}                 & - \W_{12}
	\end{bmatrix}
	\begin{bmatrix}
		\rho_0 \\ \rho_1 \\ \rho_2
	\end{bmatrix}
	\label{eq:masterstatbar}
	\\
	\frac{d}{dt}
	\begin{bmatrix}
		\rho_0 \\ \rho_1 \\ \rho_2
	\end{bmatrix}
	&=
	\begin{bmatrix}
		-\W_{10} & \W_{01}          & 0         \\
		\W_{10}  & -\W_{01}-\W_{21} & \W_{12}   \\
		0        & \W_{21}          & - \W_{12}
	\end{bmatrix}
	\begin{bmatrix}
		\rho_0^\a \\ 	\rho_1^\a \\ 	\rho_2^\a
	\end{bmatrix}
	\label{eq:masterrespbar}	
\end{align}
by absorbing the degeneracy factors into the rates with an overbar.
The corresponding formulas for the response part of the current [\Eq{eq:current} with rates \eq{eq:Walpha2} or \eq{eq:Walpha1}] read
\begin{align}
	I_{N^\alpha}^\a
	& = \Bra{N} W^{\alpha} \Ket{\rho^\a}
	\\
	&= \begin{bmatrix}
		0 & 1 & 2
	\end{bmatrix}
	\begin{bmatrix}
		-\W_{10}^\alpha & \W_{01}^\alpha                 & 0               \\
		\W_{10}^\alpha  & -\W_{01}^\alpha-\W_{21}^\alpha & \W_{12}^\alpha   \\
		0              & \W_{21}^\alpha                 & - \W_{12}^\alpha
	\end{bmatrix}
	\begin{bmatrix}
		\rho_0^\a \\ \rho_1^\a \\ \rho_2^\a
	\end{bmatrix}
	\notag
	\\
	&=
	\W_{10}^\alpha \rho_0^\a + (-\W_{01}^\alpha+\W_{21}^\alpha)\rho_1^\a - \W_{12}^\alpha \rho_2^\a
\end{align}
Using trace normalization, these equations can be reduced to formulas involving only $2 \times 2$ matrices and vectors.
From \Eq{eq:masterstatbar} we eliminate $\rho_1 = 1 - \rho_0 - \rho_2$
\begin{align}
	\begin{bmatrix}
		- \W_{01} \\ - \W_{21}
	\end{bmatrix}
	&=
	\begin{bmatrix}
		-\W_{10}- \W_{01} & - \W_{01}          \\
		- \W_{21}         & - \W_{21} -\W_{12}
	\end{bmatrix}
	\begin{bmatrix}
		\rho_0  \\ \rho_2
	\end{bmatrix}
\end{align}
and from \Eq{eq:masterrespbar} we eliminate $\rho_1^\a =  - \rho_0^\a - \rho_2^\a$
\begin{align}
	\frac{d}{dt}
	\begin{bmatrix}
		\rho_0  \\ \rho_2
	\end{bmatrix}	
	&=
	\begin{bmatrix}
		-\W_{10}- \W_{01} & - \W_{01}          \\
		- \W_{21}         & - \W_{21} -\W_{12}
	\end{bmatrix}
	\begin{bmatrix}
		\rho_0^\a  \\ 	\rho_2^\a
	\end{bmatrix}	
\end{align}
Similarly, the response-current formula reduces to
\begin{align}
	I_{N^\alpha}^\a
	&= \begin{bmatrix}
		-1 & 1 
	\end{bmatrix}
	\begin{bmatrix}
		-\W_{10}^\alpha- \W_{01}^\alpha & - \W_{01}^\alpha          \\
		- \W_{21}^\alpha         & - \W_{21}^\alpha -\W_{12}^\alpha
	\end{bmatrix}
	\begin{bmatrix}
		\rho_0^\a \\  \rho_2^\a
	\end{bmatrix}
	\notag
	\\
	&=
	( \W_{10}^\alpha + \W_{01}^\alpha ) \rho_0^\a  - ( \W_{21}^\alpha+\W_{12}^\alpha ) \rho_2^\a
\end{align}

Solving these three equations [This amounts to the calculation of the pseudo inverse $W^{-1}$ in \Eq{eq:DeltaN}] one obtains after some algebra an expression which can be written as
$I_{N^\alpha}^\a = A^\alpha d{R}/dt$ where $A^\alpha$ is the pumping connection.
The result \Eq{eq:Falpha-result} given in the main text then follows from $\F^\alpha = \nabla \times A^\alpha$.
The gradients in this expression can be evaluated more explicitly to give
\begin{widetext}\begin{subequations}\begin{align}
	\F^\alpha
	=
	\nabla \left\{ \frac{1}{ [\W_{10} \W_{12} + \W_{10} {\W_{21}} + {\W_{01}} \W_{12}]^3 }
	\begin{bmatrix}
	- (\W_{10}^{\alpha} + {\W_{01}^{\alpha} - \W_{21}^{\alpha}}) \W_{12} - (\W_{10}^{\alpha} + \W_{12}^{\alpha} ) {\W_{21}}
	\\
	- ({\W_{01}^{\alpha} - \W_{21}^{\alpha}} - \W_{12}^{\alpha}) \W_{10} + (\W_{10}^{\alpha} + \W_{12}^{\alpha} ) {\W_{01}}
	\end{bmatrix}
	\right\}
	\\
	\cdot \times
	\begin{bmatrix}
	(\nabla \W_{12} {\W_{01}}) \W_{10} [\W_{12} + {\W_{21}}] - \W_{12} {\W_{01}} \nabla \W_{10} [\W_{12} + {\W_{21}}]
	\\
	(\nabla \W_{10} {\W_{21}}) [\W_{10} + {\W_{01}}] \W_{12} - \W_{10} {\W_{21}} \nabla [\W_{10} + {\W_{01}}] \W_{12}
	\end{bmatrix}
\end{align}\end{subequations}
where $\cdot \times$ indicates that the scalar product of the column vectors and the cross product of the derivative operators $\nabla$.
Antisymmetrization gives the most explicit result for $\F:=(\F^\R-\F^\L)/2=\F^\R$:
\begin{align}
	\F
	 =
	&
\small{
	\nabla \left\{ \frac{1}{ [\W_{10} \W_{12} + \W_{10} {\W_{21}} + {\W_{01}} \W_{12}]^3 }
	\begin{bmatrix}
	- \tfrac{1}{2} (\W_{01}^\R - \W_{01}^\L) \W_{12} - \tfrac{1}{2} (\W_{10}^\R -\W_{10}^\L) (\W_{12} + \W_{21})
	+ (\W_{21}^\R \W_{12}^\L - \W_{21}^\L \W_{12}^\R)
	\\
	\phantom{-}
	\tfrac{1}{2} (\W_{21}^\R - \W_{21}^\L) \W_{10} + \tfrac{1}{2} (\W_{12}^\R - \W_{12}^\L)  (\W_{10} + \W_{01})
	 - (\W_{01}^\R \W_{10}^\L - \W_{01}^\L \W_{10}^\R)
	\end{bmatrix}
	\right\}
}	
	\notag
	\\
	& \cdot \times
\small{
	\begin{bmatrix}
		\big( \nabla \W_{12} {\W_{01}} \big) \W_{10} [\W_{12} + {\W_{21}}] - \W_{12} {\W_{01}} \nabla \big( \W_{10} [\W_{12} + {\W_{21}}] \big)
		\\
		\big( \nabla \W_{10} {\W_{21}} \big) [\W_{10} + {\W_{01}}] \W_{12} - \W_{10} {\W_{21}} \nabla \big( [\W_{10} + {\W_{01}}] \W_{12} \big)
	\end{bmatrix}
}
\end{align}
\end{widetext}


\section{Effective driving parameters\label{app:effective}}

Here, as an example, we derive the first relation of \eq{eq:relation-B}
\begin{align}
	\F_{U,V_\b}(\epsilon)
	&
	\approx
	\M_{\B^{+}} (\epsilon + U-\mu^\L, \epsilon-\mu^\R )
	\approx
	\tfrac{1}{2} \F_{\epsilon,V_\b}(U),
	\label{eq:last}
\end{align}
in order to indicate where the variety of prefactors in the relations between different curvatures components come from.
From the fact that the $\B^{+}$ mechanism dominates the response, one expects that the curvature is a function of the distance of the upper (lower) addition energy to the left (right) electrochemical potential. This can be written in two ways: either as a function of
$(U,V_\b)$ with fixed $\epsilon$
\begin{align}
	&
	\M_{\B^{+}} (\epsilon + U-\mu^\L, \epsilon-\mu^\R )
	\\
	&
	=
	\M_{\B^{+}} ( \epsilon + U/2-\mu, \epsilon-(\mu - \tfrac{1}{2} V_\b) )
	 =
	4 \F_{U,V_\b}(\epsilon)
\end{align}
with the inverse of the Jacobian $|\partial(\epsilon+U/2-\mu,\epsilon-\mu +V_\b/2)/ \partial(U,V)|=
\tfrac{1}{4}$,
or 
as function of $(\epsilon,V_\b)$ with fixed $U$
\begin{align}
	\M_{\B^{+}} ( \epsilon + U/2-\mu, \epsilon-(\mu - \tfrac{1}{2} V_\b) )
	& =
	2 \F_{\epsilon,V_\b}(U),
\end{align}
now using the inverse of $|\partial(\epsilon+U/2-\mu,\epsilon-\mu +V_\b/2)/ \partial(\epsilon,V)|=\tfrac{1}{2}$.
Thus, if the curvature components stem from a common mechanism they must be related as in \Eq{eq:last}.

Note that these and similar relations in the main text only hold true when the considered mechanism is well separated from others.
In each case they were verified on the analyicaly computed curvature components in the proper physical limits.

\end{document}